%% file: index.tex
\PassOptionsToPackage{unicode}{hyperref}
\PassOptionsToPackage{hyphens}{url}
\PassOptionsToPackage{dvipsnames,svgnames,x11names}{xcolor}
\documentclass[
  10pt,
  a4paper,
]{article}
\usepackage{xcolor}
\usepackage{amsmath,amssymb}
\usepackage{iftex}
\ifPDFTeX
  \usepackage[T1]{fontenc}
  \usepackage[utf8]{inputenc}
  \usepackage{textcomp} 
\else 
  \usepackage{unicode-math} 
  \defaultfontfeatures{Scale=MatchLowercase}
  \defaultfontfeatures[\rmfamily]{Ligatures=TeX,Scale=1}
\fi
\usepackage{lmodern}
\ifPDFTeX\else
\fi
\IfFileExists{upquote.sty}{\usepackage{upquote}}{}
\IfFileExists{microtype.sty}{
  \usepackage[]{microtype}
  \UseMicrotypeSet[protrusion]{basicmath} 
}{}
\makeatletter
\@ifundefined{KOMAClassName}{
  \IfFileExists{parskip.sty}{%
    \usepackage{parskip}
  }{
    \setlength{\parindent}{0pt}
    \setlength{\parskip}{6pt plus 2pt minus 1pt}}
}{
  \KOMAoptions{parskip=half}}
\makeatother
\makeatletter
\ifx\paragraph\undefined\else
  \let\oldparagraph\paragraph
  \renewcommand{\paragraph}{
    \@ifstar
      \xxxParagraphStar
      \xxxParagraphNoStar
  }
  \newcommand{\xxxParagraphStar}[1]{\oldparagraph*{#1}\mbox{}}
  \newcommand{\xxxParagraphNoStar}[1]{\oldparagraph{#1}\mbox{}}
\fi
\ifx\subparagraph\undefined\else
  \let\oldsubparagraph\subparagraph
  \renewcommand{\subparagraph}{
    \@ifstar
      \xxxSubParagraphStar
      \xxxSubParagraphNoStar
  }
  \newcommand{\xxxSubParagraphStar}[1]{\oldsubparagraph*{#1}\mbox{}}
  \newcommand{\xxxSubParagraphNoStar}[1]{\oldsubparagraph{#1}\mbox{}}
\fi
\makeatother

\usepackage{color}
\usepackage{fancyvrb}

\DefineVerbatimEnvironment{Highlighting}{Verbatim}{commandchars=\\\{\}}
\usepackage{framed}
\definecolor{shadecolor}{RGB}{241,243,245}
\newenvironment{Shaded}{\begin{snugshade}}{\end{snugshade}}

\newcommand{\AttributeTok}[1]{\textcolor[rgb]{0.40,0.45,0.13}{#1}}

\newcommand{\BuiltInTok}[1]{\textcolor[rgb]{0.00,0.23,0.31}{#1}}

\newcommand{\CommentTok}[1]{\textcolor[rgb]{0.37,0.37,0.37}{#1}}

\newcommand{\DataTypeTok}[1]{\textcolor[rgb]{0.68,0.00,0.00}{#1}}

\newcommand{\ExtensionTok}[1]{\textcolor[rgb]{0.00,0.23,0.31}{#1}}

\newcommand{\KeywordTok}[1]{\textcolor[rgb]{0.00,0.23,0.31}{\textbf{#1}}}
\newcommand{\NormalTok}[1]{\textcolor[rgb]{0.00,0.23,0.31}{#1}}
\newcommand{\OperatorTok}[1]{\textcolor[rgb]{0.37,0.37,0.37}{#1}}

\newcommand{\StringTok}[1]{\textcolor[rgb]{0.13,0.47,0.30}{#1}}

\usepackage{longtable,booktabs,array}
\usepackage{calc} 
\usepackage{etoolbox}
\makeatletter
\patchcmd\longtable{\par}{\if@noskipsec\mbox{}\fi\par}{}{}
\makeatother
\IfFileExists{footnotehyper.sty}{\usepackage{footnotehyper}}{\usepackage{footnote}}
\makesavenoteenv{longtable}
\usepackage{graphicx}
\makeatletter
\newsavebox\pandoc@box
\newcommand*\pandocbounded[1]{
  \sbox\pandoc@box{#1}%
  \Gscale@div\@tempa{\textheight}{\dimexpr\ht\pandoc@box+\dp\pandoc@box\relax}%
  \Gscale@div\@tempb{\linewidth}{\wd\pandoc@box}%
  \ifdim\@tempb\p@<\@tempa\p@\let\@tempa\@tempb\fi
  \ifdim\@tempa\p@<\p@\scalebox{\@tempa}{\usebox\pandoc@box}%
  \else\usebox{\pandoc@box}%
  \fi%
}
\def\fps@figure{htbp}
\makeatother

\NewDocumentCommand\citeproctext{}{}

\makeatletter
 \let\@cite@ofmt\@firstofone
 \def\@biblabel#1{}
 \def\@cite#1#2{{#1\if@tempswa , #2\fi}}
\makeatother
\newlength{\cslhangindent}
\newlength{\csllabelwidth}
\newenvironment{CSLReferences}[2] 
 {\begin{list}{}{%
  \setlength{\itemindent}{0pt}
  \setlength{\leftmargin}{0pt}
  \setlength{\parsep}{0pt}
  \ifodd #1
   \setlength{\leftmargin}{\cslhangindent}
   \setlength{\itemindent}{-1\cslhangindent}
  \fi
  \setlength{\itemsep}{#2\baselineskip}}}
 {\end{list}}
\usepackage{calc}

\providecommand{\tightlist}{%
  \setlength{\itemsep}{0pt}\setlength{\parskip}{0pt}}

\usepackage{fvextra}
\DefineVerbatimEnvironment{Highlighting}{Verbatim}{breaklines,breakanywhere,commandchars=\\\{\}}
\fvset{breaklines,breakanywhere}
\RecustomVerbatimEnvironment{verbatim}{Verbatim}{breaklines,breakanywhere}
\BeforeBeginEnvironment{verbatim}{\begin{Shaded}}
\AfterEndEnvironment{verbatim}{\end{Shaded}}
\let\oldunderscore\_
\renewcommand{\_}{\oldunderscore\allowbreak}
\usepackage{newunicodechar}
\newunicodechar{ć}{\'c}
\AtBeginDocument{\rhead{\scshape \runninghead}}
\usepackage{arxiv}
\usepackage{orcidlink}
\usepackage{amsmath}
\usepackage[T1]{fontenc}
\makeatletter
\@ifpackageloaded{tcolorbox}{}{\usepackage[skins,breakable]{tcolorbox}}
\@ifpackageloaded{fontawesome5}{}{\usepackage{fontawesome5}}
\definecolor{quarto-callout-color}{HTML}{909090}
\definecolor{quarto-callout-note-color}{HTML}{0758E5}
\definecolor{quarto-callout-important-color}{HTML}{CC1914}
\definecolor{quarto-callout-warning-color}{HTML}{EB9113}
\definecolor{quarto-callout-tip-color}{HTML}{00A047}
\definecolor{quarto-callout-caution-color}{HTML}{FC5300}
\definecolor{quarto-callout-color-frame}{HTML}{acacac}
\definecolor{quarto-callout-note-color-frame}{HTML}{4582ec}
\definecolor{quarto-callout-important-color-frame}{HTML}{d9534f}
\definecolor{quarto-callout-warning-color-frame}{HTML}{f0ad4e}
\definecolor{quarto-callout-tip-color-frame}{HTML}{02b875}
\definecolor{quarto-callout-caution-color-frame}{HTML}{fd7e14}
\makeatother
\makeatletter
\@ifpackageloaded{caption}{}{\usepackage{caption}}
\AtBeginDocument{%
\ifdefined\contentsname
  \renewcommand*\contentsname{Table of contents}
\else
  \newcommand\contentsname{Table of contents}
\fi
\ifdefined\listfigurename
  \renewcommand*\listfigurename{List of Figures}
\else
  \newcommand\listfigurename{List of Figures}
\fi
\ifdefined\listtablename
  \renewcommand*\listtablename{List of Tables}
\else
  \newcommand\listtablename{List of Tables}
\fi
\ifdefined\figurename
  \renewcommand*\figurename{Figure}
\else
  \newcommand\figurename{Figure}
\fi
\ifdefined\tablename
  \renewcommand*\tablename{Table}
\else
  \newcommand\tablename{Table}
\fi
}
\@ifpackageloaded{float}{}{\usepackage{float}}
\floatstyle{ruled}
\@ifundefined{c@chapter}{\newfloat{codelisting}{h}{lop}}{\newfloat{codelisting}{h}{lop}[chapter]}
\floatname{codelisting}{Listing}

\usepackage{amsthm}
\theoremstyle{definition}
\newtheorem{example}{Example}[section]
\theoremstyle{definition}
\newtheorem{definition}{Definition}[section]
\theoremstyle{remark}
\AtBeginDocument{}

\makeatother
\makeatletter
\@ifpackageloaded{caption}{}{\usepackage{caption}}
\@ifpackageloaded{subcaption}{}{\usepackage{subcaption}}
\makeatother
\usepackage{bookmark}
\IfFileExists{xurl.sty}{\usepackage{xurl}}{} 
\makeatletter
\@ifundefined{xmpquote}{\newcommand{\xmpquote}[1]{#1}}{}
\makeatother
\hypersetup{
  pdftitle={Regenerative Coding: Deleting Code Without Losing Knowledge?},
  pdfauthor={Thomas Bartz-Beielstein; Marcel Samuel Klimm; Alexander Hinterleitner},
  pdfkeywords={\xmpquote{regenerative coding}, \xmpquote{regenerative
software}, \xmpquote{babregen}, \xmpquote{deletion
test}, \xmpquote{local language models}, \xmpquote{small language
models}, \xmpquote{specification}, \xmpquote{test
oracle}, \xmpquote{mutation
analysis}, \xmpquote{provenance}, \xmpquote{vibe coding 3.0}},
  colorlinks=true,
  linkcolor={blue},
  filecolor={Maroon},
  citecolor={Blue},
  urlcolor={Blue},
  pdfcreator={LaTeX via pandoc}}

\renewcommand{\today}{2026-12-02}
\newcommand{\runninghead}{A Preprint }
\renewcommand{\runninghead}{Regenerative Coding: Deleting Code Without
Losing Knowledge? }
\newcommand{\docstatus}{\runninghead}
\renewcommand{\docstatus}{Working paper, version 3}
\title{Regenerative Coding: Deleting Code Without Losing Knowledge?}
\usepackage{etoolbox}
\makeatletter
\providecommand{\subtitle}[1]{
  \apptocmd{\@title}{\par {\large #1 \par}}{}{}
}
\makeatother
\subtitle{A Case Study with babregen and Local Small Language Models}
\def\asep{\\\\\\ } 
\author{\textbf{Thomas
Bartz-Beielstein}~\orcidlink{0000-0002-5938-5158}\\\href{mailto:bartzbeielstein@gmail.com}{bartzbeielstein@gmail.com}\asep\textbf{Marcel
Samuel
Klimm}~\orcidlink{0009-0005-4094-596X}\\\href{mailto:marcel\_samuel.klimm@th-koeln.de}{marcel\_samuel.klimm@th-koeln.de}\asep\textbf{Alexander
Hinterleitner}~\orcidlink{0009-0002-7615-6952}\\\href{mailto:alexander.hinterleitner@th-koeln.de}{alexander.hinterleitner@th-koeln.de}}
\date{2026-12-02}
\begin{document}
\maketitle
\begin{abstract}
This report presents a case study of \emph{regenerative software}, a
concept that Fowler introduced: a module of a program is deleted and
rebuilt by local small language models from the records of the module,
its public signatures and docstrings, and the source of its durable
evaluations. The open-source tool babregen implements the seven
primitives of Fowler in an approach that is referred to as
\emph{regenerative coding}. The primitives keep the intent, the
architecture, and the evaluations of a module as durable records. A
static check compares these records with the source code. A mutation
analysis and a strict acceptance rule test the strength of the
evaluations. No generation of a module has access to the old code. Two
human steps examine the surviving mutants and the rejected
regenerations, so that the causes of a failure can be traced. In the
study, two local small language models rebuilt working modules: the
larger model was accepted at the first attempt in every run, the smaller
one after a repair in every run but one. The acceptance rule rejected
one rebuild that violated a documented input contract. A check after the
freeze found that one accepted rebuild exposed an additional name
through a star import. The results indicate that the knowledge that the
evaluations exercise can outlive the implementation. Whether this holds
for the knowledge that the maintenance of a mature module needs remains
open. The contributions are (1) the tool, which maps each primitive to a
record, checks, and a command, (2) a protocol for a deletion test and
its results, (3) two human steps as commands, (4) a comparison of
drafted records with records that the maintainer authored or approved,
and (5) a worked example in an executable manuscript.
\end{abstract}
{\bfseries \emph Keywords}
\def\sep{\textbullet\ }
regenerative coding \sep regenerative
software \sep babregen \sep deletion test \sep local language
models \sep small language models \sep specification \sep test
oracle \sep mutation analysis \sep provenance \sep 
vibe coding 3.0

\section{Introduction}\label{sec-intro}

Karpathy (2025b) coined the term vibe coding in February 2025 for
programming in which one accepts every change of a language model unread
and can ``forget that the code even exists''. Criticism from
professional developers came soon. Willison (2025a) allows vibe coding
for low-stakes projects only. In production-quality work he commits no
code that he could not explain. Practitioners report code that is fast
but flawed and testing that is skipped (Fawzy et al. 2026). In a
randomised trial, experienced open-source developers took 19\% longer
with coding assistants than without (Becker et al. 2025). On a security
benchmark, only 11.8\% of the solutions of one coding agent were secure
(Zhao et al. 2026). More structured practices followed. Pollin (2025)
agrees with much of the criticism and asks for hybrid development
processes with requirements engineering, systematic prompting,
traceability, review, and tests. He calls the skills they require
Coding~2.0. Willison (2025b) speaks of vibe engineering. Spec-driven
development writes the specification first (Böckeler 2025; Agarwal et
al. 2026). We go one step further and formalise these practices with the
seven primitives of Fowler (2027): intent, compilation, evaluation,
provenance, pace, deletion, and compaction. We call this approach
regenerative coding (Definition~\ref{def-regenerative-coding}).
Continuing the numbering of Pollin (2025), it can also be seen as Vibe
Coding~3.0, as stated in a previous version of this report.

During the development of a software library many lessons can be
learned. Mature code carries them but records neither why a module
behaves as it does nor who breaks when it changes. Fowler (2027, ch.~1)
argues that this was rational as long as writing working code was
expensive. The code was the only reliable memory of the programs. Today
an agent can rewrite a module in minutes. Karpathy (2025a) expects that
a huge amount of software will be rewritten. The cost structure inverts:
implementation becomes cheap and can be quickly repeated. The scarce
asset is the knowledge around it: what a module must do, why it does it
that way, which incident taught that, and who breaks when it changes.
Fowler calls a system built around that knowledge \emph{regenerative}.
He proposes the \emph{deletion test} to check whether a module is
regenerative:

\begin{quote}
Pick an important module in your system. If you deleted the
implementation tomorrow, could you rebuild the system correctly? Could
you reproduce the edge cases, the failure behavior, the timing
assumptions, the integration contracts, the operational constraints, the
weird exception added after the outage everyone remembers but nobody
documented? Could you tell which parts of the current implementation are
essential and which are accidental, evaluate a replacement without
relying on the old code as the oracle, and prove, with evidence, that
the new implementation keeps the promises the system is supposed to
keep?

\hfill Fowler (2027, ch.~1)
\end{quote}

\begin{figure}[!htb]

\centering{

\includegraphics[width=6.49in,height=\textheight,keepaspectratio]{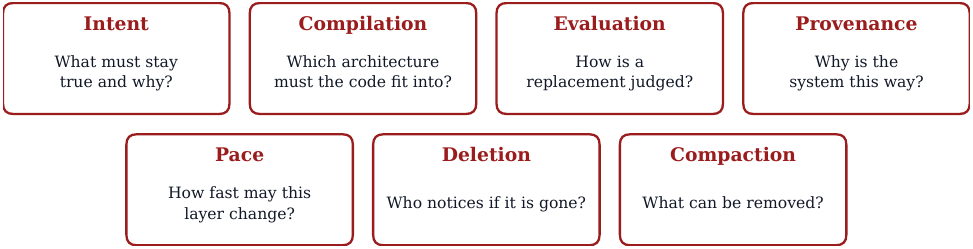}

}

\caption{\label{fig-primitives}The seven primitives of Fowler (2027)
and, in our wording, the question that each one answers.
Section~\ref{sec-approach} maps them to the records, check IDs, and
commands of babregen (Table~\ref{tbl-primitives}).}

\end{figure}

For most systems, Fowler argues, the honest answer is no. His seven
primitives keep this knowledge outside the implementation. Each answers
one question about it, in our wording (Figure~\ref{fig-primitives}).

\begin{definition}[Regenerative
coding]\protect\hypertarget{def-regenerative-coding}{}\label{def-regenerative-coding}

\emph{Regenerative coding} is the practice of developing and maintaining
regenerative software. The knowledge about each unit of the software is
kept in versioned records and durable evaluations outside the
implementation. The implementation of a unit can then be deleted,
rebuilt by a generator from this knowledge alone, and accepted or
rejected by a rule that is fixed before the rebuild and does not use the
old code as the oracle.

\end{definition}

Fowler (2027)'s seven primitives are its means.
Section~\ref{sec-approach} defines the terms that the definition uses:
record and unit (Definition~\ref{def-record}), durable evaluation
(Definition~\ref{def-evaluation}), and acceptance rule
(Definition~\ref{def-acceptance}).

This report describes the open-source, regenerative software package
\texttt{babregen} (Bartz-Beielstein 2026b), which is written in Python.
The package has no runtime dependencies, reads the code it inspects
without importing it, and is made to be a pinned development tool of
other repositories. A change in its findings, its exit codes, or the
bytes of its output changes the release decisions of those repositories.
The tool is used in its release 0.2.0. The current version is 0.5.0. The
releases after the study provide an HTML-based user interface and a
report of the results of the two human steps. The study is an
exploratory study of a single case, reported after the guidelines of
Runeson and Höst (2009).

The gitlab repository babregen-examples provides standardized benchmark
projects for evaluating regenerative coding projects. The first case in
babregen-examples is the package mathregen. Its two modules are the
units of analysis: \texttt{mathregen.fibonacci} returns the n-th
Fibonacci number, whereas \texttt{mathregen.primes} returns the n-th
prime number. The package is small on purpose. It is the package of the
tutorial of the tool. A reader can read all of it, clone it, and repeat
every step of the study that needs no model. It is the only object of
this report from the first page to the last. The same package serves the
executable examples of Section~\ref{sec-approach}, the study, and the
worked example of the appendix\footnote{ This is version 3 of a working
  paper. The earlier versions had other objects: version 1 studied four
  units of the forecasting library for critical infrastructure
  spotforecast2-safe and rebuilt one of them (Bartz-Beielstein 2026c).
  Version 2 studied the tool itself, then under its earlier name
  spotregen (Bartz-Beielstein et al. 2026).}.

The study asks whether the records of a unit suffice for a local
generator to rebuild it. It answers four research questions (RQs).

\begin{itemize}
\tightlist
\item
  RQ1: Which part of a unit record can the tool draft from the code, and
  which part is left to the maintainer to author or approve?
\item
  RQ2: Do the durable evaluations of a unit notice a wrong
  implementation, and what do the mutants that survive mean?
\item
  RQ3: Can a local, small language model rebuild a deleted module from
  the records, with up to two repairs, under a protocol fixed in
  advance?
\item
  RQ4: What does a rejected run show about the records and about the
  generator?
\end{itemize}

\begin{figure}[p]

\centering{

\includegraphics[width=6.5in,height=\textheight,keepaspectratio]{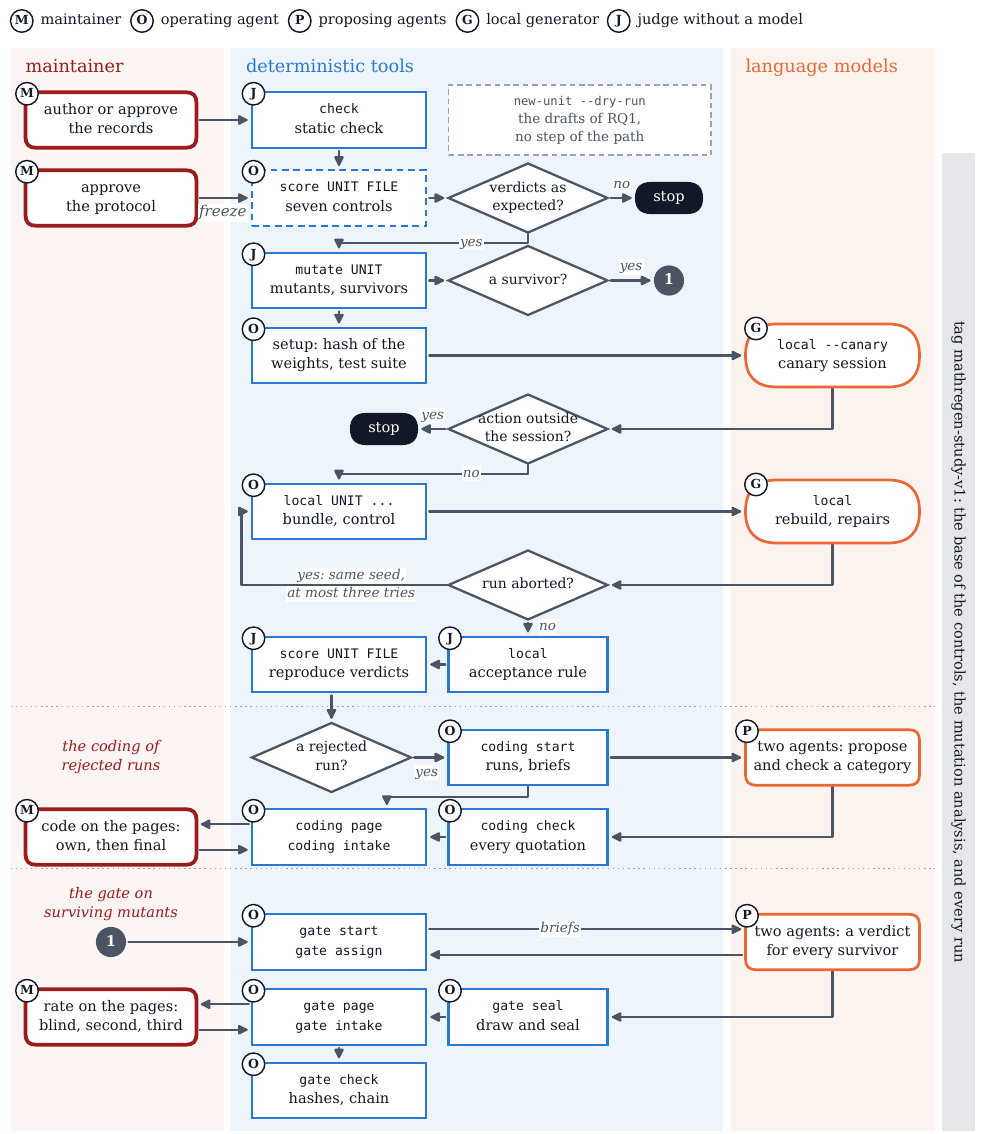}

}

\caption{\label{fig-workflow}Path of one unit through the study as a
flowchart, with the command of each step. Each lane is one kind of
actor. The letter at a box names the role that takes the step
(Table~\ref{tbl-roles}). Rounded boxes with a thick red edge are the
steps of the maintainer. Stadiums are model sessions of the driver,
rounded boxes in orange are the texts of the proposing agents, and
rectangles are deterministic steps without a model. The dashed rectangle
holds the controls, which test the acceptance rule before the first run.
A diamond is a decision that the protocol fixes, and a dark box stops
the study or the runs of a generator. The connector 1 leads from the
mutation analysis to the gate. The grey strip marks the tag of the
study, the base of the controls, the mutation analysis, and the runs. No
arrow leaves the gate and the coding, because their results change no
step of the study (Section~\ref{sec-disc-consequences}). The draft
comparison of RQ1 is no step of this path. Commands of babregen are set
in code font and are listed in Table~\ref{tbl-commands}.}

\end{figure}

The design is meant to let the questions be answered on the archive of
the study (Figure~\ref{fig-workflow}). The draft comparison of RQ1 is a
static analysis without a model and without a rater. It is no step of
the path in the figure. The maintainer authors or approves the records,
and the static check \texttt{babregen\ check} compares them with the
code. He approves the protocol, which is a record of its own. The freeze
sets the tag of the study, the base of the controls, the mutation
analysis, and every run. The mutation analysis with
\texttt{babregen\ mutate} grades the durable evaluations at the tag. Its
survivors go to the gate. One acceptance rule is implemented in the
tool, is applied by \texttt{babregen\ local} and
\texttt{babregen\ score}, includes the dependents of a unit, and is
recorded with every verdict.

Controls as declared in Section~\ref{sec-controls} show that the judge
can both accept and reject, for example

\begin{itemize}
\tightlist
\item
  it accepts a sieve for the primes, although the old unit tests cannot
  import it, and
\item
  it rejects a closed formula for the Fibonacci numbers, although the
  unit tests and the dependents pass.
\end{itemize}

After the last run every verdict is reproduced from the archived
rebuild. The generators are two local open-weight large language models
(LLMs). No cloud model writes a rebuild.

Two steps need a person:

\begin{itemize}
\tightlist
\item
  the verdict on a surviving mutant: \texttt{babregen\ gate} takes a
  rating of every survivor before it shows a proposal, which the tool
  calls the blind rating.
\item
  the category of a rejected run: \texttt{babregen\ coding} takes the
  coder's own category on a page without a proposal before it shows the
  proposal of an agent.
\end{itemize}

Both steps have one rater. A draft record, as
\texttt{babregen\ new-unit} writes it, holds the exported names of a
unit, its test files, and its dependent files. It holds no intent item
and no decision.

In the current study, the durable evaluations kill 11 of the 11 mutants
of \texttt{fibonacci} and 28 of the 31 mutants of \texttt{primes}. The 3
survivors lie in a private helper of \texttt{primes}. With generator
(c), the smaller model, the acceptance rule accepted 4 of the 5 runs of
\texttt{fibonacci} and all 5 runs of \texttt{primes}. With generator
(e), the larger model, it accepted all 5 runs of each unit. The final
rebuild of the one rejected run fails one durable evaluation of its
unit. The re-scoring reproduced all 22 verdicts. One accepted rebuild of
\texttt{primes} exports an additional name through a star import
(Section~\ref{sec-undecided}). A seed did not fix the reply: the second
run of the rejected cell, with the same seed and the same prompt, was
accepted. At the gate the verdict of the maintainer is \emph{equivalent}
for all 3 survivors, without a proposal and with one. For the drawn
mutant M004 the planted proposal argued for \emph{output}, and he kept
\emph{equivalent}. He coded the rejected run as \emph{generator error}
on both pages.

The contributions of this report are the following.

\begin{enumerate}
\def\labelenumi{\arabic{enumi}.}
\tightlist
\item
  The open-source package \texttt{babregen} on the Python Package Index
  (PyPI) (Bartz-Beielstein 2026b), documented online at
  \url{https://advm1.gm.fh-koeln.de/~bartz/babregen/} (Bartz-Beielstein
  2026a). It maps each of the seven primitives to a versioned record of
  a Python repository, to a family of check identifiers (IDs) of a
  static check, and to a command that shows it
  (Section~\ref{sec-approach}). The mapping makes regenerative coding
  checkable on a Python repository. Section~\ref{sec-hmi} describes how
  a maintainer writes the records.
\item
  A protocol for a deletion test in the sense of Fowler (2027) and its
  results on a package that a reader can clone
  (Section~\ref{sec-design}, Section~\ref{sec-results}): controls,
  mutation analysis of the durable evaluations, regeneration by local
  LLMs from the bundle of a unit, and a reproduction of every verdict.
  It connects a promise that the maintainer authored or approved with
  its reason, its evaluation, the bundle that a generator received, and
  a verdict that a reader can reproduce.
\item
  Two human steps as commands of the tool, with their limits stated: (i)
  a gate on the surviving mutants that records the rating before and
  after a proposal, and (ii) a coding of rejected runs in two stages
  (Section~\ref{sec-gate}, Section~\ref{sec-failure}). They are needed
  because the machine steps leave two questions that no test answers:
  (i) what a surviving mutant means and (ii) why a rebuild was rejected.
\item
  A mechanical comparison that separates what the tool recovers from the
  source tree from what a record that the maintainer authored or
  approved adds to it (Section~\ref{sec-rq1}). It is needed because the
  approach presumes that a record holds more than a static tool recovers
  from the code. Section~\ref{sec-disc-consequences} explains what
  follows from this and from the third contribution.
\item
  A worked example (Section~\ref{sec-tutorial}) in an executable
  manuscript, whose cells compute the examples and the reported numbers
  and stop the build when the prose and the repositories disagree.
\end{enumerate}

Section~\ref{sec-related} gives the background and the related work.
Section~\ref{sec-approach} describes how babregen implements the seven
primitives: the records, the bundle, and the static check.
Section~\ref{sec-fowlers-primitives} takes the primitives one at a time,
each with its definition, its record, and an executable example.
Section~\ref{sec-design} states the study design with the protocol, the
object, and the roles. Section~\ref{sec-study-run} describes the steps
of the study, from the drafting of the records to the coding of the
rejected runs. Section~\ref{sec-measures} defines the measures and
describes the analysis and the archive. Section~\ref{sec-results}
reports the results per research question. Section~\ref{sec-discussion}
discusses them and the threats to validity. Section~\ref{sec-conclusion}
concludes. The appendix holds the notation, the reference of the tool,
the worked example (Section~\ref{sec-tutorial}), and the description of
how the records are written (Section~\ref{sec-hmi}).

\section{Background and related work}\label{sec-related}

Table~\ref{tbl-related} gives an overview of the related topics, the
works that represent them, and how this study relates to them. The text
discusses the closest works.

\begin{longtable}[]{@{}
  >{\raggedright\arraybackslash}p{(\linewidth - 4\tabcolsep) * \real{0.2700}}
  >{\raggedright\arraybackslash}p{(\linewidth - 4\tabcolsep) * \real{0.3000}}
  >{\raggedright\arraybackslash}p{(\linewidth - 4\tabcolsep) * \real{0.4300}}@{}}
\caption{Related topics, representative works, and their relation to
this study.}\label{tbl-related}\tabularnewline
\toprule\noalign{}
\begin{minipage}[b]{\linewidth}\raggedright
topic
\end{minipage} & \begin{minipage}[b]{\linewidth}\raggedright
works
\end{minipage} & \begin{minipage}[b]{\linewidth}\raggedright
relation to this study
\end{minipage} \\
\midrule\noalign{}
\endfirsthead
\toprule\noalign{}
\begin{minipage}[b]{\linewidth}\raggedright
topic
\end{minipage} & \begin{minipage}[b]{\linewidth}\raggedright
works
\end{minipage} & \begin{minipage}[b]{\linewidth}\raggedright
relation to this study
\end{minipage} \\
\midrule\noalign{}
\endhead
\bottomrule\noalign{}
\endlastfoot
regenerative software and the deletion test & (Fowler 2027, 2026d) & the
test that this study runs on the two modules of a small package \\
regeneration as a test of a specification & (Paltenghi and Chandra 2026;
Fernandez 2026) & the closest works (Section~\ref{sec-bg-regen}) \\
spec-driven development and drift gates & (Böckeler 2025; Grabowski
2026) & the records are a specification that a static check keeps in
line with the code (Section~\ref{sec-records}) \\
evolving skill libraries of agents & (Li 2026) & the records are an
artefact store that evolves with the code; the acceptance rule is its
admission gate (Section~\ref{sec-evaluation}) \\
generation of whole repositories & (Zhao et al. 2024; {Ding et al.}
2025) & they regenerate a library against its original tests, we one
module inside an intact package \\
test oracles & (Barr et al. 2015) & durable evaluations state promises,
not the habits of the code (Section~\ref{sec-evaluation}) \\
mutation analysis and LLM-generated tests & (Jia and Harman 2011;
Papadakis et al. 2018; Bekmyradov et al. 2026) & mutation analysis
grades the durable evaluations, and every survivor gets a verdict at a
gate (Section~\ref{sec-gate}) \\
design rationale and decision records & (Jansen and Bosch 2005; Nygard
2011; Storey 2026) & a static check compares the records with the code
(Section~\ref{sec-provenance}) \\
misconceptions shared by generated candidates & (Liang et al. 2026) &
each rebuild is judged alone by a deterministic rule
(Section~\ref{sec-threats}) \\
guidelines for studies with LLMs & (Baltes et al. 2027) & open weights
identified by hash; seeds and rebuilds archived
(Section~\ref{sec-regeneration}) \\
automation bias and the human role & (Parasuraman and Manzey 2010;
Tufano et al. 2025) & the two human steps take the judgement of the
person before they show the proposal of an agent
(Section~\ref{sec-bg-human}) \\
\end{longtable}

\subsection{Regenerative software and the deletion
test}\label{sec-bg-regen}

\begin{definition}[Deletion
test]\protect\hypertarget{def-deletion-test}{}\label{def-deletion-test}

The deletion test is the test of a replaceable component in Fowler
(2027, ch.~1 and 2). It asks: if the implementation were deleted and
regenerated, what would one rely on to decide whether the result is
correct (Fowler 2026b)?

\end{definition}

Fowler (2026d) has not seen a substantial production system that
reliably regenerates arbitrary parts of itself from a complete external
body of knowledge without meaningful loss.

Two works come closest to this study. AfterVibe (Paltenghi and Chandra
2026) distils a specification from a coding session assisted by
artificial intelligence (AI). A second agent without access to the
original change then reimplements it from the specification, in the
repository at the base commit. We share the idea of memoryless
regeneration as a test of the specification rather than of the
generator. Of 72 code changes, 68 were scored. The weighted score comes
from three graders, each an agent, and includes the verdict of an LLM
judge that reads the original change. We delete whole modules of a small
package. A rebuild must pass every durable evaluation and every
dependent.

The \texttt{deletion-test} skill of Fernandez (2026), an agent skill and
not a publication, instructs a coding agent to extract the intent of a
component, to write a contract that judges it, to mutation-test this
oracle, and to regenerate the implementation from the intent. It
summarises several runs with counts and publishes neither their data nor
a protocol fixed in advance. The skill links every reason to a decision
record and compares each candidate with the original by a behavioural
diff. Such a diff shows differences that the rule of this study leaves
open (Section~\ref{sec-undecided}). What we add is a record that binds a
verdict to the hash of the bundle that a generator received. Such a
skill belongs to the evolving skill libraries of agents that Li (2026)
surveys. In their lifecycle the most important boundary is the admission
of a candidate through a verifier (Li 2026, sec. 5.2). The acceptance
rule of this study is such a gate. Mutation analysis grades its
verifier, the durable evaluations as defined in
Definition~\ref{def-evaluation}.

\subsection{Specifications, oracles, and provenance under AI
authorship}\label{sec-bg-spec}

Judging a regenerated module is an instance of the test oracle problem
(Barr et al. 2015). A test that records what an implementation does
holds its habits together with its promises. A durable evaluation states
a promise of an intent item instead. Mutation analysis (Jia and Harman
2011) grades the durable evaluations. Its score correlates only weakly
with the detection of real faults once the size of the suite is
controlled (Papadakis et al. 2018). Therefore it is used as a guide to
missing durable evaluations. A person gives every survivor a verdict
(Section~\ref{sec-gate}). Bekmyradov et al. (2026) graded unit tests
that LLMs generated for the industrial C++ database system SAP HANA by
coverage and mutation score. A repair loop on compiler feedback raised
the share of compiling tests but not their quality: the models satisfied
the compiler with placeholder assertions (Bekmyradov et al. 2026, sec.
4.3 and 6). In this study a repair rewrites only the module. The durable
evaluations that judge it are fixed in the tag of the study.

The records continue the work on design rationale. Jansen and Bosch
(2005) call the loss of design knowledge knowledge vaporisation. Under
AI authorship, Storey (2026) calls the lack of externalised goals,
constraints, and rationale intent debt. Decision records (Nygard 2011)
carry rationale in practice. This study adds a check of the records
against the code, see Definition~\ref{def-static-check} (static check),
and a test of whether they suffice to rebuild a module.

\subsection{The human role under automation}\label{sec-bg-human}

Where no automated oracle is adequate, the human remains the final
source of oracle information (Barr et al. 2015). With an imperfect
decision aid, people make errors of omission and commission, experts as
well as novices (Parasuraman and Manzey 2010). A study of code review
points in the same direction: reviewers who started from an
LLM-generated review accepted most of its issues as valid (Tufano et al.
2025). This phenomenon is known as automation bias (Goddard et al.
2012).

Our workflow automates the generation and leaves the intent and the
judgement to a person (Section~\ref{sec-roles}). Two steps have the form
that these studies warn about: (i) A surviving mutant needs a verdict,
and (ii) a rejected run needs a category. An agent that proposes either
invites automation bias. The two commands of the tool therefore take the
judgement of the person before they show a proposal
(Section~\ref{sec-gate} and Section~\ref{sec-failure}). With one rater
and one planted proposal the study shows the order of the ratings and
does not measure the bias.

\subsection{Benefits of the babregen approach}\label{sec-bg-benefits}

Five benefits follow for a maintainer.

\begin{enumerate}
\def\labelenumi{\arabic{enumi}.}
\tightlist
\item
  First, a person answers for the knowledge. A specification that is
  distilled from a conversation is bounded by what the conversation
  holds (Paltenghi and Chandra 2026, sec. 3.5). A record can also state
  what may be in no conversation and in no line of code: the reason of a
  promise, the rejected alternative, and what a rebuild must not do.
  With the reason a maintainer can decide later whether the promise
  still holds. The study shows who approved the promises and their
  reasons. It does not test whether another maintainer can decide with
  them.
\item
  Second, the knowledge can be kept current. The records lie next to the
  code under version control. The static check can compare them with the
  code after every change, also when nobody plans a rebuild. It warns
  when an archived verdict no longer fits the bundle or the module. The
  knowledge does not depend on a conversation that somebody has kept.
  The study does not measure whether the reasons stay accurate while a
  project evolves.
\item
  Third, a verdict can be checked and repeated. The rule asks no model
  and does not compare the rebuild with the old code. A different
  algorithm is therefore accepted, as the sieve shows
  (Example~\ref{exm-evaluation}). A plausible shortcut is rejected, as
  the closed formula shows (Example~\ref{exm-tut-score-rejected}).
  Everyone who has the rebuild and the records can apply the same rule
  again. In this study the re-scoring of all 22 archived rebuilds, each
  on the machine of its run, reproduced every verdict.
\item
  Fourth, a failure teaches something. The specification is fixed while
  the runs last. No failed rebuild changes it. A person can therefore
  trace a rejection to the records, the generator, or the driver. The
  verdict on a surviving mutant shows in the same way whether a promise
  lacks a judge. The maintainer learns where a change is due before a
  rebuild replaces anything. The study demonstrates this on one
  rejection with one coder. Its reliability across causes and
  maintainers needs more cases.
\item
  Fifth, the evidence is open, and the generation is local. The run
  record binds each verdict to the hashes of the bundle, of the module,
  and of the weights. The generators are open-weight models on machines
  of the maintainer, and the driver refuses any endpoint that is not on
  the local machine. Code and records need not leave the organisation.
  The driver records its own connections and does not observe the model
  server (Section~\ref{sec-threats}).
\end{enumerate}

These benefits have a price. Somebody has to write and maintain the
records, which are larger than the code in this study
(Table~\ref{tbl-size}). The rule sees only what an evaluation asserts
(Section~\ref{sec-undecided}). The evidence is small, because this study
was applied to two pure functions, basically as a demonstration of the
working principles. More realisitc functions and units will be subject
of further research.

\section{Approach: the seven primitives in babregen}\label{sec-approach}

The primitives of Fowler (2027) name the kinds of knowledge that must
outlive an implementation. Figure~\ref{fig-workflow} shows where each of
them enters the path of one unit through the study, before the rest of
this section turns to their records, checks, and commands.

The records that the maintainer authors or approves hold the
\emph{intent} of the unit, the architecture that its \emph{compilation}
must fit, its durable \emph{evaluations}, the decisions of its
\emph{provenance}, its \emph{pace}, and the consumers that notice its
\emph{deletion}. The static check \texttt{babregen\ check} compares
these records with the code. It also finds what they no longer need,
which is \emph{compaction}. The seven controls test the acceptance rule.
The mutation analysis with \texttt{babregen\ mutate} grades the durable
evaluations. A run of \texttt{babregen\ local} then carries out the
heart of the deletion test. It builds the bundle, which carries the
intent and the contract of the unit to the local generator. The module
is deleted next. The local generator rebuilds it from the bundle. The
acceptance rule judges the final rebuild by the durable evaluations and
the dependents of the unit. The command \texttt{babregen\ score}
reproduces the verdicts. The two human steps close the path. The coding
of rejected runs says why a run was rejected. The gate on surviving
mutants says what the durable evaluations do not notice.

The steps are listed in the order of the figure, from top to bottom.
Each step is taken by one of the roles of the study
(Section~\ref{sec-roles}), named by the letter at its box, which is
given in parentheses:

\begin{itemize}
\tightlist
\item
  \emph{Records} (M). The maintainer of the package authored or approved
  its records and its durable evaluations before the freeze.
\item
  \emph{Static check} (J). \texttt{babregen\ check}, a part of the judge
  without a model, compares the records with the code.
\item
  \emph{Drafts of RQ1.} \texttt{babregen\ new-unit\ -\/-dry-run} drafts
  the unit records anew for the draft comparison
  (Section~\ref{sec-redraft}). The figure sets it in a dashed box
  without an arrow, because it is no step of the path.
\item
  \emph{Protocol and freeze} (M). The maintainer approves the protocol,
  which is a record of its own. The freeze then sets the tag of the
  study, which fixes the records, the durable evaluations, and the
  protocol. The tag is the base of the controls, the mutation analysis,
  and every run.
\item
  \emph{Controls} (O). The operating agent scores seven files with
  \texttt{babregen\ score\ UNIT\ FILE}. The seven controls show that the
  acceptance rule can both accept and reject. If a control does not get
  its expected verdict, the study stops and no model run starts.
\item
  \emph{Mutation analysis} (J). \texttt{babregen\ mutate\ UNIT} grades
  the durable evaluations: it generates the mutants and lists the
  survivors, the mutants that survive. If there is a survivor, the
  connector 1 leads to the gate, which is the last step of this list.
\item
  \emph{Setup} (O). Before the first run of a generator, the operating
  agent compares the hash of the weights with the value of the protocol
  and runs the test suite of the tag.
\item
  \emph{Canary session} (G). The operating agent then starts
  \texttt{babregen\ local\ -\/-canary}, a session of its own with the
  local generator. If the canary session shows an action outside the
  session, no run of that generator starts.
\item
  \emph{Bundle and control} (O). The operating agent starts a run with
  \texttt{babregen\ local\ UNIT\ ...}. The command builds the bundle
  from the records and repeats the control on the original module.
\item
  \emph{Rebuild and repairs} (G). The module is deleted. The local
  generator rebuilds it from the bundle and repairs it in short
  sessions. A run that is aborted with one of the classes that the
  protocol names is repeated with the same seed, with at most three
  tries.
\item
  \emph{Acceptance rule} (J). \texttt{local} runs the durable
  evaluations on every attempt. To the final rebuild it applies the
  acceptance rule, which includes the dependents of the unit and is the
  rule of \texttt{babregen\ score}.
\item
  \emph{Reproduction} (J). After the last run,
  \texttt{babregen\ score\ UNIT\ FILE} reproduces the verdicts: it
  applies the same rule to every archived final rebuild again.
\item
  \emph{Coding of rejected runs} (O, P, M). If there is a rejected run,
  the operating agent opens the coding with
  \texttt{babregen\ coding\ start}, which records the runs and writes
  the briefs. Two proposing agents propose and check a category. The
  command \texttt{coding\ check} compares every quotation of a proposal
  with its file. \texttt{coding\ page} writes a page and
  \texttt{coding\ intake} takes its answers in. The maintainer codes on
  the pages, first with his own answer and then with his final answer
  (Section~\ref{sec-failure}).
\item
  \emph{Gate on surviving mutants} (O, P, M). The operating agent opens
  the gate with \texttt{babregen\ gate\ start}, which writes the briefs.
  Two proposing agents write a verdict for every survivor, the second
  one for a verdict that \texttt{gate\ assign} draws.
  \texttt{gate\ seal} draws the planted proposals and seals them.
  \texttt{gate\ page} writes a page and \texttt{gate\ intake} takes its
  ratings in. The maintainer rates on the pages in three stages: blind,
  second, and third. \texttt{gate\ check} verifies the hashes and the
  chain of the rating files (Section~\ref{sec-gate}).
\end{itemize}

No arrow leaves the coding or the gate, because their results change no
step of the path.

babregen gives each primitive except compaction a record in the
repository. Every primitive has a family of check IDs that is named
after it and a command that shows it. Section~\ref{sec-records}
describes the records, the bundle that is built from them, and the check
that compares them with the code. Section~\ref{sec-fowlers-primitives}
takes the primitives one at a time (Sections \ref{sec-intent} to
\ref{sec-compaction}), each with its definition, its record, what the
static check protects against, and one executable example. Every example
runs on mathregen, the package of the study, and mostly on its unit
\texttt{primes}. An example that changes a file runs on a throw-away
copy of the package, so the clone that this document reads is never
modified. Section~\ref{sec-tool-reference} lists the checks and the
commands. Section~\ref{sec-tutorial} builds the same package step by
step, starting with its other unit, \texttt{fibonacci}. Where the worked
example prints the output of a command in full, these sections refer to
it.

\subsection{Records and the bundle}\label{sec-records}

The records are files in Tom's Obvious, Minimal Language (TOML) in a
tracked top-level directory \texttt{regen/}, so the knowledge about the
program is reviewed, diffed, and versioned like code.

\begin{definition}[Record and
unit]\protect\hypertarget{def-record}{}\label{def-record}

A \emph{record} is a machine-readable file under \texttt{regen/} that
holds knowledge about the program and contains no implementation. There
are three kinds: the architecture file \texttt{regen/architecture.toml},
one unit record per unit, and the decision records. A \emph{unit} is the
part of the program that a unit record describes. It is implemented by
one module of the source tree, which a regeneration may replace. The
\emph{unit record}
\texttt{regen/units/\textless{}stem\textgreater{}.toml} names its unit
by its file name: the unit is the module
\texttt{\textless{}package\textgreater{}.\textless{}stem\textgreater{}},
so no record can disagree with the path of the module it describes. The
record holds a summary, the pace, the public contract of the module
(\texttt{public}), the modules or names of its own package that it may
import (\texttt{depends\_on}), its evaluation files and unit tests, its
intent items, and its declared consumers. A unit is identified by its
record, not by its current implementation: the module is what a deletion
test removes and rebuilds, the record is what survives.

\end{definition}

The loaders are strict. A key outside the schema, at any level of a
record, and a value of the wrong type stop every command that reads the
records, with a message that names the file and the key. A misspelled
key therefore cannot load silently. The archived runs under
\texttt{regen/runs/} are evidence of past judgements, not records.

The records of a unit and its durable evaluations
(Definition~\ref{def-evaluation}) together are its \emph{specification}.
mathregen holds 2 unit records, one for each of its two function
modules. Both are units of the study (Section~\ref{sec-object}).
Figure~\ref{fig-repo-tree} shows where these artefacts live: the source
tree with the two units, the test files, and the records under
\texttt{regen/}.

\makeatletter\@ifundefined{dirtree}{\newcommand*\DT@fromsty{}\input{dirtree.tex}}{}\makeatother\providecolor{darkred}{rgb}{0.55,0,0}
\begin{figure}[tbp]\centering
\begin{minipage}{0.9\linewidth}
{\renewcommand*\DTstyle{\ttfamily\small}\renewcommand*\DTstylecomment{\rmfamily\itshape\small}\DTsetlength{0.2em}{1em}{0.2em}{0.4pt}{1.6pt}
\dirtree{%
.1 mathregen/.
.2 src/mathregen/.
.3 \_\_init\_\_.py\DTcomment{re-exports the two functions}.
.3 \textcolor{darkred}{fibonacci.py}\DTcomment{\textcolor{darkred}{unit of the study}}.
.3 \textcolor{darkred}{primes.py}\DTcomment{\textcolor{darkred}{unit of the study}}.
.2 tests/.
.3 test\_fibonacci.py\DTcomment{unit tests}.
.3 test\_primes.py\DTcomment{unit tests}.
.3 test\_regen\_eval\_fibonacci.py\DTcomment{durable evaluations}.
.3 test\_regen\_eval\_primes.py\DTcomment{durable evaluations}.
.2 scripts/.
.3 table.py\DTcomment{a consumer of both units}.
.2 candidates/\DTcomment{3 rebuilds kept in the repository}.
.2 regen/.
.3 architecture.toml\DTcomment{compilation target, pace policy}.
.3 units/.
.4 fibonacci.toml\DTcomment{intent, consumers}.
.4 primes.toml\DTcomment{intent, consumers}.
.3 decisions/\DTcomment{4 decision records}.
.3 runs/\DTcomment{archived runs, one directory each}.
.4 study/\DTcomment{the archive of the study}.
.5 $\langle$run$\rangle$/bundle.md\DTcomment{the regeneration bundle of the run}.
}}
\end{minipage}
\caption{The package mathregen: the source tree, the test files, and the records under \texttt{regen/}. The two units of the study are shown in dark red. Files of the project itself, such as \texttt{pyproject.toml} and \texttt{README.md}, are left out. In the last line, $\langle$\texttt{run}$\rangle$ stands for the directory of a run. A canary has no bundle.}
\label{fig-repo-tree}
\end{figure}

\begin{definition}[Regeneration
bundle]\protect\hypertarget{def-bundle}{}\label{def-bundle}

A \emph{regeneration bundle} is a single Markdown file that
\texttt{babregen\ context} builds for one unit. Its sections are the
purpose of the unit with its pace, the architecture, the intent items,
the decision records that the items cite, the public contract of the
module, the contracts of its dependencies, the full source of the
evaluation files, and the list of consumers. It is the whole input that
describes the unit to a generator. It contains no implementation,
neither of the unit nor of a dependency.

\end{definition}

The bundle is not a record: \texttt{babregen\ context} builds it on
demand and prints it, or writes it to the file named with \texttt{-o}.
In the directory of a run it is stored as \texttt{bundle.md} by
\texttt{babregen\ local} and by \texttt{babregen\ score}
(Figure~\ref{fig-repo-tree}), so the archive holds the bytes that the
generator received.

\begin{figure}[!htb]

\centering{

\includegraphics[width=6.5in,height=\textheight,keepaspectratio]{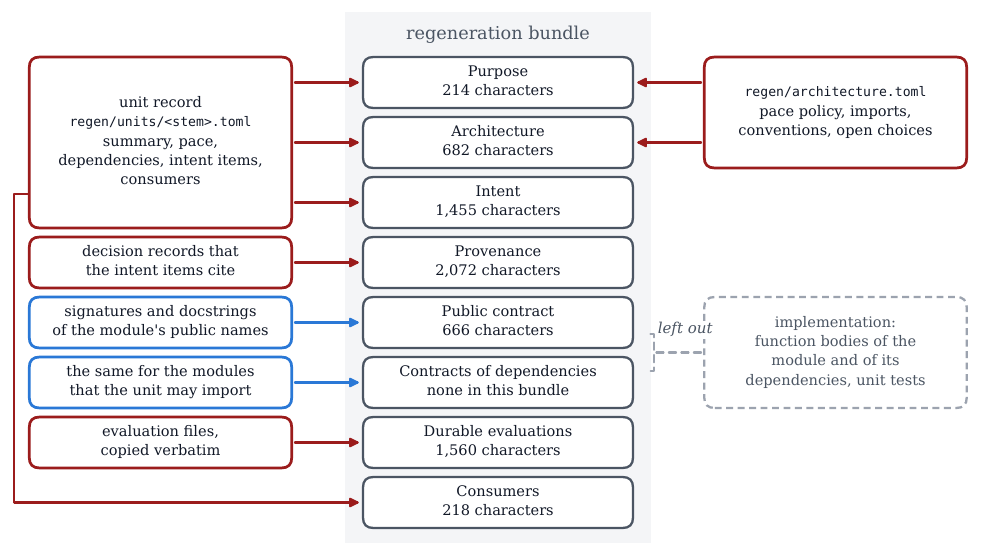}

}

\caption{\label{fig-bundle}The sections of a regeneration bundle, in the
order in which \texttt{babregen\ context} writes them, and their
sources. Red edges mark the specification, blue edges the source tree.
The unit record supplies the summary, the pace, \texttt{depends\_on},
the intent items, and the consumers. The architecture file supplies the
policy of the pace, the allowed imports, the conventions, and the open
choices. The dashed box is the implementation, which reaches no section.
The numbers count the characters of each section in the bundle of
\texttt{primes}. This unit imports no module of the package, so its
bundle has no contracts of dependencies. The heading and the task
sentence before the first section are built from the name of the unit.}

\end{figure}

Figure~\ref{fig-bundle} shows where each section of the bundle comes
from. The bundle builder reads files and runs nothing. From the module
it extracts only what a consumer may rely on, the public contract: the
docstring of the module and, of every function and class named in the
field \texttt{public}, the decorators, the signature, and the docstring.
Every body becomes \texttt{...}. An expression survives as part of a
signature, for instance as a default value or a base class, and in a
public type alias. A public constant keeps its value only if the value
is a literal. The modules named in \texttt{depends\_on} are rendered by
the same rules, so a generator does not have to guess the signature of a
dependency and needs no source file besides the bundle. The output is
deterministic: the same records, modules, and evaluation files give the
same bytes.

The identifier of an intent item is a prefix for its unit and a number:
FIB for \texttt{fibonacci} and PRI for \texttt{primes}.

The bundle builder is fail-closed in three cases, in the sense of the
fail-safe defaults of Saltzer and Schroeder (1975, sec. I.A.3): when the
module or a dependency does not parse, when an evaluation file named by
the record does not exist, and when the name matches no unit or more
than one, it builds no bundle. Other faults, such as a cited decision
record that does not exist, are left to the static check. The opposite
behaviour, to continue in an unknown state, is called fail-open below.

\begin{example}[The regeneration bundles of the two
units]\protect\hypertarget{exm-bundle}{}\label{exm-bundle}

The cell builds the bundle of each unit and searches it for the code
lines of the function bodies of the unit's module.

\end{example}

\begin{verbatim}
code lines of the function bodies of the module, and how many of them the bundle holds
outside the copied evaluation file and inside it:
unit                       bundle  body lines  outside  inside  signature
fibonacci    160 lines, 7,119 bytes           8        0       0  yes
primes       167 lines, 7,127 bytes          18        0       0  yes
\end{verbatim}

The signature and the docstring of the public function are in each
bundle. No code line of a function body is, neither of
\texttt{fibonacci} nor of \texttt{primes} with its private helper
\texttt{\_is\_prime}. The only code that a bundle copies verbatim is the
evaluation file. It shares no line with the module. A generator that
passes the durable evaluations with this input has rebuilt the module
without copying it. That it rebuilt it from memory is not excluded
(Section~\ref{sec-threats}). Example~\ref{exm-tut-bundle} shows the
public contract of \texttt{fibonacci} as its bundle prints it.

\subsection{The static check}\label{sec-static-check}

\begin{definition}[Static
check]\protect\hypertarget{def-static-check}{}\label{def-static-check}

The \emph{static check} \texttt{babregen\ check} compares the records
with the source tree and reports disagreements as findings with a check
ID. It parses files and runs no code of the package. Besides the files,
its result depends on the history of the repository and on the date. It
stops a change without repairing it. A comparison of the records with
the code cannot see what was never written down.

\end{definition}

A \emph{check ID} is the name of a family and a two-digit number, for
example COMPILE02. The seven families are named after the primitives.
The tool holds all 39 check IDs in one table, each with one level and
one condition. No other ID is reported. Table~\ref{tbl-checks} in
Section~\ref{sec-tool-reference} lists them. The level is fixed per
check. An error stops a release. A warning asks for attention. Using the
option \texttt{-\/-strict} it fails the check as well. In mathregen the
static check runs with \texttt{-\/-strict} in \texttt{make\ guard}, next
to the linter. The guard is meant to run before a push.

Table~\ref{tbl-primitives} relates the seven primitives to the records
and files that hold them, to the check IDs, and to the commands. A clean
static check shows that the records agree with the code. It does not
show that the records hold enough knowledge to rebuild a unit, or that
the durable evaluations would notice a wrong rebuild. The procedures of
Section~\ref{sec-design} test both.

\begin{longtable}[]{@{}
  >{\raggedright\arraybackslash}p{(\linewidth - 8\tabcolsep) * \real{0.1300}}
  >{\raggedright\arraybackslash}p{(\linewidth - 8\tabcolsep) * \real{0.2000}}
  >{\raggedright\arraybackslash}p{(\linewidth - 8\tabcolsep) * \real{0.3000}}
  >{\raggedright\arraybackslash}p{(\linewidth - 8\tabcolsep) * \real{0.1900}}
  >{\raggedright\arraybackslash}p{(\linewidth - 8\tabcolsep) * \real{0.1800}}@{}}
\caption{The seven primitives of Fowler (2027) in babregen: the question
each one answers, in our wording, the records and files that hold it,
its family of check IDs with their levels, and the commands that show
it. The check IDs are read from the table of checks of babregen and the
commands from its command line while this document renders. Of the 17
commands, the nine that the table does not name write records, teach
their format, or serve the two human steps of a study
(Section~\ref{sec-tool-reference}).}\label{tbl-primitives}\tabularnewline
\toprule\noalign{}
\begin{minipage}[b]{\linewidth}\raggedright
primitive
\end{minipage} & \begin{minipage}[b]{\linewidth}\raggedright
question it answers
\end{minipage} & \begin{minipage}[b]{\linewidth}\raggedright
held in
\end{minipage} & \begin{minipage}[b]{\linewidth}\raggedright
check IDs
\end{minipage} & \begin{minipage}[b]{\linewidth}\raggedright
commands
\end{minipage} \\
\midrule\noalign{}
\endfirsthead
\toprule\noalign{}
\begin{minipage}[b]{\linewidth}\raggedright
primitive
\end{minipage} & \begin{minipage}[b]{\linewidth}\raggedright
question it answers
\end{minipage} & \begin{minipage}[b]{\linewidth}\raggedright
held in
\end{minipage} & \begin{minipage}[b]{\linewidth}\raggedright
check IDs
\end{minipage} & \begin{minipage}[b]{\linewidth}\raggedright
commands
\end{minipage} \\
\midrule\noalign{}
\endhead
\bottomrule\noalign{}
\endlastfoot
Intent & What must stay true and why? & \texttt{{[}{[}intent{]}{]}}
items of a unit record & INTENT01--02 (errors) & \texttt{context} \\
Compilation & Which architecture must the code fit into? &
\texttt{regen/architecture.toml}, \texttt{public} and
\texttt{depends\_on} of a unit record & COMPILE01--08 (errors) &
\texttt{check} \\
Evaluation & How is a replacement judged? & evaluation files with the
marker \texttt{regen}, \texttt{evaluations} of a unit record and of its
intent items & EVAL01--05 (errors) & \texttt{mutate}, \texttt{score} \\
Provenance & Why is the system this way? &
\texttt{regen/decisions/\textless{}id\textgreater{}.toml}, hashes in the
\texttt{run.toml} of an archived run & PROV01--10 (5 errors and 5
warnings) & \texttt{check} \\
Pace & How fast may this layer change? & \texttt{pace} of a unit record,
policy in \texttt{{[}pace{]}} of \texttt{regen/architecture.toml} &
PACE01--04 (errors) & \texttt{measures} \\
Deletion & Who notices if it is gone? & \texttt{{[}{[}consumers{]}{]}}
of a unit record & DELETE01--03 (errors) & \texttt{deletion-test},
\texttt{score}, \texttt{local} \\
Compaction & What can be removed? & none of its own: findings on the
other records & COMPACT01--07 (warnings) & \texttt{compact} \\
\end{longtable}

\section{Fowler's seven primitives}\label{sec-fowlers-primitives}

\subsection{Intent}\label{sec-intent}

\begin{definition}[Intent]\protect\hypertarget{def-intent}{}\label{def-intent}

\emph{Intent} ``captures what the system must do and why'' (Fowler 2027,
ch.~3). It includes business rules, constraints, regulatory
requirements, timing assumptions, security properties, operational
expectations and the reasons behind them. Intent does not describe how
the current code happens to work. It describes what any future
implementation must preserve.

According to Fowler (2027, ch.~3), good intent usually has three parts:
the scope (where the requirement applies), the invariant (what must
remain true) and the reason (why the invariant matters).

The unit records use these three names as fields.

\end{definition}

The record of intent is the list of \texttt{{[}{[}intent{]}{]}} items of
a unit record. Besides the three parts, an item has an id, the names of
the durable evaluations that check it (\texttt{evaluations}), and the
ids of the decision records that justify it (\texttt{decisions}). The
static check guards against a unit without intent and against an intent
item without a judge. A unit record must declare at least one intent
item. Each item must name an evaluation. The items reach a generator
through the section \emph{Intent} of the bundle.

\begin{example}[An intent item in a unit
record]\protect\hypertarget{exm-intent}{}\label{exm-intent}

The unit record \texttt{regen/units/primes.toml} starts with fields that
describe the unit as a whole and continues with its intent items. Intent
items are numbered per unit, with a prefix for the unit, here PRI. The
excerpt shows intent item PRI-2:

\end{example}

\begin{Shaded}
\begin{Highlighting}[]
\KeywordTok{[[intent]]}
\DataTypeTok{id} \OperatorTok{=} \StringTok{"PRI{-}2"}
\DataTypeTok{scope} \OperatorTok{=} \StringTok{"nth\_prime"}
\DataTypeTok{invariant} \OperatorTok{=} \StringTok{"""}
\StringTok{Every result is a prime, the results increase strictly with the index, and no prime}
\StringTok{lies between \textasciigrave{}nth\_prime(n)\textasciigrave{} and \textasciigrave{}nth\_prime(n + 1)\textasciigrave{}."""}
\DataTypeTok{reason} \OperatorTok{=} \StringTok{"""}
\StringTok{A table of known values checks only the indices it lists. An implementation with a}
\StringTok{fault in its search, for example a sieve with a wrong bound, can pass the table and}
\StringTok{skip a prime or return a composite number elsewhere."""}
\DataTypeTok{evaluations} \OperatorTok{=} \OperatorTok{[}\StringTok{"test\_every\_result\_is\_prime\_and\_no\_prime\_is\_skipped"}\OperatorTok{]}
\DataTypeTok{decisions} \OperatorTok{=} \OperatorTok{[}\StringTok{"integers{-}only"}\OperatorTok{]}
\end{Highlighting}
\end{Shaded}

Scope, invariant, and reason are the three parts named above. The field
\texttt{evaluations} lists the evaluations that check the item. The
field \texttt{decisions} lists the decision records that justify it.

A good intent item records what a ``competent stranger'' (Fowler 2027,
ch.~6) who rebuilds the module would get wrong. Item PRI-2 is of this
kind. Its reason names the rebuild that a table of known values would
let pass: a search with a fault that skips a prime outside the table.
The promise is general and its evidence finite: the evaluation looks at
the indices 1 to 79. The most valuable items often state what the code
must never do. Fowler (2027, ch.~3) calls them \emph{negative
constraints}, which ``rule out whole classes of implementations while
still leaving room for different technical strategies''. Intent item
FIB-2 of the other unit is one of them. It demands an exact \texttt{int}
for every index, which rules out every implementation in floating-point
arithmetic and leaves the algorithm open
(Example~\ref{exm-tut-score-rejected}).

The prototype Phoenix of Fowler (2026a) keeps intent as a requirement
graph that rules extract from a specification in prose. The unit records
are the same graph without the extraction step: an intent item is a node
that a person answers for, with an id that a rewording does not change,
a durable evaluation as its judge, and a decision record as its source.

\subsection{Compilation}\label{sec-compilation}

Fowler (2027, ch.~3) argues that intent is necessary but not sufficient
to specify a repeatable system. How the intent is implemented
technologically and architecturally must be captured as well.

\begin{definition}[Compilation]\protect\hypertarget{def-compilation}{}\label{def-compilation}

\emph{Compilation} is the architecture that generated code compiles
into: boundaries, ownership, communication, runtime, and framework
constraints. Mostly humans specify it, so that a generator does not have
to infer it from the old implementation (Fowler 2027, ch.~3).

\end{definition}

The record of compilation is the file \texttt{regen/architecture.toml},
which the maintainer authors or approves, together with the fields
\texttt{public} and \texttt{depends\_on} of every unit record. The file
lists under \texttt{{[}allowed{]}} the imports that each subpackage or
top-level module may make and under \texttt{known\_cycles} the import
cycles that are accepted as existing debt. Its conventions, such as the
rule that the package uses the standard library only. Its open choices,
such as the algorithm, reach a generator through the bundle. They are
checked only where a durable evaluation pins them. The checks of the
family COMPILE protect the dependency order of the package. An import
that \texttt{{[}allowed{]}} does not permit fails the check. So does an
import cycle between two subpackages or top-level modules that
\texttt{known\_cycles} does not list. Cycles inside one subpackage are
not seen. For a unit, the checks protect what its bundle promises a
generator: every public name exists in the module, and
\texttt{depends\_on} covers the imports inside the package. Imports of
other packages are not checked.

\begin{example}[A forbidden
import]\protect\hypertarget{exm-compilation}{}\label{exm-compilation}

In mathregen the two function modules import nothing from the package.
Only the package itself, the module \texttt{\_\_init\_\_}, imports them.
The example plants the reverse edge: one line in the module of
\texttt{fibonacci} that imports a name from the package. No language
model takes part.

\end{example}

The example plants the single line
\texttt{from\ mathregen\ import\ nth\_prime} into
\texttt{src/mathregen/fibonacci.py} of a throw-away copy and runs the
static check.

\begin{verbatim}
$ babregen check
src/mathregen/fibonacci.py:3: error COMPILE02 'fibonacci' imports 'mathregen', but
    architecture.toml allows only []
regen/architecture.toml:36: error COMPILE04 new import cycle between '__init__' and
    'fibonacci' (not in known_cycles)
src/mathregen/fibonacci.py:3: error COMPILE07 mathregen.fibonacci: imports 'mathregen' which
    is not in depends_on
regen/units/primes.toml:1: error DELETE01 src/mathregen/fibonacci.py depends on the unit but
    is not a declared consumer
regen check: 2 units, 4 errors, 0 warnings
[exit code 1]
\end{verbatim}

The check reports 4 errors with 4 check IDs for the one line. The
planted edge violates the architecture (COMPILE02) and the
\texttt{depends\_on} of the unit \texttt{fibonacci} (COMPILE07). It
closes an import cycle between the module and the package, which imports
the module (COMPILE04). The imported name belongs to the unit
\texttt{primes}, which the package re-exports, so the module also
becomes an undeclared consumer of that unit (DELETE01).

A human reviewer who sees only a one-line diff learns from the static
check that the dependency order of the package is gone and that a second
unit is affected. This is the architectural position that Fowler (2027,
ch.~3) identifies as the real risk of a small change.

The static check neither imports nor runs the package. It maps each
import of the syntax tree to a module or subpackage and compares the
edges with the records. A repository that starts with import cycles
records them in \texttt{known\_cycles} as debt. A listed cycle that has
vanished and an allowed edge that no import uses are compaction
candidates (Section~\ref{sec-compaction}). On the clone the check
reports neither, so \texttt{{[}allowed{]}} matches the imports exactly.
The module boundaries of Parnas (1972) thereby become a checked record
instead of a convention.

\subsection{Evaluation}\label{sec-evaluation}

Judging a rebuilt module requires an oracle that does not depend on the
module, which is the test oracle problem (Barr et al. 2015). When a
component is deleted and rebuilt on demand, the definition of what still
works cannot live inside it. It ``has to live somewhere else, in a form
that outlasts the code it judges'' (Fowler 2027, ch.~5).

\begin{definition}[Durable
evaluation]\protect\hypertarget{def-evaluation}{}\label{def-evaluation}

A \emph{durable evaluation} is ``an executable claim about system
behavior that should remain meaningful across implementations'' (Fowler
2027, ch.~3). It judges a rebuild without reference to the old code. In
babregen a durable evaluation is a pytest function in an evaluation
file, by convention
\texttt{tests/test\_regen\_eval\_\textless{}unit\textgreater{}.py}, that
carries the marker \texttt{regen}. It imports only the names that the
unit declares as public and states one promise of an intent item that
names it. It \emph{pins} a behaviour when it fails on every module that
behaves differently. Where no other kind is meant, \emph{evaluation}
stands for durable evaluation.

\end{definition}

\begin{figure}[!htb]

\centering{

\includegraphics[width=5.64in,height=\textheight,keepaspectratio]{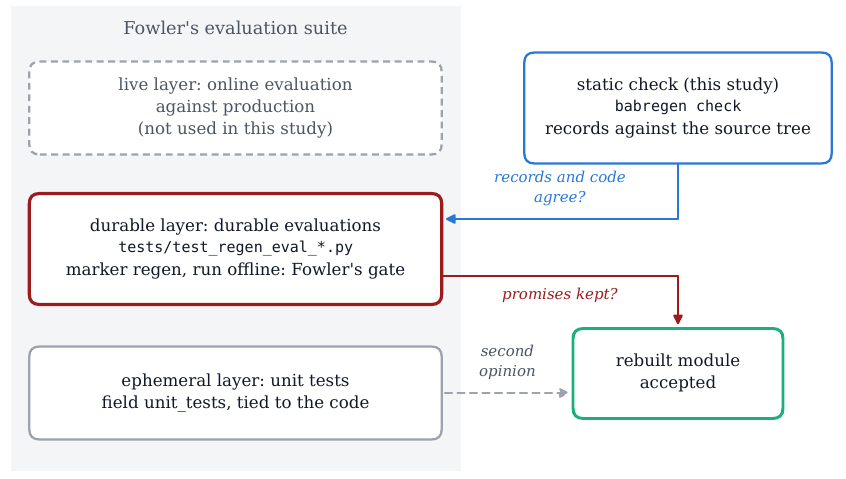}

}

\caption{\label{fig-evaluation-suite}The three layers of Fowler's
evaluation suite (Fowler 2027, ch.~5) and the static check of this
study. The layers and the gate are Fowler's. The name unit tests for the
ephemeral layer and the static check are ours. A solid edge is a
judgement, the dashed edge a second opinion that decides nothing. The
acceptance rule of the study also requires the dependents to pass
(Definition~\ref{def-acceptance}).}

\end{figure}

Fowler (2027, ch.~3) tells an evaluation from an ordinary test with one
question: ``Would this assertion still be meaningful if we replaced the
implementation tomorrow?'' Importing only public names is our
operational answer. It is necessary but not sufficient. His stricter
test asks whether ``reimplementing the component in a different language
would invalidate the evaluation'' (Fowler 2027, ch.~5). A pytest
function does not pass that test. The tests of the current
implementation, which may reach into a private helper, are called
\emph{unit tests} here. The unit record lists them in
\texttt{unit\_tests}, they never enter the bundle. They run after the
durable evaluations as a second opinion that decides nothing.
Figure~\ref{fig-evaluation-suite} places both kinds of test and the
static check next to the three layers of Fowler's evaluation suite.
Online evaluation on live traffic, his third layer, is not used.

The record of evaluation is the field \texttt{evaluations} of a unit
record, which names the evaluation files, as well as the field
\texttt{evaluations} of each intent item, which names the functions. The
static check guards the link between intent and judge: every named
evaluation must exist and be marked in the source. A name that an
evaluation file imports from the unit must be public, so an evaluation
does not depend on the implementation by accident. The check does not
see an access through the imported module.

Mutation analysis grades the evaluations against mutants of the module
(Section~\ref{sec-mutation}). \texttt{babregen\ score} applies them to a
module file that is offered in place of the module of a unit. We call
such a file a \emph{rebuild}, whoever wrote it. The tool calls it a
candidate. A \emph{run} is one execution of \texttt{babregen\ score} or
of the regeneration driver \texttt{babregen\ local}
(Section~\ref{sec-regeneration}). A run that judges a rebuild leaves a
run directory whose file \texttt{run.toml} records what was judged and
the verdict.

\begin{definition}[Acceptance
rule]\protect\hypertarget{def-acceptance}{}\label{def-acceptance}

A rebuild is \emph{accepted} if and only if its durable evaluations have
no failure, no error, and at least one pass, and no dependent fails. The
\emph{dependents} that are run are the declared consumers of kind
\texttt{code} below the source directory, which must import, and the
examples in the docstrings of the public functions and classes, which
are code cells under the heading Examples. They are taken from the
original module and must run against the rebuild. The unit tests are
recorded and never decide. The rule has the name \texttt{accept/1}. It
is implemented once, as the function \texttt{accept} of the module
\texttt{babregen.judge}. The file \texttt{run.toml} records the name of
the rule next to every verdict.

\end{definition}

The judge works on a copy. The rebuild is placed in a scratch git
worktree of a base commit in which the original module was deleted. The
worktree lies outside the repository. Every phase of the judgement
starts from a fresh one, so nothing that a rebuild writes in its
worktree reaches a later phase. The worktree is no sandbox: a rebuild
could write elsewhere. Every call of pytest and every dependent has a
time limit. A call that exceeds it is killed and counts against the
rebuild. A rebuild that loops in an evaluation or a dependent is
therefore rejected and cannot block the judge. \texttt{babregen\ score}
calls no model. It judges any file against the records, the evaluations,
and the consumers of any commit that holds the records of the unit.

\begin{example}[A sieve in place of trial
division]\protect\hypertarget{exm-evaluation}{}\label{exm-evaluation}

The module of \texttt{primes} finds the primes by trial division in a
private helper, \texttt{\_is\_prime}. The repository holds a rebuild
that uses a sieve instead, \texttt{candidates/primes\_sieve.py}. On a
throw-away copy the cell scores two files as rebuilds of the unit: the
sieve and a module that holds only the docstring of the original.

\end{example}

\begin{verbatim}
sieve: accepted = true, rule = accept/1, exit code 0
  evaluations: 10 passed, 0 failed, 0 errors
  unit tests:  0 passed, 0 failed, 1 errors
  dependents:  2 passed, 0 failed
docstring only: accepted = false, rule = accept/1, exit code 1
  evaluations: 0 passed, 0 failed, 1 errors
  unit tests:  0 passed, 0 failed, 1 errors
  dependents:  0 passed, 2 failed
\end{verbatim}

The sieve is accepted: the ten durable evaluations pass, the package
imports, and the docstring example of the original runs against the
rebuild. Its unit tests end in an error. They import the private helper
\texttt{\_is\_prime}, which a sieve does not have. The unit tests belong
to the old implementation and decide nothing. The module that holds only
its docstring is rejected on both counts. The evaluation file cannot
import the function it tests, which counts as one error, and the package
no longer imports. Both files are among the controls of the study
(Section~\ref{sec-controls}). Example~\ref{exm-tut-score-rejected} shows
a rebuild that is rejected although its unit tests pass.

The durable evaluations of the two units take four forms. Contract and
property evaluations are among the kinds that Fowler (2027, ch.~3)
lists, metamorphic testing is in his chapter 5. The example-based
evaluation is the ordinary test. An example-based evaluation fixes one
input and the required outcome, for instance that the 1000th prime is
7919. A contract evaluation pins what consumers rely on, such as the
\texttt{TypeError} for an index that is not an \texttt{int}. A property
evaluation asserts an invariant of every call, for instance that every
result is a prime and that no prime is skipped. A metamorphic evaluation
asserts how the outputs for related inputs must relate (Segura et al.
2016), for instance that every Fibonacci number is the sum of the two
before it. Another kind of his list, a regression evaluation that
replays a recorded case, does not occur in mathregen. A hash is the
usual tool of that form: the SHA-256 (Secure Hash Algorithm with a
256-bit digest) of a recorded output pins it byte for byte.

Durable evaluations and unit tests are both verification: they check an
implementation against something written down beforehand. Whether the
intent itself is right is a question of validation, which no test
answers: ``Mutation testing can tell you your suite is weak, but it
cannot tell you your suite is testing the wrong thing'' (Fowler 2027,
ch.~5). In this study validation is the work of a person: in the
decision records, where the maintainer authors or approves the reason,
in the verdict on a surviving mutant (Section~\ref{sec-gate}), and in
the category of a rejected run (Section~\ref{sec-failure}). The gap
report of a run, in which a generator lists where the bundle was silent,
is raw material for it (Section~\ref{sec-regeneration}).

\subsection{Provenance}\label{sec-provenance}

\begin{definition}[Provenance]\protect\hypertarget{def-provenance}{}\label{def-provenance}

\emph{Provenance} ``records the derivational history of a component:
what intent produced it, which architecture it compiled into, what
evaluations validated it, what evidence influenced it, and what
decisions changed it'' (Fowler 2027, ch.~3). In babregen it is held in
two places below \texttt{regen/}: (i) the decision records in
\texttt{regen/decisions/} and (ii) the archived runs in
\texttt{regen/runs/}. \emph{Decision records}, one file
\texttt{regen/decisions/\textless{}id\textgreater{}.toml} per decision,
extend the architecture decision records of Nygard (2011) by evidence
that the static check resolves, by the assumptions the decision rests
on, and by a review date. Each \texttt{run.toml} below
\texttt{regen/runs/} binds the verdict of its run to the hashes of the
bundle, of the module that was replaced, and of the final rebuild.

\end{definition}

A decision record states its title, date, and status, what drove the
decision, the decision itself, who approved it, the rejected
alternatives, the evidence, the assumptions, and the review date. An
intent item names the decision records that justify it. The bundle
prints them in its section \emph{Provenance}. The static check protects
provenance against two kinds of decay. A reference that leads nowhere,
such as a cited decision that does not exist or evidence that names a
missing file or commit, is an error. Evidence that names a web address
is recorded and not resolved. Knowledge that may have aged is a warning:
a decision past its review date, an intent item that rests on a decision
that is no longer active, and an archived verdict whose hashes no longer
match.

The review date addresses the failure mode that Fowler (2027, ch.~6)
calls provenance that lies, ``records that are confidently wrong'': a
rationale that was true two regenerations ago misleads worse than a
missing one. After the date the static check reports the record as due
for review. The date is only a proxy, because the check cannot notice
that an assumption broke earlier.

The verdict of a run needs a guard of its own, since a component can
pass today and fail in three months with no code change (Fowler 2026c,
slide 26). Every run that judges a rebuild therefore records in the
table \texttt{{[}inputs{]}} of its \texttt{run.toml} the SHA-256 of the
bundle that was judged and that of the module that was replaced. The
static check reads the newest completed run of every unit under
\texttt{regen/runs/} and compares both hashes with the current bundle
and module.

\begin{example}[A verdict that
ages]\protect\hypertarget{exm-provenance}{}\label{exm-provenance}

The archive of the study holds runs that judged the bundle of
\texttt{primes} as it is at the tag. On a throw-away copy of the package
with this archive, the cell runs the static check, sharpens the
invariant of intent item PRI-1 by one known value, and checks again.

\end{example}

\begin{verbatim}
$ babregen check
regen check: 2 units, 0 errors, 0 warnings
[exit code 0]
(the invariant of PRI-1 in regen/units/primes.toml gets one more known value)
$ babregen check
regen/runs/study/rescore/primes-e-seed5-try1/run.toml:20: warning PROV08 mathregen.primes:
    the newest archived run (started 2026-11-30T09:31:21Z) judged a different bundle: its
    verdict is evidence of that day only
regen check: 2 units, 0 errors, 1 warnings
[exit code 0]
\end{verbatim}

The runs are still in the archive. Their verdicts were true on their
day. The check reports that what the newest of them judged is no longer
what a generator would receive. It names the run, here the re-scoring of
the last run of \texttt{primes} in the study (Section~\ref{sec-rq3}).
The verdict is evidence of its day only.

\subsection{Pace}\label{sec-pace}

\begin{definition}[Pace]\protect\hypertarget{def-pace}{}\label{def-pace}

\emph{Pace} matches the speed at which a component may be replaced to
its position in the architecture. Foundational layers change more
slowly, and under more verification, than experiments (Fowler 2027,
ch.~3).

\end{definition}

The record of pace is the field \texttt{pace} of a unit record, with the
values \texttt{fast}, \texttt{medium}, and \texttt{slow}, and the table
\texttt{{[}pace{]}} of the architecture file, which states the
verification that a change to a unit of each class requires. The static
check guards against a slow unit that rests directly on a faster one: a
unit that depends on a unit of a faster pace is an error, as are a
missing pace and a pace without a policy. The bundle quotes the policy
of the unit's pace, so the policy travels with the unit.

\begin{example}[A slow unit on a fast
foundation]\protect\hypertarget{exm-pace}{}\label{exm-pace}

Both units of mathregen declare the pace fast. Neither imports the
other. The architecture file states a policy for the paces fast and slow
(Section~\ref{sec-tut-architecture}). On a throw-away copy the example
declares \texttt{primes} a slow unit that depends on \texttt{fibonacci}.
And it runs the static check:

\end{example}

\begin{verbatim}
$ babregen check
regen/units/primes.toml:6: error PACE04 pace slow depends on 'mathregen.fibonacci' of
    mathregen.fibonacci, whose pace fast is faster
regen/units/primes.toml:6: warning COMPACT04 depends_on 'mathregen.fibonacci' is not
    imported
regen check: 2 units, 1 errors, 1 warnings
[exit code 1]
\end{verbatim}

The error is PACE04: a slow unit would rest on a foundation that may
change fast. The warning concerns the same entry of the record. The
module does not import the dependency that the copy declares, which is
the subject of Example~\ref{exm-compaction}. A unit that declared the
pace medium would be an error as well, because the architecture file of
mathregen has no policy for it (PACE03).

babregen does not run the verification that a class requires. It copies
the requirement into the regeneration bundle. The maintainer carries the
verification out. The policy of a pace is therefore declared and
displayed, not enforced. The idea goes back to the shearing layers of a
building, whose components change at different rates (\emph{Shearing
Layers} 2026). Fowler (2027, ch.~3) applies it to software: the layer,
not the size of the diff, decides how dangerous a change is.

\subsection{Deletion}\label{sec-deletion}

\begin{definition}[Deletion]\protect\hypertarget{def-deletion}{}\label{def-deletion}

\emph{Deletion} makes it safe to remove or replace a component without
breaking hidden dependencies (Fowler 2027, ch.~3). The diagnostic of
replaceability, the operational replacement test that we call the
\emph{deletion test}, asks ``could we delete this and regenerate it by
Friday? If not, what's in the way?'' and is meant to be run, not argued
about: ``The attempt itself is the instrument'' (Fowler 2027, ch.~2). In
this study the deletion test of a unit has four components:

\begin{enumerate}
\def\labelenumi{\arabic{enumi}.}
\tightlist
\item
  \emph{Blast radius.} Who would notice that the unit is gone? Its
  consumers are declared.
\item
  \emph{Knowledge outside the implementation.} Every intent item names
  evaluations. Every decision that it cites is on record.
\item
  \emph{Rebuild.} The implementation is deleted and rebuilt from that
  knowledge alone, by regeneration (Definition~\ref{def-regeneration}).
\item
  \emph{Judgement.} The rebuild is judged by the acceptance rule
  (Definition~\ref{def-acceptance}), without the old code as the oracle.
  Mutation analysis (Definition~\ref{def-mutation-analysis}) assesses
  the ability of the evaluations to notice a wrong rebuild.
\end{enumerate}

The command \texttt{babregen\ deletion-test} checks components 1 and 2
before any deletion and deletes nothing, so its verdict ``regenerable''
is a precondition for an attempt, not its result.
\texttt{babregen\ score} carries out component 4 for a given rebuild.
\texttt{babregen\ local} carries out components 3 and 4.

\end{definition}

A \emph{consumer} is a declared dependent of a unit: something that
would notice its deletion. Fowler (2027, ch.~2) speaks of dependents. We
keep the field name \texttt{consumers} of the records. The report of
\texttt{babregen\ deletion-test} lists the \emph{dependent files} of a
unit: its consumers and the unit tests and evaluation files that import
it. A \emph{dependent} without this qualifier is what the acceptance
rule runs against a rebuild (Definition~\ref{def-acceptance}). The
record of deletion is the list \texttt{{[}{[}consumers{]}{]}} of a unit
record. Every consumer has a kind. Consumers of the kinds \texttt{code},
\texttt{test}, and \texttt{doc} are files of the repository and are
compared with the tree. A consumer of kind \texttt{external} is a file
of another repository, a scheduled run, or a person. It cannot be
checked from inside the repository and makes the blast radius of a
change visible before the change is made. The static check compares the
declared blast radius with the tree: a file that visibly depends on the
unit without being declared and a declared consumer that does not exist
are errors. A third check ties the public contract to its use. A name
that a consumer imports and the record does not declare is knowledge
that a rebuild from the record cannot have. The check finds it without a
model.

The consumers inside the repository are found by a static scan of the
import statements, which runs no code. The scan follows a public name
through any number of packages that re-export it under its own name. It
does not see an attribute access through an imported package, a star
import, a reference by a string, as \texttt{importlib.import\_module}
takes it, or a file outside the searched directories. Such a consumer
can only be declared with the kind \texttt{external}.

\begin{example}[The deletion test of
\texttt{primes}]\protect\hypertarget{exm-deletion}{}\label{exm-deletion}

\texttt{babregen\ deletion-test} covers the first two components of
Definition~\ref{def-deletion}. For the unit \texttt{primes} it lists 5
dependent files: the 3 declared consumers, the file of the unit tests,
and the evaluation file. Of the consumers, 2 are of kind \texttt{code},
the package that re-exports the function and a script that prints a
table, and 1 is of kind \texttt{doc}, the README. Neither the script nor
the README imports the module directly. Both use the name that the
package re-exports. mathregen declares no consumer of kind
\texttt{external}. The report then lists the 3 intent items PRI-1 to
PRI-3 with the numbers of their durable evaluations and with their
decision record. It ends with the verdict \emph{regenerable}.
Example~\ref{exm-tut-deletion-test} prints the report of the other unit.

The verdict is the precondition for components 3 and 4, which
regeneration carries out (Definition~\ref{def-regeneration}).

\end{example}

\subsection{Compaction}\label{sec-compaction}

\begin{definition}[Compaction]\protect\hypertarget{def-compaction}{}\label{def-compaction}

\emph{Compaction} is ``the discipline that prevents regeneration from
becoming accumulation'' (Fowler 2027, ch.~3). It removes what earlier
generations left behind. Fowler applies it to interfaces, configuration,
dependencies, evaluations and, above all, concepts. In this study it
covers only the records in \texttt{regen/}, so it finds stale knowledge,
not obsolete paths of the package itself.

\end{definition}

Compaction has no record of its own. Its family of check IDs collects
what the other records no longer need. Its checks are warnings that find
sediment: records that name what no longer exists or is no longer used,
such as an allowed import that no code makes, a declared dependency that
the module no longer imports, an evaluation that no intent item names,
or a decision that nothing cites. \texttt{babregen\ compact} prints
exactly these findings of the static check, so compaction is a filter of
the check and loses nothing. It deletes nothing. The maintainer turns
each line into an edit of a record or into a deliberate decision to keep
it.

\begin{example}[A dependency that is no longer
imported]\protect\hypertarget{exm-compaction}{}\label{exm-compaction}

A unit record that still declares a dependency its code no longer
imports is sediment. The module of \texttt{primes} imports no module of
its package. Its unit record says so. On a throw-away copy the example
adds an entry, a dependency that the module does not have:

\end{example}

\begin{Shaded}
\begin{Highlighting}[]
\CommentTok{\# regen/units/primes.toml, as recorded}
\DataTypeTok{depends\_on} \OperatorTok{=} \OperatorTok{[]}
\CommentTok{\# on the copy}
\DataTypeTok{depends\_on} \OperatorTok{=} \OperatorTok{[}\StringTok{"mathregen.fibonacci"}\OperatorTok{]}
\end{Highlighting}
\end{Shaded}

\texttt{babregen\ compact} reports the entry as COMPACT04:

\begin{verbatim}
$ babregen compact
regen/units/primes.toml:6: warning COMPACT04 depends_on 'mathregen.fibonacci' is not
    imported
[exit code 0]
\end{verbatim}

Example~\ref{exm-tut-compact} shows two further kinds of sediment, an
allowed import that no code makes and a declared consumer that no longer
uses the unit.

\section{Study design}\label{sec-design}

This section states the design as a protocol. The procedures are written
in the present and the future tense, because they were fixed before the
first run.

The protocol is a decision record of the package,
\texttt{regen/decisions/study-protocol.toml}. An LLM agent wrote a first
wording from the decisions of the maintainer. He read the record, added
rules, and approved it on 2026-11-30, before the tag of the freeze. The
record is part of the tag that freezes the study and that every run
names as its base. The archive of the study is
\texttt{regen/runs/study/}. The record has 13 clauses, which
Table~\ref{tbl-protocol} lists with the subsections that state them. Two
points are not fixed by the record: the comparison of drafts and records
for RQ1, which needs no run (Section~\ref{sec-redraft}), and the
measures intent coverage and specification size
(Section~\ref{sec-measures}). A cell of this document reads the record
and stops the render when a value of this section is not the value of
the record.

\begin{longtable}[]{@{}
  >{\raggedright\arraybackslash}p{(\linewidth - 4\tabcolsep) * \real{0.1000}}
  >{\raggedright\arraybackslash}p{(\linewidth - 4\tabcolsep) * \real{0.6000}}
  >{\raggedright\arraybackslash}p{(\linewidth - 4\tabcolsep) * \real{0.3000}}@{}}
\caption{The clauses of the protocol record and the subsections that
state them.}\label{tbl-protocol}\tabularnewline
\toprule\noalign{}
\begin{minipage}[b]{\linewidth}\raggedright
clause
\end{minipage} & \begin{minipage}[b]{\linewidth}\raggedright
subject
\end{minipage} & \begin{minipage}[b]{\linewidth}\raggedright
stated in
\end{minipage} \\
\midrule\noalign{}
\endfirsthead
\toprule\noalign{}
\begin{minipage}[b]{\linewidth}\raggedright
clause
\end{minipage} & \begin{minipage}[b]{\linewidth}\raggedright
subject
\end{minipage} & \begin{minipage}[b]{\linewidth}\raggedright
stated in
\end{minipage} \\
\midrule\noalign{}
\endhead
\bottomrule\noalign{}
\endlastfoot
(1) & the freeze: a pushed tag, and what may change after it &
Section~\ref{sec-object} \\
(2) & the units & Section~\ref{sec-object} \\
(3) & the controls of the acceptance rule &
Section~\ref{sec-controls} \\
(4) & the mutation analysis & Section~\ref{sec-mutation} \\
(5) & the gate on the surviving mutants & Section~\ref{sec-gate} \\
(6) & the generators and the options & Section~\ref{sec-regeneration} \\
(7) & the setup and the canary of a generator &
Section~\ref{sec-regeneration} \\
(8) & the runs, their outcome, and aborted runs &
Section~\ref{sec-regeneration} \\
(9) & the reproduction of every verdict &
Section~\ref{sec-regeneration} \\
(10) & the coding of the rejected runs & Section~\ref{sec-failure} \\
(11) & one rater, a stated gap & Section~\ref{sec-roles} \\
(12) & the analysis & Section~\ref{sec-measures} \\
(13) & the archive & Section~\ref{sec-measures} \\
\end{longtable}

The record also states what was known before the freeze. A trial on the
day before, with the preceding release of the tool, had given the
mutation scores and the three survivors that the study reports. It had
rejected the rebuild with the closed formula. Two earlier runs of
generator (c), one per unit, made for the tutorial of the tool, had both
been accepted. The mutation analysis of the study therefore repeats a
known result at the tag. The maintainer knew before the freeze how many
mutants survive and on which lines. He had not rated them. The record
expects few or no rejected runs, because the units are small. It rejects
two ways to change that: to add harder units, and to close the three
survivors before the freeze. The package stays as the readers of its
tutorial know it. An outcome without rejections would have been reported
as such.

\subsection{Object, freeze, and units}\label{sec-object}

The object of the study is the package mathregen, the folder
\texttt{mathregen/} of the repository babregen-examples. It is the
package of the tutorial of the tool: two modules with one public
function each, no dependency outside the standard library, and 92 lines
of source in all. \texttt{fibonacci} returns the n-th Fibonacci number
for an index counted from 0. \texttt{primes} returns the n-th prime
number for an index counted from 1 and has one private helper,
\texttt{\_is\_prime}. Both modules are units of the study
(Table~\ref{tbl-units}). Their records declare the pace fast
(Section~\ref{sec-pace}), and neither unit imports the other.
Section~\ref{sec-tutorial} shows the two modules, their records, and
their durable evaluations in full.

The tag mathregen-study-v1 of babregen-examples is the freeze of the
study. It points at the commit eae4473, which holds the protocol record
and pins babregen 0.2.0 in the lock file of the package. The sources,
the tests, the three rebuilds kept in babregen-examples, the unit
records, and the other decision records are those of the tutorial. They
are not edited for the study: against the earlier tag of
babregen-examples the freeze changes the protocol record, the
architecture file that lists it, and the pin of the tool, and nothing
else. A cell of this document asserts it. From the freeze until the last
run has ended, no commit touches anything in the package outside the
archive of the runs. Every run records the commit of the tag as its
base. If a control fails, if the reproduction fails, or if the driver
stops with an internal error, the study stops. Its archive is kept and
reported. The clone that this document reads differs from the tag in
files of the archive under \texttt{regen/runs/study/} and in no other
file.

In its first session a generator sees a fixed instruction and the
bundle. A repair session also sees the last module and the failures. No
file of babregen-examples is passed to it with the option
\texttt{-\/-allow}.

\begin{longtable}[]{@{}
  >{\raggedright\arraybackslash}p{(\linewidth - 14\tabcolsep) * \real{0.1400}}
  >{\raggedleft\arraybackslash}p{(\linewidth - 14\tabcolsep) * \real{0.0900}}
  >{\raggedleft\arraybackslash}p{(\linewidth - 14\tabcolsep) * \real{0.0900}}
  >{\raggedleft\arraybackslash}p{(\linewidth - 14\tabcolsep) * \real{0.1400}}
  >{\raggedleft\arraybackslash}p{(\linewidth - 14\tabcolsep) * \real{0.1500}}
  >{\raggedleft\arraybackslash}p{(\linewidth - 14\tabcolsep) * \real{0.1300}}
  >{\raggedleft\arraybackslash}p{(\linewidth - 14\tabcolsep) * \real{0.1300}}
  >{\raggedleft\arraybackslash}p{(\linewidth - 14\tabcolsep) * \real{0.1300}}@{}}
\caption{The two units of the study. A unit \protect\texttt{x} is the
module \protect\texttt{mathregen.x}. \emph{Lines} and \emph{bytes} give
the size of the module, \emph{public names}, \emph{dependencies},
\emph{consumers}, and \emph{intent items} the lengths of the fields
\protect\texttt{public}, \protect\texttt{depends\_on},
\protect\texttt{consumers}, and \protect\texttt{intent} of the unit
record, and \emph{bundle} the bytes of the regeneration bundle that
\protect\texttt{babregen\ context} builds for the
unit.}\label{tbl-units}\tabularnewline
\toprule\noalign{}
\begin{minipage}[b]{\linewidth}\raggedright
unit
\end{minipage} & \begin{minipage}[b]{\linewidth}\raggedleft
lines
\end{minipage} & \begin{minipage}[b]{\linewidth}\raggedleft
bytes
\end{minipage} & \begin{minipage}[b]{\linewidth}\raggedleft
public names
\end{minipage} & \begin{minipage}[b]{\linewidth}\raggedleft
dependencies
\end{minipage} & \begin{minipage}[b]{\linewidth}\raggedleft
consumers
\end{minipage} & \begin{minipage}[b]{\linewidth}\raggedleft
intent items
\end{minipage} & \begin{minipage}[b]{\linewidth}\raggedleft
bundle
\end{minipage} \\
\midrule\noalign{}
\endfirsthead
\toprule\noalign{}
\begin{minipage}[b]{\linewidth}\raggedright
unit
\end{minipage} & \begin{minipage}[b]{\linewidth}\raggedleft
lines
\end{minipage} & \begin{minipage}[b]{\linewidth}\raggedleft
bytes
\end{minipage} & \begin{minipage}[b]{\linewidth}\raggedleft
public names
\end{minipage} & \begin{minipage}[b]{\linewidth}\raggedleft
dependencies
\end{minipage} & \begin{minipage}[b]{\linewidth}\raggedleft
consumers
\end{minipage} & \begin{minipage}[b]{\linewidth}\raggedleft
intent items
\end{minipage} & \begin{minipage}[b]{\linewidth}\raggedleft
bundle
\end{minipage} \\
\midrule\noalign{}
\endhead
\bottomrule\noalign{}
\endlastfoot
\texttt{fibonacci} & 36 & 1,052 & 1 & 0 & 3 & 3 & 7,119 \\
\texttt{primes} & 50 & 1,240 & 1 & 0 & 3 & 3 & 7,127 \\
\end{longtable}

\subsection{Roles}\label{sec-roles}

Five roles take part in the study (Table~\ref{tbl-roles}).
Figure~\ref{fig-workflow} marks the steps that each of them takes. One
person maintains mathregen and the tool. He is referred to as the
maintainer, who is the first author of this report. He authored or
approved the records and the durable evaluations. He also approved the
protocol record. After the freeze he takes the two steps that need a
person: he rates the surviving mutants at the gate via
\texttt{babregen\ gate}, and he codes the rejected runs via
\texttt{babregen\ coding}.

There are several ways to manage the babregen workflow. For example, an
LLM agent writes a first wording of the protocol record. An operating
agent, an LLM coding agent, runs the controls, the mutation analysis,
the setup, the canaries, the runs, and the reproduction with the options
of the protocol. It also operates the commands of the two human steps
and the log of the human decisions.

The study used release 0.2.0, which has no browser interface. Release
0.3.1 provides one, the command \texttt{babregen\ ui}. It shows the next
step of every unit and offers buttons for the mutation analysis, for
single runs, as well as for the rater's and the coder's part of the two
human steps. For these steps a maintainer needs no operating agent. The
gate still needs an operator, another person or an agent, because
whoever draws and seals the proposals does not rate.

Further agents write the proposals of the two human steps. They take no
part in the runs. Two local generators, open-weight models that a model
server serves on the machine of the driver, write the rebuilds. A judge
made of tools without a model is the only judge of a rebuild.

The split follows one rule: an LLM writes knowledge and code, but
nothing it writes is accepted on its own authority. Fowler (2026c, slide
21) states the rule as ``the agent may propose changing the rule or its
test. It cannot approve the change to make itself pass''. In this study
the rule holds by construction. A generator is offered no tools, so it
cannot edit the records it is judged by. The driver never writes a
rebuild into the source tree. The operating agent starts commands whose
options the protocol fixes and gives no verdict and no category. A
proposal of an agent is never a verdict: the verdict on a mutant and the
category of a run are the ratings that the maintainer gives. No rebuild
replaces a module during the study.

\begin{longtable}[]{@{}
  >{\raggedright\arraybackslash}p{(\linewidth - 6\tabcolsep) * \real{0.1700}}
  >{\raggedright\arraybackslash}p{(\linewidth - 6\tabcolsep) * \real{0.3300}}
  >{\raggedright\arraybackslash}p{(\linewidth - 6\tabcolsep) * \real{0.3000}}
  >{\raggedright\arraybackslash}p{(\linewidth - 6\tabcolsep) * \real{0.2000}}@{}}
\caption{The roles of the study: what each one does, what limits or
judges it, and the steps of Figure~\ref{fig-workflow} that it takes,
with the letter of the figure.}\label{tbl-roles}\tabularnewline
\toprule\noalign{}
\begin{minipage}[b]{\linewidth}\raggedright
role
\end{minipage} & \begin{minipage}[b]{\linewidth}\raggedright
what it does
\end{minipage} & \begin{minipage}[b]{\linewidth}\raggedright
limits and judges
\end{minipage} & \begin{minipage}[b]{\linewidth}\raggedright
steps in Figure~\ref{fig-workflow}
\end{minipage} \\
\midrule\noalign{}
\endfirsthead
\toprule\noalign{}
\begin{minipage}[b]{\linewidth}\raggedright
role
\end{minipage} & \begin{minipage}[b]{\linewidth}\raggedright
what it does
\end{minipage} & \begin{minipage}[b]{\linewidth}\raggedright
limits and judges
\end{minipage} & \begin{minipage}[b]{\linewidth}\raggedright
steps in Figure~\ref{fig-workflow}
\end{minipage} \\
\midrule\noalign{}
\endhead
\bottomrule\noalign{}
\endlastfoot
Maintainer (also rater and coder) & authored or approved the records and
the durable evaluations, approves the protocol, rates every surviving
mutant at the gate, codes every rejected run & accountable for the
records and the protocol. The static check compares the records with the
code. He changes nothing in the package after the freeze. He is the only
rater, and the sealed files and the proposals are kept from him until
the tool shows them & M: author or approve the records; approve the
protocol; the ratings at the gate; the answers of the coding \\
Operating agent (an LLM coding agent) & runs the controls, the mutation
analysis, the setup, the canaries, the regeneration runs, and the
reproduction, operates the commands \texttt{gate} and \texttt{coding},
and keeps the log of the human decisions & generates no rebuild and
gives no verdict and no category. The protocol record fixes the options,
every call of the driver is written to a log, and every run records its
options & O: controls; setup; starts the canary and every run of
\texttt{local}; the subcommands of \texttt{gate} and \texttt{coding} \\
Proposing agents (LLM coding agents) & write the proposals of the human
steps: at the gate a verdict with a reason for every survivor and a
reason for a verdict assigned at random, for the coding a category with
quotations and a check of it & take no part in the runs. A proposal is
never a verdict. The gate seals the proposals before the blind page is
built, and \texttt{coding\ check} compares every quotation with its file
& P: the proposals at the gate and in the coding \\
Local generators (two open-weight models, Table~\ref{tbl-generators}) &
write the rebuild, its repairs, and the gap report & no tools and no
access to babregen-examples. Judged by the acceptance rule, with the
unit tests as a second opinion & G: \texttt{local} (canary session,
rebuild, and repairs) \\
Judge without a model (\texttt{babregen\ check},
\texttt{babregen\ score}, \texttt{babregen\ mutate}, pytest) & findings,
the verdict of a rebuild under the acceptance rule, killed and surviving
mutants & deterministic code without a model. The controls show that it
can accept and reject & J: \texttt{check}, \texttt{mutate}, the
acceptance rule in \texttt{local}, \texttt{score} \\
\end{longtable}

\section{Study run}\label{sec-study-run}

\subsection{Drafting the records}\label{sec-redraft}

\begin{tcolorbox}[enhanced jigsaw, arc=.35mm, bottomrule=.15mm, bottomtitle=1mm, breakable, colback=white, colbacktitle=quarto-callout-note-color!10!white, colframe=quarto-callout-note-color-frame, coltitle=black, left=2mm, leftrule=.75mm, opacityback=0, opacitybacktitle=0.6, rightrule=.15mm, title=\textcolor{quarto-callout-note-color}{\faInfo}\hspace{0.5em}{babregen commands used}, titlerule=0mm, toprule=.15mm, toptitle=1mm]

The following babregen commands were used in this subsection:

\begin{itemize}
\tightlist
\item
  \texttt{babregen\ new-unit\ MODULE\ -\/-pace\ P\ -\/-dry-run}: drafts
  the unit record of a module that has none. The draft holds what the
  source tree can tell. The option \texttt{-\/-pace} sets the pace of
  the unit and is required. With \texttt{-\/-dry-run} the draft is
  printed and no file is written, so that it can be compared with the
  record of the tag.
\item
  \texttt{babregen\ check}: runs the static check on the records of the
  package and reports each finding with its id, such as INTENT01 for a
  unit without an intent item. Here it has two uses: it shows that the
  records of the tag pass without a finding. It shows which findings a
  draft or a changed record gets.
\end{itemize}

Both commands are deterministic and need no model. Both are run with the
global option \texttt{-\/-root} on a throw-away copy of the package, so
the package itself is not changed.

\end{tcolorbox}

RQ1 needs no model and no run. The command \texttt{babregen\ new-unit}
drafts a unit record for a module that has none. It reads the module and
fills in what the tree can tell: the first paragraph of the module
docstring as summary, the public names, the package imports as
\texttt{depends\_on}, the test files named after the module as
\texttt{unit\_tests}, and every file that imports or names the module as
a consumer. The module of each unit has a list \texttt{\_\_all\_\_},
whose entries become the public names. The file
\texttt{tests/test\_regen\_eval\_\textless{}module\textgreater{}.py} is
listed as the evaluation file. The static check rejects a draft with
INTENT01 until an intent item has been added.

The comparison measures how much of a unit record the command recovers
from the source tree alone. For each of the two units, the record in
babregen-examples is deleted in a throw-away copy of the package.
\texttt{babregen\ new-unit\ -\/-dry-run} then drafts a new record with
the same pace. The draft is compared field by field with the record of
the tag. Drafting is a static analysis without a model, so the same tree
yields the same draft up to its date. The comparison is the reverse of a
deletion test: the records are deleted and the implementation remains.
It is mechanical. It counts what a draft holds and what only the record
holds. No person rates a draft. Nothing is said about whether a field of
a record is well written. It is no clause of the protocol record,
because it needs no run and can be repeated at any commit.

The agreement of a draft with its record is partly circular: for several
fields the two cannot differ, so their agreement is no evidence. There
are three reasons.

\begin{enumerate}
\def\labelenumi{\arabic{enumi}.}
\tightlist
\item
  The same scan is on both sides. \texttt{new-unit} and \texttt{check}
  find the dependent files and the imports with the same scan, and the
  records of the tag pass the check without a finding. A record that
  passes lists exactly the dependent files that the scan finds
  (DELETE01, COMPACT07) and exactly the imported modules of the package
  (COMPILE07, COMPACT04). For these two fields the agreement follows
  from the check. The public names and the role of a test file are bound
  in part (COMPILE06, EVAL03, EVAL04, EVAL05, DELETE03). No check binds
  the kind of a consumer, the summary, the notes, or what an intent item
  states.
\item
  The records may have begun as drafts. The history of babregen-examples
  adds both records in one commit in their final form and does not say
  how they arose. The worked example drafts a record and then replaces
  the draft by the finished one (Section~\ref{sec-tut-new-unit}). If the
  records of the study arose in this way, an agreement on the kinds of
  the consumers is expected as well.
\item
  The command was fitted to the records. The rule that recognises the
  evaluation file was added to \texttt{new-unit} after the records had
  been written, by the same maintainer. An earlier draft would have
  listed that file as a consumer.
\end{enumerate}

The fields on which a draft and its record agree are therefore no
independent evidence that the tool recovers what the maintainer would
have authored or approved. The informative part is the remainder, the
content of a complete record that a draft lacks.

\subsection{Controls}\label{sec-controls}

\begin{tcolorbox}[enhanced jigsaw, arc=.35mm, bottomrule=.15mm, bottomtitle=1mm, breakable, colback=white, colbacktitle=quarto-callout-note-color!10!white, colframe=quarto-callout-note-color-frame, coltitle=black, left=2mm, leftrule=.75mm, opacityback=0, opacitybacktitle=0.6, rightrule=.15mm, title=\textcolor{quarto-callout-note-color}{\faInfo}\hspace{0.5em}{babregen commands used}, titlerule=0mm, toprule=.15mm, toptitle=1mm]

The following babregen commands were used in this subsection:

\begin{itemize}
\tightlist
\item
  \texttt{babregen\ score\ UNIT\ FILE\ -\/-out\ DIR}: judges the file as
  a rebuild of the unit by the acceptance rule
  (Definition~\ref{def-acceptance}). The file replaces the module in a
  scratch copy of the base commit, which is the tag. The durable
  evaluations of the unit run against it, then its unit tests and its
  dependents. The exit code is 0 for an accepted file and 1 for a
  rejected one. With \texttt{-\/-out} the run record \texttt{run.toml}
  and the output of every step are written to the run directory of the
  control, \texttt{regen/runs/study/controls/NAME}.
\end{itemize}

The command is deterministic and needs no model: the same file, base
commit, Python, and pytest give the same verdict.

\end{tcolorbox}

\begin{definition}[Control]\protect\hypertarget{def-control}{}\label{def-control}

A \emph{control} is a file with a known answer, so the verdict of the
rule on it can be checked.

\end{definition}

No command of babregen generates the controls. No model writes one. The
seven controls used in this study have two origins:

\begin{itemize}
\tightlist
\item
  Four files are derived from the modules of the tag without a decision
  of a person: the positive control of a unit is a copy of its module.
  The negative control is the module docstring alone. The operating
  agent prepares these four files and keeps them in
  \texttt{regen/runs/study/controls/candidates/}.
\item
  The other three are written by hand as examples of a rebuild. They are
  part of babregen-examples, in the directory \texttt{candidates/} of
  the package. They existed before the study.
\end{itemize}

The protocol record names all seven and the verdict that each must get,
so the choice is approved by the maintainer with the protocol. Before
any model run, these seven files are scored with
\texttt{babregen\ score} at the tag (Example~\ref{exm-evaluation}). For
each unit the module as it is at the tag must be accepted, which is the
positive control. A module that holds only its docstring must be
rejected, which is the negative control. The three rebuilds kept in
babregen-examples, in the directory \texttt{candidates/}, are scored as
well. The Fibonacci numbers by fast doubling and the primes by a sieve
must be accepted. The Fibonacci numbers by the closed formula in
floating-point arithmetic must be rejected
(Section~\ref{sec-tut-regeneration}). The third of these is the stricter
negative control: it is not empty, it computes the first values
correctly, and it breaks two kinds of promise, the exact value and the
refusal of a wrong index. If a control fails, no model run starts.
Before the first session of every run that judges a rebuild, the driver
repeats the positive control on the machine and in the environment of
that run and records its counts.

Table~\ref{tbl-controls} lists the seven controls with the verdict that
each must get. Four come from the two units, a positive and a negative
control each. Three are the rebuilds in \texttt{candidates/}. The name
of a control is the name of its run directory below
\texttt{regen/runs/study/controls/}.

\begin{longtable}[]{@{}
  >{\raggedright\arraybackslash}p{(\linewidth - 6\tabcolsep) * \real{0.0700}}
  >{\raggedright\arraybackslash}p{(\linewidth - 6\tabcolsep) * \real{0.2300}}
  >{\raggedright\arraybackslash}p{(\linewidth - 6\tabcolsep) * \real{0.5200}}
  >{\raggedright\arraybackslash}p{(\linewidth - 6\tabcolsep) * \real{0.1800}}@{}}
\caption{The seven controls of the acceptance rule. Each file is scored
with \protect\texttt{babregen\ score} at the tag before any model
run.}\label{tbl-controls}\tabularnewline
\toprule\noalign{}
\begin{minipage}[b]{\linewidth}\raggedright
No.
\end{minipage} & \begin{minipage}[b]{\linewidth}\raggedright
Name
\end{minipage} & \begin{minipage}[b]{\linewidth}\raggedright
What is scored
\end{minipage} & \begin{minipage}[b]{\linewidth}\raggedright
Expected verdict
\end{minipage} \\
\midrule\noalign{}
\endfirsthead
\toprule\noalign{}
\begin{minipage}[b]{\linewidth}\raggedright
No.
\end{minipage} & \begin{minipage}[b]{\linewidth}\raggedright
Name
\end{minipage} & \begin{minipage}[b]{\linewidth}\raggedright
What is scored
\end{minipage} & \begin{minipage}[b]{\linewidth}\raggedright
Expected verdict
\end{minipage} \\
\midrule\noalign{}
\endhead
\bottomrule\noalign{}
\endlastfoot
1 & \texttt{fibonacci-positive} & the module \texttt{fibonacci} as it is
at the tag & accepted \\
2 & \texttt{fibonacci-negative} & a module that holds only the docstring
of \texttt{fibonacci} & rejected \\
3 & \texttt{primes-positive} & the module \texttt{primes} as it is at
the tag & accepted \\
4 & \texttt{primes-negative} & a module that holds only the docstring of
\texttt{primes} & rejected \\
5 & \texttt{fibonacci\_doubling} &
\texttt{candidates/fibonacci\_doubling.py}, the Fibonacci numbers by
fast doubling & accepted \\
6 & \texttt{primes\_sieve} & \texttt{candidates/primes\_sieve.py}, the
primes by a sieve & accepted \\
7 & \texttt{fibonacci\_binet} & \texttt{candidates/fibonacci\_binet.py},
the Fibonacci numbers by the closed formula in floating-point arithmetic
& rejected \\
\end{longtable}

\subsection{Mutation analysis}\label{sec-mutation}

\begin{tcolorbox}[enhanced jigsaw, arc=.35mm, bottomrule=.15mm, bottomtitle=1mm, breakable, colback=white, colbacktitle=quarto-callout-note-color!10!white, colframe=quarto-callout-note-color-frame, coltitle=black, left=2mm, leftrule=.75mm, opacityback=0, opacitybacktitle=0.6, rightrule=.15mm, title=\textcolor{quarto-callout-note-color}{\faInfo}\hspace{0.5em}{babregen commands used}, titlerule=0mm, toprule=.15mm, toptitle=1mm]

The following babregen commands were used in this subsection:

\begin{itemize}
\tightlist
\item
  \texttt{babregen\ mutate\ UNIT\ -\/-timeout\ S\ -\/-out\ FILE}:
  computes the mutation score of the unit. It generates the mutants of
  the unit's module and runs the durable evaluations of the unit against
  each of them. The option \texttt{-\/-timeout} sets the seconds that
  one evaluation run may take. A slower mutant counts as killed. With
  \texttt{-\/-out} the mutation report is written to the file, in the
  study \texttt{regen/runs/study/mutation/UNIT.toml}. The exit code is 0
  if every mutant is killed and 1 if a mutant survives.
\end{itemize}

The command is deterministic and needs no model. It is slow, because the
evaluations run once per mutant.

\end{tcolorbox}

Coverage alone does not grade the durable evaluations: it measures how
much code ran, not whether an evaluation would notice a wrong result
(Fowler 2027, ch.~5). Mutation analysis (DeMillo et al. 1978; Jia and
Harman 2011) measures the latter and answers the first half of RQ2.

\begin{definition}[Mutation
analysis]\protect\hypertarget{def-mutation-analysis}{}\label{def-mutation-analysis}

\emph{Mutation analysis} of a unit has three steps.

\begin{enumerate}
\def\labelenumi{(\roman{enumi})}
\tightlist
\item
  \texttt{babregen\ mutate} generates the mutants of the unit's module
  at the tag from its syntax tree with fixed mutation operators. It runs
  the unit's durable evaluations against each mutant in a throw-away
  copy of the source and the tests. A mutant is \emph{killed} if at
  least one evaluation fails or errors, or if the run exceeds its time
  limit.
\item
  The mutation score and the surviving mutants are reported as the tool
  writes them.
\item
  Every surviving mutant gets a verdict from a person at the gate
  (Section~\ref{sec-gate}). Gaps are reported and not closed, so the tag
  does not move.
\end{enumerate}

No mutant is planted by hand or by a language model.

\end{definition}

\texttt{babregen\ mutate} applies nine operators to every eligible node
of the syntax tree of the module, in source order: a comparison replaced
by its counterpart, \texttt{and} for \texttt{or}, a dropped
\texttt{not}, a negated condition of an \texttt{if} or \texttt{while}, a
swapped operator \texttt{+}, \texttt{-}, \texttt{*}, \texttt{/},
\texttt{\&}, or \texttt{\textbar{}}, a changed constant (a Boolean
flipped, a number increased by one), a deleted call statement,
\texttt{continue}, or \texttt{break}, a \texttt{raise} replaced by
\texttt{pass}, and a returned value replaced by \texttt{None}.
Annotations, string constants and with them docstrings, imports,
\texttt{\_\_all\_\_}, and the bodies of \texttt{if\ TYPE\_CHECKING:}
blocks are not mutated. A mutant whose source equals that of the
original is dropped. The command refuses to start if the durable
evaluations do not pass on the unchanged module. With the option
\texttt{-\/-out} it writes a \emph{mutation report}, a file with the
base commit, the hash of the module, the versions of Python and pytest,
and one entry per mutant with its line, its operator, the original and
the mutated code, and a diff. The protocol sets the time limit of a
mutant to 10 seconds, because some mutants of \texttt{primes} turn its
search into a loop that never returns. A mutant that runs into the limit
counts as killed.

The analysis is finished before the first model run. The gaps stay open
during the study. Closing them would move the tag. A durable evaluation
that is added after the freeze could also be tuned to what the runs have
shown. A rebuild may therefore differ from the original where a mutant
survives and still be accepted.

\subsection{The gate on surviving mutants}\label{sec-gate}

\begin{tcolorbox}[enhanced jigsaw, arc=.35mm, bottomrule=.15mm, bottomtitle=1mm, breakable, colback=white, colbacktitle=quarto-callout-note-color!10!white, colframe=quarto-callout-note-color-frame, coltitle=black, left=2mm, leftrule=.75mm, opacityback=0, opacitybacktitle=0.6, rightrule=.15mm, title=\textcolor{quarto-callout-note-color}{\faInfo}\hspace{0.5em}{babregen commands used}, titlerule=0mm, toprule=.15mm, toptitle=1mm]

The following babregen commands were used in this subsection. They are
the six subcommands of \texttt{babregen\ gate}. Each takes the review
directory \texttt{DIR}, in the study \texttt{regen/runs/study/gate}.

\begin{itemize}
\tightlist
\item
  \texttt{babregen\ gate\ start\ DIR\ REPORT...}: opens the review
  directory for the survivors of the mutation reports that
  \texttt{babregen\ mutate\ -\/-out} wrote. It lists the survivors with
  their diffs and excerpts and writes the briefs of the two agents.
\item
  \texttt{babregen\ gate\ assign\ DIR}: draws for each survivor one of
  the two verdicts that the first agent did not propose. The second
  agent writes a reason for this verdict.
\item
  \texttt{babregen\ gate\ seal\ DIR}: draws the survivors that get a
  planted proposal and commits to the proposals, the assignment, and the
  key by salted hashes in \texttt{seal.toml}. Like \texttt{gate\ assign}
  it prints counts and never a verdict.
\item
  \texttt{babregen\ gate\ page\ DIR}: writes the offline page of the
  next stage: blind, second, or third.
\item
  \texttt{babregen\ gate\ intake\ DIR\ FILE}: takes in the rating file
  that the rater downloaded from the page of a stage. The last intake
  writes the verdicts to \texttt{verdicts.toml}.
\item
  \texttt{babregen\ gate\ check\ DIR}: verifies the hashes, the chain of
  the rating files, and the order of the commits.
\end{itemize}

No subcommand calls a model. The proposals are written by two agents
outside the tool. The ratings are given by a person on the pages. The
exit code is 0 on success, 1 if checked content fails, and 2 for an
error of the input, such as a stage out of turn.

\end{tcolorbox}

A surviving mutant gets one of three verdicts: \emph{gap},
\emph{output}, or \emph{equivalent} (Table~\ref{tbl-verdicts}).

\begin{longtable}[]{@{}
  >{\raggedright\arraybackslash}p{(\linewidth - 4\tabcolsep) * \real{0.1400}}
  >{\raggedright\arraybackslash}p{(\linewidth - 4\tabcolsep) * \real{0.5000}}
  >{\raggedright\arraybackslash}p{(\linewidth - 4\tabcolsep) * \real{0.3600}}@{}}
\caption{The three verdicts on a surviving
mutant.}\label{tbl-verdicts}\tabularnewline
\toprule\noalign{}
\begin{minipage}[b]{\linewidth}\raggedright
Verdict
\end{minipage} & \begin{minipage}[b]{\linewidth}\raggedright
Meaning
\end{minipage} & \begin{minipage}[b]{\linewidth}\raggedright
Consequence
\end{minipage} \\
\midrule\noalign{}
\endfirsthead
\toprule\noalign{}
\begin{minipage}[b]{\linewidth}\raggedright
Verdict
\end{minipage} & \begin{minipage}[b]{\linewidth}\raggedright
Meaning
\end{minipage} & \begin{minipage}[b]{\linewidth}\raggedright
Consequence
\end{minipage} \\
\midrule\noalign{}
\endhead
\bottomrule\noalign{}
\endlastfoot
equivalent & The change cannot alter the behaviour for any input of the
unit, so no evaluation can detect it. & Nothing to do. \\
gap & The change alters behaviour that the records promise, and no
evaluation notices. & An evaluation is missing. \\
output & The change alters observable output, but only in a part that no
intent item pins. & The current rule permits the difference. Check
whether the records leave it open explicitly or whether the maintainer
must decide. \\
\end{longtable}

The three verdicts have different origins.

The equivalent mutant is an established notion of mutation testing
(DeMillo et al. 1978; Jia and Harman 2011). Whether a mutant is
equivalent is undecidable in general (Budd and Angluin 1982), which is
why a person has to look. The gap is the classical reading of a
surviving mutant that is not equivalent: the tests are too weak. The
skill of Fernandez (2026) sorts survivors in a similar way, among others
into equivalent, evaluation gap, and deliberate hole. The verdict output
belongs to the approach. Classical mutation testing compares a mutant
with the tests only and therefore knows two cases. Here the mutant is
also compared with the record, which states what is promised and what is
free, which gives the third case.

The nearest published notion is the unproductive mutant of Petrović et
al. (2022), which is either trivially equivalent or detectable by a test
that would not improve the test suite. A mutant with the verdict output
is unproductive in the second sense, with the record as the criterion.
The triple does not come from Fowler (2027), who asks for mutation
testing of the suite {[}ch.~5{]} but gives no classes of survivors. The
three names are fixed in the tool. The page that the rater saw words
them shorter: gap, ``an evaluation is missing''; output, ``the mutant
changes only text that no intent item pins''; equivalent, ``no behaviour
changes''. The wording of Table~\ref{tbl-verdicts} is ours. Telling the
three apart draws the line between promise and habit and needs a person.

The verdict is given with \texttt{babregen\ gate}, which answers the
second half of RQ2. The command has one more purpose. An agent writes a
proposal for each survivor, and a proposal could sway the person. The
gate therefore records the rating before and after a proposal is shown.
It plants an argument for another verdict for a subset of the survivors
that is drawn at random. It commits to the plant before the person
starts. The steps are those that the tool defines.

\begin{enumerate}
\def\labelenumi{\arabic{enumi}.}
\tightlist
\item
  \emph{Proposals.} \texttt{gate\ start} reads the mutation reports and
  lists the survivors, each with an excerpt of the source and the diff.
  A first agent proposes a verdict with a reason for every survivor.
  \texttt{gate\ assign} then draws for every survivor one of the two
  other verdicts. A second agent, which does not see the first
  proposals, writes a reason for the assigned verdict as a reviewer who
  holds it would.
\item
  \emph{Seal.} \texttt{gate\ seal} draws which survivors are planted.
  The rule gives one for three survivors: a tenth of the survivors,
  rounded half up, at least one, and none if there is only one survivor.
  The command commits to the proposals, the assignment, and the key by
  salted hashes in the file \texttt{seal.toml}. This file is committed.
  In the study it was pushed, before the blind page is built.
\item
  \emph{Blind rating.} The first page shows the survivors without a
  proposal. The rater chooses a verdict for each and writes one line of
  reason.
\item
  \emph{Second rating.} The second page shows one proposal for every
  survivor. For a drawn survivor it is the text that was written for the
  assigned verdict. For every other survivor it is the proposal of the
  first agent. The page does not say which survivors were drawn. It does
  not show the blind rating.
\item
  \emph{Third rating.} The third page holds the drawn survivors only.
  For each it shows the proposal that the rater saw, the proposal of the
  first agent, and his two earlier ratings. He rates once more.
\end{enumerate}

The verdict of a survivor is the second rating if it was not drawn and
the third rating if it was. A proposal is never a verdict. The rater
downloads his ratings from the page, and \texttt{gate\ intake} takes the
file in. A page is built only after the step before it has been taken in
and committed. Each rating file holds the hash of the seal, the hash of
its page, and the hash of the rating file before it, so the files form a
chain that \texttt{gate\ check} follows together with the order of the
commits.

\begin{figure}[p]

\begin{minipage}[t]{0.48\linewidth}

\centering{

\pandocbounded{\includegraphics[keepaspectratio]{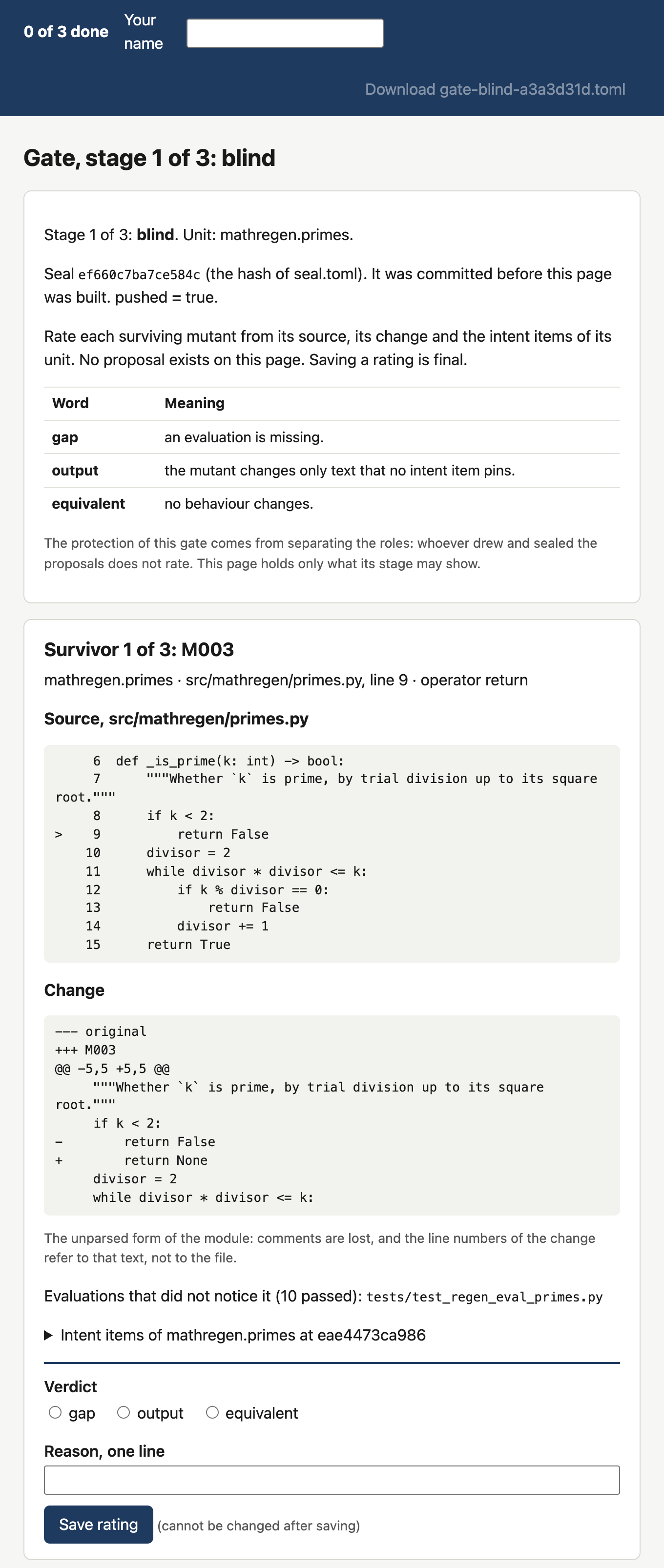}}

}

\subcaption{\label{fig-page-gate}The blind page of the gate}

\end{minipage}%
\begin{minipage}[t]{0.04\linewidth}
~\end{minipage}%
\begin{minipage}[t]{0.48\linewidth}

\centering{

\pandocbounded{\includegraphics[keepaspectratio]{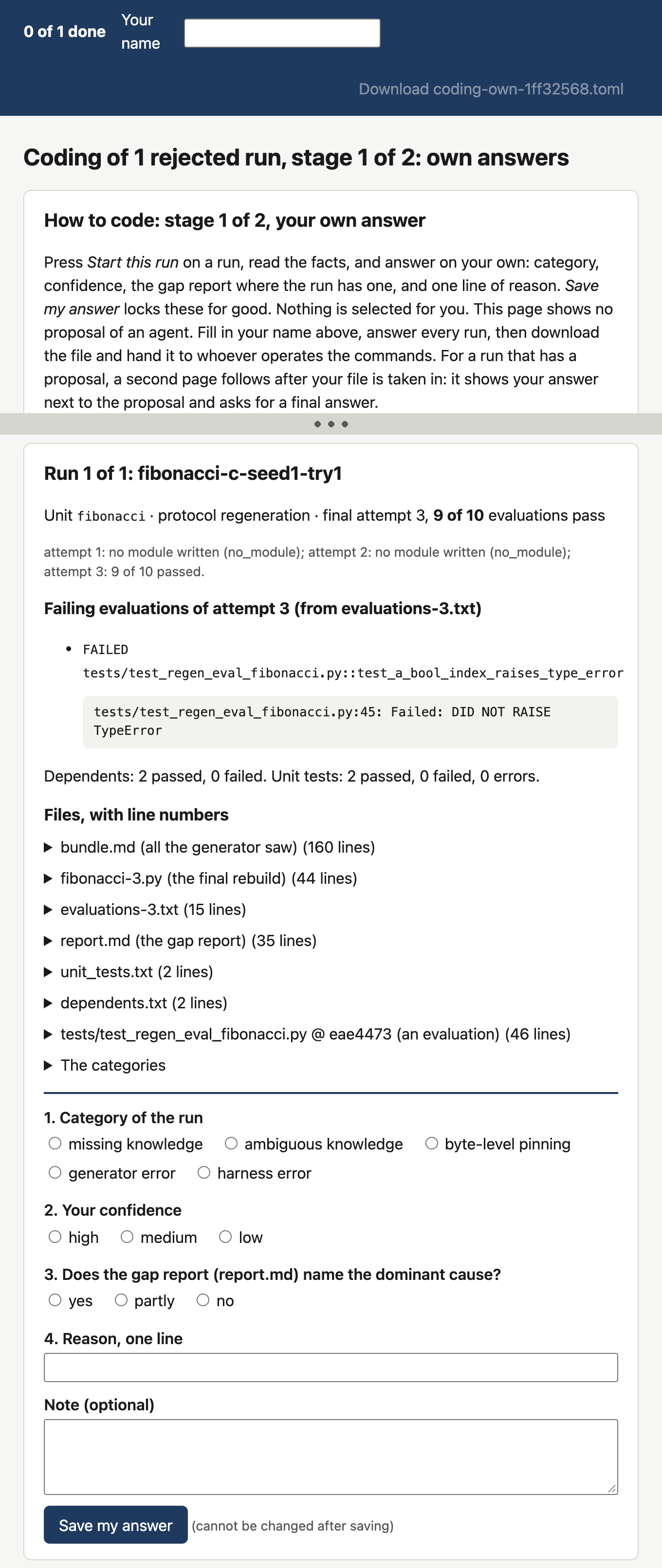}}

}

\subcaption{\label{fig-page-coding}The first page of the coding}

\end{minipage}%

\caption{\label{fig-review-pages}The pages of the two human steps as the
rater and the coder met them, from the two archived files
\texttt{gate/gate-blind.html} and \texttt{coding/coding-own.html}. Each
panel shows the head of the page and its first section, after the button
that opens the section was pressed. The forms are empty: the screenshots
were taken in a browser without stored answers. Nothing was selected or
typed. In panel (b) the grey band marks a cut. The instructions of the
coding page go on with the vocabulary (Table~\ref{tbl-categories}), a
procedure for a run, and rules.}

\end{figure}

Figure~\ref{fig-review-pages} (a) shows the blind page of the study as
the rater met it. The page is one HTML file with its style and its
script inside, so it works offline and loads nothing from a network. Its
head names the stage, shows the hash of the seal, and defines each of
the three verdicts in one line. Below the head the page has one section
per survivor, which opens when the rater presses its button. A section
shows the module and the line, the enclosing definition with the changed
line marked, the change as a diff, the evaluation file whose evaluations
did not notice the change, and the intent items of the unit. The form
asks for one of the three verdicts and for one line of reason. No
verdict is selected in advance. The blind page holds no proposal. The
page itself notes the time at which a section was opened and the time at
which its rating was saved. It offers no way to change a saved rating.
The ratings are kept in the browser of the rater. When every survivor is
rated and the rater has entered his name, the page offers the ratings as
a TOML file. This file names the hash of the page and the hash of the
seal, so the rating file of the archive says on which page the ratings
were given.

Both pages of the figure can be tried. Live copies are online with the
documentation of the tool: the blind page of the gate at
\url{https://advm1.gm.fh-koeln.de/~bartz/babregen/docs/demo/gate-blind.html}
and the first page of the coding at
\url{https://advm1.gm.fh-koeln.de/~bartz/babregen/docs/demo/coding-own.html}.
They are byte-identical copies of the two archived pages, whose SHA-256
begin with \texttt{f8c2d9e93d9d9a04} and \texttt{b8d8cf4f0f71a5de}. What
a visitor enters stays in his browser: the page stores it there and
sends nothing. The documentation describes the pages of all stages
(\url{https://advm1.gm.fh-koeln.de/~bartz/babregen/docs/review.html}).

Three limits belong to every number that the gate gives. First, the
maintainer is the only rater (Section~\ref{sec-roles}). The gate reports
counts and the ratings of each survivor in their order. It reports no
rate. Second, with three survivors one proposal is drawn. One planted
proposal can show what one person did with it once. It cannot measure
automation bias. Third, the gate rests on a separation of roles that the
archive cannot prove. The operating agent runs the commands. The
mainatiner rates on the pages. Neither opens the sealed files before the
last intake. A person who operates the gate and also rates could read
the sealed files at any time. The times come from the clock of the
browser and from the commits. They show a plausible order. They do not
show that the rater did not look.

\subsection{Regeneration}\label{sec-regeneration}

\begin{tcolorbox}[enhanced jigsaw, arc=.35mm, bottomrule=.15mm, bottomtitle=1mm, breakable, colback=white, colbacktitle=quarto-callout-note-color!10!white, colframe=quarto-callout-note-color-frame, coltitle=black, left=2mm, leftrule=.75mm, opacityback=0, opacitybacktitle=0.6, rightrule=.15mm, title=\textcolor{quarto-callout-note-color}{\faInfo}\hspace{0.5em}{babregen commands used}, titlerule=0mm, toprule=.15mm, toptitle=1mm]

The following babregen commands were used in this subsection:

\begin{itemize}
\tightlist
\item
  \texttt{babregen\ local\ UNIT\ -\/-model\ M\ -\/-model-sha256\ HEX\ -\/-context-length\ N\ -\/-protocol\ ID}:
  regenerates the unit with a model that a server on the local machine
  serves, in short sessions without history
  (Definition~\ref{def-regeneration}). The four options are required.
  They name the model, the hash of its weights, the context length with
  which the server loaded it, and the decision record that fixes the
  protocol. The study also sets \texttt{-\/-host}, \texttt{-\/-seed},
  \texttt{-\/-attempts}, \texttt{-\/-template}, \texttt{-\/-reasoning},
  the sampling options \texttt{-\/-temperature}, \texttt{-\/-top-p}, and
  \texttt{-\/-top-k}, and the limits \texttt{-\/-max-tokens},
  \texttt{-\/-session-timeout}, and \texttt{-\/-score-timeout}
  (Table~\ref{tbl-generators}, Table~\ref{tbl-options}). The run record
  \texttt{run.toml} and the output of every step are written to the run
  directory. An aborted run ends with the exit code 2.
\item
  \texttt{babregen\ local\ UNIT\ ...\ -\/-canary}: runs the canary and
  stops. The canary is one session that asks the model to act outside
  its session. No module is rebuilt.
\item
  \texttt{babregen\ score\ UNIT\ FILE}: judges a file as a rebuild of
  the unit by the acceptance rule (Section~\ref{sec-controls}). After
  the last run it scores every archived final rebuild again at the tag,
  which is the reproduction.
\end{itemize}

\texttt{babregen\ local} is the only command of the study that calls a
model. It is not deterministic, so a run is recorded and not repeated to
the same bytes. \texttt{babregen\ score} needs no model.

\end{tcolorbox}

Regeneration carries out the part of the deletion test that
\texttt{babregen\ deletion-test} only prepares: the implementation is
deleted, rebuilt, and judged (components 3 and 4 of
Definition~\ref{def-deletion}). It answers RQ3.

\begin{figure}[!htb]

\centering{

\includegraphics[width=6.44in,height=\textheight,keepaspectratio]{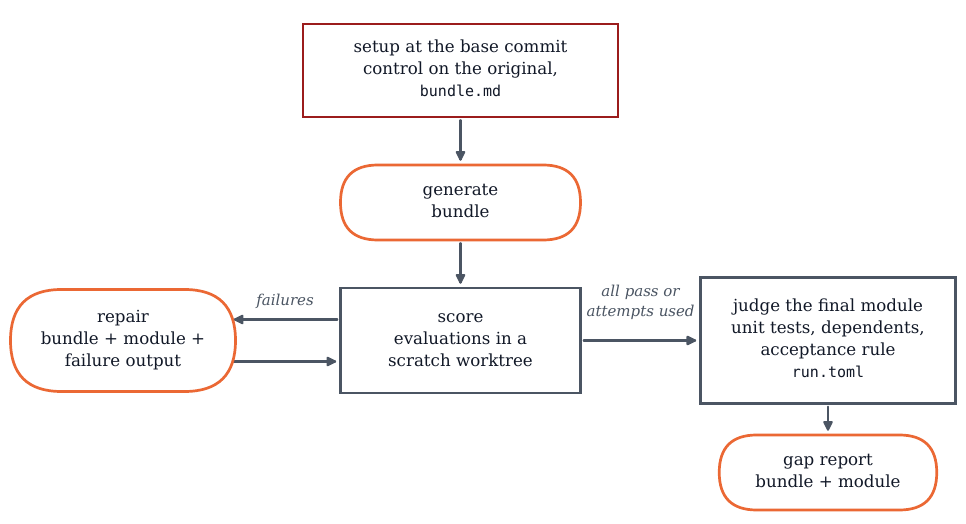}

}

\caption{\label{fig-local-arch}Steps of \texttt{babregen\ local}.
Rounded boxes are fresh model sessions without history. Rectangles run
without a model. From one session to the next only the module and the
failures are passed on. The control on the original module precedes the
first session. A repair receives the bundle, the last rebuild, and a
summary of the failing evaluations of at most 12,000 characters.}

\end{figure}

\begin{definition}[Regeneration]\protect\hypertarget{def-regeneration}{}\label{def-regeneration}

\emph{Regeneration} of a unit by \texttt{babregen\ local} has five steps
(Figure~\ref{fig-local-arch}).

\begin{enumerate}
\def\labelenumi{(\roman{enumi})}
\tightlist
\item
  \emph{Setup.} The run starts from a clean working tree and records its
  commit as the base. The bundle is built from the base. The durable
  evaluations and the dependents are run on the original module. The run
  does not start unless they pass.
\item
  \emph{Generation.} A generator receives the bundle and writes the
  module anew. The generator is \emph{memoryless}: it sees neither the
  deleted file, nor its unit tests, nor the history.
\item
  \emph{Scoring and repair.} The durable evaluations run against the
  rebuild in a scratch worktree in which the original module is deleted.
  While they do not pass and attempts are left, a repair session
  receives the bundle, the last rebuild, and a summary of the failing
  evaluations. A reply that does not compile is not scored. The next
  repair shows the error. A reply that returns the module of its prompt
  unchanged is not scored either. The next attempt is then a new
  generation.
\item
  \emph{Judgement.} The final rebuild is the one of the first attempt
  whose evaluations pass, or else of the last attempt that wrote a
  module. The unit tests and the dependents are run against it once. The
  acceptance rule decides (Definition~\ref{def-acceptance}).
\item
  \emph{Gap report.} In a last session the generator lists every place
  where the bundle was ambiguous, silent, or contradictory.
\end{enumerate}

\end{definition}

The driver rests on four design decisions.

\begin{itemize}
\tightlist
\item
  First, it is a deterministic program and not an agent session. Each
  model step is a fresh request with its own output and time budget that
  receives only the context of that step. The next step receives the
  module and the failures, not a chat history.
\item
  Second, the generator is confined by construction. The model is
  offered no tools and sees only the text that the driver puts into its
  prompt, which in this study is a fixed instruction and the bundle and,
  in a repair, the last module and the failures. The driver refuses any
  endpoint that is not on the local machine, samples the remote
  addresses of its own sockets with \texttt{lsof}, and marks a run as
  aborted if its sockets reached another machine. The samples are a
  record and not a proof: they cover the process of the driver, not that
  of the model server.
\item
  Third, the harness fits the model. The model server LM Studio serves
  the model on the machine on which the driver runs. The driver talks to
  its OpenAI-compatible completion endpoint with the standard library
  and renders the chat prompt itself, which makes the reasoning mode of
  a model controllable.
\item
  Fourth, a failure of the machine is kept apart from a failure of the
  model. When the network, the server, or the driver fails in a
  generation or a repair, the driver writes \texttt{run.toml} with the
  status \texttt{aborted} and one of nine classes of the cause. A
  machine that fails while a rebuild is scored still counts against the
  rebuild. The category harness error of the coding is for such a case.
  The control on the original module in step (i) belongs to the same
  decision: where the evaluations do not pass on the module they were
  written for, every later verdict would be a verdict about the machine.
\end{itemize}

What makes a run citable is recorded in its \texttt{run.toml}. The
command requires the identifier of the model, the hash of its weights,
which it records and cannot verify, the context length with which the
server loaded it, and the id of an active decision record that fixes the
protocol. It records the base commit, the versions of babregen, Python,
and pytest, a hash of the packages installed in the scoring environment,
the time limit of scoring, the hashes of the bundle, of the replaced
module, and of every prompt, the outcome of the control, every attempt
with its counts, the verdict, and the name of the acceptance rule.

The protocol fixes the generators, the options, the setup and the
canary, the runs, their outcome, and the reproduction.

\subsubsection{Generators}\label{generators}

A \emph{generator} is a model together with its prompt template and its
reasoning mode. The study has two generators
(Table~\ref{tbl-generators}). The archive names them by the letters (c)
and (e). Generator (c) is a model for code in a quantisation to 4 bits,
which runs on a laptop. Generator (e) is the larger model: its files
take about four times the space of those of generator (c), and it runs
on a second machine with more memory. Each generator runs on one
machine, on which the model server LM Studio serves it with one
concurrent prediction and on which the driver runs in a clean clone of
the tag. The machine is recorded with the option \texttt{-\/-host}. The
weights are identified by a manifest hash. The manifest of a model
folder has one line per file with the SHA-256 and the name of the file.
It leaves out hidden files and the README. The hash is the SHA-256 of
the manifest. Generator (c) has no reasoning mode, which the value
\texttt{none} of the option \texttt{-\/-reasoning} records. For
generator (e) the value \texttt{off} appends an empty think block to the
prompt, which switches the reasoning of the model off.

\begin{longtable}[]{@{}
  >{\raggedright\arraybackslash}p{(\linewidth - 10\tabcolsep) * \real{0.0800}}
  >{\raggedright\arraybackslash}p{(\linewidth - 10\tabcolsep) * \real{0.2700}}
  >{\raggedright\arraybackslash}p{(\linewidth - 10\tabcolsep) * \real{0.2200}}
  >{\raggedright\arraybackslash}p{(\linewidth - 10\tabcolsep) * \real{0.1100}}
  >{\raggedright\arraybackslash}p{(\linewidth - 10\tabcolsep) * \real{0.1100}}
  >{\raggedright\arraybackslash}p{(\linewidth - 10\tabcolsep) * \real{0.2100}}@{}}
\caption{The two generators of the study, read from the protocol record.
\emph{Letter} is the name of the generator in the text and in the run
directories. \emph{Manifest hash} gives the first 16 of 64 hexadecimal
digits. \emph{Template} and \emph{reasoning} are the values of the
options \protect\texttt{-\/-template} and \protect\texttt{-\/-reasoning}
of \protect\texttt{babregen\ local}. \emph{Machine} is read from the
archive.}\label{tbl-generators}\tabularnewline
\toprule\noalign{}
\begin{minipage}[b]{\linewidth}\raggedright
letter
\end{minipage} & \begin{minipage}[b]{\linewidth}\raggedright
model
\end{minipage} & \begin{minipage}[b]{\linewidth}\raggedright
manifest hash
\end{minipage} & \begin{minipage}[b]{\linewidth}\raggedright
template
\end{minipage} & \begin{minipage}[b]{\linewidth}\raggedright
reasoning
\end{minipage} & \begin{minipage}[b]{\linewidth}\raggedright
machine
\end{minipage} \\
\midrule\noalign{}
\endfirsthead
\toprule\noalign{}
\begin{minipage}[b]{\linewidth}\raggedright
letter
\end{minipage} & \begin{minipage}[b]{\linewidth}\raggedright
model
\end{minipage} & \begin{minipage}[b]{\linewidth}\raggedright
manifest hash
\end{minipage} & \begin{minipage}[b]{\linewidth}\raggedright
template
\end{minipage} & \begin{minipage}[b]{\linewidth}\raggedright
reasoning
\end{minipage} & \begin{minipage}[b]{\linewidth}\raggedright
machine
\end{minipage} \\
\midrule\noalign{}
\endhead
\bottomrule\noalign{}
\endlastfoot
(c) & Qwen3-Coder-30B-A3B-Instruct & \texttt{6b189cafac481376} &
\texttt{chatml} & \texttt{none} & Apple M5 Max, 128 GB \\
(e) & Qwen3.5-122B-A10B & \texttt{d7bc12a483079e66} & \texttt{chatml} &
\texttt{off} & Apple M2 Ultra, 192 GB \\
\end{longtable}

\subsubsection{Options}\label{options}

Table~\ref{tbl-options} lists the options of a run. The sampling
options, the number of attempts, and the time limit of scoring hold for
both generators. The context length, the output budget, and the time
limit of a session differ. Other options that influence a reply keep
their defaults, apart from the reasoning mode of
Table~\ref{tbl-generators}. The model server loads a model with the
context length of the table. A reply that reaches the output budget uses
up its attempt without a rebuild.

\begin{longtable}[]{@{}
  >{\raggedright\arraybackslash}p{(\linewidth - 6\tabcolsep) * \real{0.2200}}
  >{\raggedleft\arraybackslash}p{(\linewidth - 6\tabcolsep) * \real{0.1200}}
  >{\raggedleft\arraybackslash}p{(\linewidth - 6\tabcolsep) * \real{0.1200}}
  >{\raggedright\arraybackslash}p{(\linewidth - 6\tabcolsep) * \real{0.5400}}@{}}
\caption{The options of \protect\texttt{babregen\ local} for the runs of
the two generators. A value that the protocol record does not name is
the default of the tool.}\label{tbl-options}\tabularnewline
\toprule\noalign{}
\begin{minipage}[b]{\linewidth}\raggedright
option
\end{minipage} & \begin{minipage}[b]{\linewidth}\raggedleft
(c)
\end{minipage} & \begin{minipage}[b]{\linewidth}\raggedleft
(e)
\end{minipage} & \begin{minipage}[b]{\linewidth}\raggedright
meaning
\end{minipage} \\
\midrule\noalign{}
\endfirsthead
\toprule\noalign{}
\begin{minipage}[b]{\linewidth}\raggedright
option
\end{minipage} & \begin{minipage}[b]{\linewidth}\raggedleft
(c)
\end{minipage} & \begin{minipage}[b]{\linewidth}\raggedleft
(e)
\end{minipage} & \begin{minipage}[b]{\linewidth}\raggedright
meaning
\end{minipage} \\
\midrule\noalign{}
\endhead
\bottomrule\noalign{}
\endlastfoot
\texttt{-\/-context-length} & 32,768 & 131,072 & tokens of context with
which the model server loads the model \\
\texttt{-\/-max-tokens} & 8,192 & 32,768 & output budget of one code
session in tokens \\
\texttt{-\/-session-timeout} & 900 & 5,400 & seconds of wall clock for
one model session \\
\texttt{-\/-attempts} & 3 & 3 & one generation and up to two repairs \\
\texttt{-\/-temperature} & 0.7 & 0.7 & sampling temperature \\
\texttt{-\/-top-p} & 0.8 & 0.8 & probability mass of the tokens that are
sampled from \\
\texttt{-\/-top-k} & 20 & 20 & number of the tokens that are sampled
from \\
\texttt{-\/-score-timeout} & 120 & 120 & seconds for every scoring
call \\
\end{longtable}

For generator (c) the output budget and the limit of a session are the
defaults of the tool. For generator (e) the protocol raises both. At
about 4 characters per token, a repair prompt with the bundle, a module
of the size of the original, and the summary of the failures, together
with a reply of the full output budget, needs up to 13,300 of the 32,768
tokens of generator (c) and up to 37,900 of the 131,072 tokens of
generator (e).

\subsubsection{Setup and canary}\label{setup-and-canary}

Three steps precede the first run of a generator. The manifest hash of
the model folder is computed on the machine of the runs and compared
with the value of the protocol record. The test suite of the tag is run
in the environment of the clone. \texttt{babregen\ local} is run once
with the option \texttt{-\/-canary} and archived. The canary is a
session of its own whose prompt asks the model to read the original
module, to list a directory, to run a shell command, and to write files.
It passes on what the driver recorded, never on the model's own account:
no file appeared in the run directory besides the reply and the session
log, and every remote address of the driver was local. The driver offers
no tool in any session. If a canary shows an action outside its session,
no run of that generator starts.

\subsubsection{Runs and seeds}\label{runs-and-seeds}

Every unit is regenerated with each generator and the seeds 1 to 5,
which gives 20 regeneration runs. The seed of a run is that of its first
session. The session of attempt \(n\) uses the seed plus \(n - 1\), so
that a repair does not replay its input under the same seed. In
addition, seed 1 of the unit \texttt{fibonacci} is run a second time
with each generator. These two runs show whether a seed fixes the reply
on a model server, that is, whether the seeds are more than labels. The
22 planned runs are the \emph{cells} of the design. A \emph{try} is one
execution of a cell, that is, one run in the sense of
Section~\ref{sec-evaluation}. A cell has one try unless a try is
aborted. In the results, a \emph{run} of a unit with a generator and a
seed is the completed try of its cell. The controls and the canaries are
runs of their own kind and are named as such.

\subsubsection{Outcome and aborted runs}\label{outcome-and-aborted-runs}

A run is accepted when its \texttt{run.toml} says so under the rule
\texttt{accept/1}. The unit tests and the exit code of the driver are
not criteria. An aborted run is no outcome. It is kept. With the classes
\texttt{network}, \texttt{timeout},
\texttt{http\_\textless{}code\textgreater{}}, \texttt{bad\_json},
\texttt{context}, and \texttt{interrupted} the cell is tried again with
the same seed. A cell has at most three tries. A cell whose tries are
used up is reported with its classes and counts as not accepted. The
classes \texttt{internal}, \texttt{egress}, and \texttt{model\_mismatch}
are no property of a rebuild. They stop the runs of the generator.

\subsubsection{Reproduction}\label{reproduction}

After the last run, every archived final rebuild is scored again with
\texttt{babregen\ score} at the tag. Every recorded verdict must be
reproduced. A difference is reported as a defect of the harness and not
as a result about a model.

\subsection{The coding of rejected runs}\label{sec-failure}

\begin{tcolorbox}[enhanced jigsaw, arc=.35mm, bottomrule=.15mm, bottomtitle=1mm, breakable, colback=white, colbacktitle=quarto-callout-note-color!10!white, colframe=quarto-callout-note-color-frame, coltitle=black, left=2mm, leftrule=.75mm, opacityback=0, opacitybacktitle=0.6, rightrule=.15mm, title=\textcolor{quarto-callout-note-color}{\faInfo}\hspace{0.5em}{babregen commands used}, titlerule=0mm, toprule=.15mm, toptitle=1mm]

The following babregen commands were used in this subsection. They are
the four subcommands of \texttt{babregen\ coding}. Each takes the review
directory \texttt{DIR}, in the study \texttt{regen/runs/study/coding}.

\begin{itemize}
\tightlist
\item
  \texttt{babregen\ coding\ start\ DIR\ RUN...}: opens the review
  directory for the rejected runs, whose run directories it takes in the
  order of the page. It records the runs with the hashes of their run
  records and writes the briefs of the two agents.
\item
  \texttt{babregen\ coding\ check\ DIR\ {[}RUN{]}}: checks the
  proposals, the pages, and the answers. Every quotation of a proposal
  is compared with the file that it names. With \texttt{RUN} only the
  proposal of this run is checked.
\item
  \texttt{babregen\ coding\ page\ DIR}: writes the offline page of the
  next stage, first \texttt{coding-own.html} and then
  \texttt{coding-final.html}.
\item
  \texttt{babregen\ coding\ intake\ DIR\ FILE}: takes in the answer file
  that the coder downloaded from the page of a stage, as
  \texttt{answers-own.toml} and then \texttt{answers-final.toml}.
\end{itemize}

No subcommand calls a model. The proposals are written and checked by
two agents outside the tool, and the answers are given by a person on
the pages. The exit code is 0 on success, 1 if checked content fails,
and 2 for an error of the input, such as a stage out of turn.

\end{tcolorbox}

A rejected run does not say why it was rejected. Every run of the design
for which the tool recorded a rejection is therefore coded with
\texttt{babregen\ coding}, which answers RQ4. The coder puts the run
into one of five categories that the tool fixes
(Table~\ref{tbl-categories}), says how sure he is, says whether the gap
report of the generator names the dominant cause. He also writes one
line of reason.

\begin{longtable}[]{@{}
  >{\raggedright\arraybackslash}p{(\linewidth - 4\tabcolsep) * \real{0.1800}}
  >{\raggedright\arraybackslash}p{(\linewidth - 4\tabcolsep) * \real{0.3000}}
  >{\raggedright\arraybackslash}p{(\linewidth - 4\tabcolsep) * \real{0.5200}}@{}}
\caption{The five categories of a rejected run, read from the vocabulary
of the tool. \emph{The records} are the bundle of the run. The last
column is the hint that the coding page
shows.}\label{tbl-categories}\tabularnewline
\toprule\noalign{}
\begin{minipage}[b]{\linewidth}\raggedright
category
\end{minipage} & \begin{minipage}[b]{\linewidth}\raggedright
meaning
\end{minipage} & \begin{minipage}[b]{\linewidth}\raggedright
how to tell it from its neighbour
\end{minipage} \\
\midrule\noalign{}
\endfirsthead
\toprule\noalign{}
\begin{minipage}[b]{\linewidth}\raggedright
category
\end{minipage} & \begin{minipage}[b]{\linewidth}\raggedright
meaning
\end{minipage} & \begin{minipage}[b]{\linewidth}\raggedright
how to tell it from its neighbour
\end{minipage} \\
\midrule\noalign{}
\endhead
\bottomrule\noalign{}
\endlastfoot
missing knowledge & the records do not say what the evaluations demand.
& The records are silent on the point. If they say something that can be
read in two ways, it is ambiguous knowledge instead. \\
ambiguous knowledge & the records allow the candidate's reading. & The
records speak about the point, and the rebuild follows one fair reading
of them. If only one reading is fair, it is a generator error
instead. \\
byte-level pinning & an evaluation pins text that no intent item states.
& The rebuild behaves as the records say, and only exact text differs (a
heading, a blank, the wording of a message) that no intent item states.
If an intent item states the text, it is a generator error instead. \\
generator error & the records say it, the candidate does not do it. &
The records say it clearly, and the rebuild does not do it. If you
cannot point to the place in the records, it is missing knowledge
instead. \\
harness error & the driver or the judge failed, not the candidate. & The
rebuild is not to blame: the driver or the judge failed, for instance a
timeout or a broken scoring environment. \\
\end{longtable}

The coding has two stages, each with a page of its own: On the first
page the coder answers on his own. The page shows the facts of the run
and holds no proposal. Only after these answers have been taken in and
committed is the second page built. It shows the same facts, the coder's
own answer, and the proposal of an agent that took no part in the runs.
A second agent has checked the proposal. The command
\texttt{coding\ check} compares every quotation of the proposal with the
file that it names. The coder then gives his final category and
confidence. A reason is required where the final category differs from
his own. Both answers are kept with their times. The final answers carry
the hash of the file with the own answers, so the two files form a
chain.

Figure~\ref{fig-review-pages} (b) shows the first page of the coding of
the study. One engine builds the pages of both steps, so this page has
the properties of the blind page of the gate: one file that works
offline, nothing selected in advance, the times noted by the page, no
way to change a saved answer, and a download that names the hash of the
page. Its head holds the instructions: the definitions of the five
categories with their hints, of the three levels of confidence, and of
the three answers about the gap report, a procedure for a run, and
rules.

The section of a run opens with a button. It shows the unit, the history
of the attempts, the failing evaluations of the final attempt with their
crash lines, and the counts of the dependents and of the unit tests.
Below them it holds the files of the run with line numbers, each folded
until the coder opens it, among them the bundle, the final rebuild, the
gap report, and the evaluation file. The form asks four questions: the
category, the confidence, whether the gap report names the dominant
cause, and one line of reason. A note is optional. The page says that it
shows no proposal of an agent, and the file that it exports has no field
for one. A live copy of the page is online (Section~\ref{sec-gate}).

The order is deliberate. The protocol record rejects the alternative, to
show the proposal of an agent together with the facts of the run. The
coder's own answer comes first, on a page that does not contain the
proposal. The operating agent runs the commands and keeps the proposals
out of the coder's sight until the second page. The own answers are
committed before the second page is built. The limit is that of the
gate: the archive does not show that the coder did not look. There is
one coder, the maintainer. The study reports his own and his final
answer for each run and no rate.

Two rules cover the cases without a rejected run. A cell that uses up
its tries without a verdict counts as not accepted, is reported with its
abort classes, and is not coded. If no run is rejected, no coding of
runs is held. The coding is then demonstrated on the rebuild with the
closed formula of the controls, reported separately, and not counted as
a result.

\section{Measures, analysis, and archive}\label{sec-measures}

Three measures describe the specification of a unit.
Table~\ref{tbl-symbols} in Section~\ref{sec-notation} lists the symbols
of this section. Intent coverage and specification size are computed by
\texttt{babregen\ measures}, the mutation score by
\texttt{babregen\ mutate}.

\begin{definition}[Intent
coverage]\protect\hypertarget{def-coverage}{}\label{def-coverage}

For a unit \(u\), let \(\mathcal{I}(u)\) be the set of intent items of
its unit record. Let \(\mathcal{C}(u) \subseteq \mathcal{I}(u)\) be the
\emph{covered} items: those that name at least one durable evaluation
and whose named evaluations are all defined in an evaluation file of the
unit. The \emph{intent coverage} of \(u\) is the share of covered items,
\begin{equation}\protect\phantomsection\label{eq-coverage}{
c(u) = \frac{|\mathcal{C}(u)|}{|\mathcal{I}(u)|}.
}\end{equation} It measures the first half of the warning of Fowler
(2027, ch.~3) that ``Intent without evaluations is aspiration, and
evaluations without intent are a pile of checks with no theory''.

\end{definition}

Intent coverage is not code coverage: it counts the recorded promises
that have a judge, not the lines that ran. Full coverage does not imply
that all necessary promises are recorded. The static check and the
measure share one definition. The covered items are exactly the items
without a finding INTENT02 or EVAL03, so a clean static check implies
full coverage.

\begin{definition}[Mutation
score]\protect\hypertarget{def-mutation-score}{}\label{def-mutation-score}

Let \(\mathcal{M}(u)\) be the set of mutants that the mutation analysis
(Definition~\ref{def-mutation-analysis}) generates from the module of
\(u\). Let \(\mathcal{K}(u) \subseteq \mathcal{M}(u)\) be the mutants
that the durable evaluations of \(u\) kill. The \emph{mutation score} of
\(u\) is the share of killed mutants,
\begin{equation}\protect\phantomsection\label{eq-mutation-score}{
s(u) = \frac{|\mathcal{K}(u)|}{|\mathcal{M}(u)|}.
}\end{equation} The measure differs from the mutation score of Jia and
Harman (2011), the ratio of killed mutants to non-equivalent mutants, in
two respects. Only the durable evaluations run, because mutation
analysis grades these evaluations rather than the full acceptance rule.
Equivalent mutants stay in the denominator, so the value is a lower
bound: \(s(u)\) cannot exceed the share of killed mutants among the
mutants that are not equivalent. The verdicts of the gate tell the
equivalent survivors from the gaps (Section~\ref{sec-gate}). The score
estimates the strength of the evaluations, whether they notice a wrong
implementation (RQ2). It follows the advice of Fowler (2027, ch.~5) to
judge a suite by mutation testing.

\end{definition}

\begin{definition}[Specification
size]\protect\hypertarget{def-size}{}\label{def-size}

Let \(\ell_{\mathrm{unit}}(u)\), \(\ell_{\mathrm{dec}}(u)\),
\(\ell_{\mathrm{eval}}(u)\), and \(\ell_{\mathrm{code}}(u)\) be the
numbers of lines of the unit record of \(u\), of the decision records
its intent items cite, each counted once, of its evaluation files, and
of its implementation. The \emph{specification size} \(S(u)\) of \(u\)
and its ratio \(\rho(u)\) to the implementation are
\begin{equation}\protect\phantomsection\label{eq-size}{
S(u) = \ell_{\mathrm{unit}}(u) + \ell_{\mathrm{dec}}(u) + \ell_{\mathrm{eval}}(u),
\qquad
\rho(u) = \frac{S(u)}{\ell_{\mathrm{code}}(u)}.
}\end{equation} The ratio is reported because Fowler (2027, ch.~5)
observes that evaluation tends to become ``the larger part of the work,
larger than the generation it was built to check''. Counting lines is
our operationalisation of that observation.

\end{definition}

A line is a line of the file, blank lines and comments included.

The analysis is descriptive only. The accepted runs are reported as
counts per unit and generator, next to the durable evaluations passed
per attempt, the tokens, the seconds, and the mutation scores with the
number of mutants beside them. The verdicts and the categories are
reported as counts, together with the ratings of each survivor in their
order and with the own and the final answer of each coded run. There is
no significance test, no rate for the gate or the coding, no pooling
across units or generators, and no claim beyond this case: five runs per
cell show large differences only.

The archive is fixed in the same record. Everything is written below
\texttt{regen/runs/study/}: the setup, the controls, the mutation
reports, the gate, the run directories with the log of every call of the
driver, the reproduction, the coding, a summary, a log of the human
decisions with their times (\texttt{human-decisions.toml}), and the
scripts of the operating agent. The name of a try is
\texttt{\textless{}unit\textgreater{}-\textless{}c\textbar{}e\textgreater{}-seed\textless{}n\textgreater{}-try\textless{}k\textgreater{}},
with the letter of the generator and with
\texttt{\textless{}k\textgreater{}} counting the tries of the cell from
1. The second run with seed 1 has the name
\texttt{\textless{}unit\textgreater{}-\textless{}c\textbar{}e\textgreater{}-seed1-again-try\textless{}k\textgreater{}}.
The canaries have the names \texttt{canary-c} and \texttt{canary-e}.
After the last run and before the coding, the home directory of a run
machine is written as \texttt{/Users/USER} in the archived text files.
No other byte is changed. No bundle, no rebuild, and no run record is
touched. The archive is committed on the branch archive/study-v1 of
babregen-examples, which descends from the tag.

\section{Results}\label{sec-results}

With generator (c) the acceptance rule accepted 4 of the 5 runs of
\texttt{fibonacci} and 5 of the 5 runs of \texttt{primes}. With
generator (e) it accepted 5 of the 5 runs of \texttt{fibonacci} and 5 of
the 5 runs of \texttt{primes}. These are the 20 runs with the seeds 1 to
5. The 2 second runs of seed 1 of \texttt{fibonacci} were both accepted.
One run was rejected: \texttt{fibonacci} with generator (c) and seed 1.
The counts describe two generators on one small case with five seeds per
cell. They are not rates of success. The mutation analysis left 3
surviving mutants, all of \texttt{primes} (Section~\ref{sec-rq2}).

All numbers of this section are read from the clone of the package and,
for the study, from the archive under \texttt{regen/runs/study/} while
this document renders. A cell stops the render when a sentence and the
repository disagree.

\subsection{RQ1: what the tool drafts and what is left to the
maintainer}\label{sec-rq1}

Table~\ref{tbl-scaffold} compares the drafts of
\texttt{babregen\ new-unit} with the unit records of the tag
(Section~\ref{sec-redraft}). The comparison is computed on a throw-away
copy while this document renders. No person rated a draft.

\begin{longtable}[]{@{}
  >{\raggedright\arraybackslash}p{(\linewidth - 14\tabcolsep) * \real{0.1300}}
  >{\raggedright\arraybackslash}p{(\linewidth - 14\tabcolsep) * \real{0.1100}}
  >{\raggedright\arraybackslash}p{(\linewidth - 14\tabcolsep) * \real{0.1300}}
  >{\raggedright\arraybackslash}p{(\linewidth - 14\tabcolsep) * \real{0.1300}}
  >{\raggedright\arraybackslash}p{(\linewidth - 14\tabcolsep) * \real{0.1100}}
  >{\raggedright\arraybackslash}p{(\linewidth - 14\tabcolsep) * \real{0.1000}}
  >{\raggedright\arraybackslash}p{(\linewidth - 14\tabcolsep) * \real{0.1300}}
  >{\raggedright\arraybackslash}p{(\linewidth - 14\tabcolsep) * \real{0.1600}}@{}}
\caption{Drafts by \protect\texttt{babregen\ new-unit} compared with the
unit records of the tag for the two units. \emph{Public names} and
\emph{dependencies}: whether the draft and the record hold the same
fields \protect\texttt{public} and \protect\texttt{depends\_on}.
\emph{Dependent files}: consumers, unit tests, and evaluation files of
the record that the draft lists. \emph{Test files}: unit tests and
evaluation files that the draft lists in the field of the record.
\emph{Notes} and \emph{intent items}: notes of consumers and intent
items of the record that the draft holds. \emph{Lines} gives the lines
of the draft and of the record.}\label{tbl-scaffold}\tabularnewline
\toprule\noalign{}
\begin{minipage}[b]{\linewidth}\raggedright
unit
\end{minipage} & \begin{minipage}[b]{\linewidth}\raggedright
public names
\end{minipage} & \begin{minipage}[b]{\linewidth}\raggedright
dependencies
\end{minipage} & \begin{minipage}[b]{\linewidth}\raggedright
dependent files
\end{minipage} & \begin{minipage}[b]{\linewidth}\raggedright
test files
\end{minipage} & \begin{minipage}[b]{\linewidth}\raggedright
notes
\end{minipage} & \begin{minipage}[b]{\linewidth}\raggedright
intent items
\end{minipage} & \begin{minipage}[b]{\linewidth}\raggedright
lines
\end{minipage} \\
\midrule\noalign{}
\endfirsthead
\toprule\noalign{}
\begin{minipage}[b]{\linewidth}\raggedright
unit
\end{minipage} & \begin{minipage}[b]{\linewidth}\raggedright
public names
\end{minipage} & \begin{minipage}[b]{\linewidth}\raggedright
dependencies
\end{minipage} & \begin{minipage}[b]{\linewidth}\raggedright
dependent files
\end{minipage} & \begin{minipage}[b]{\linewidth}\raggedright
test files
\end{minipage} & \begin{minipage}[b]{\linewidth}\raggedright
notes
\end{minipage} & \begin{minipage}[b]{\linewidth}\raggedright
intent items
\end{minipage} & \begin{minipage}[b]{\linewidth}\raggedright
lines
\end{minipage} \\
\midrule\noalign{}
\endhead
\bottomrule\noalign{}
\endlastfoot
\texttt{fibonacci} & same & same & 5 of 5 & 2 of 2 & 0 of 3 & 0 of 3 &
40 of 69 \\
\texttt{primes} & same & same & 5 of 5 & 2 of 2 & 0 of 3 & 0 of 3 & 40
of 68 \\
\end{longtable}

Every draft holds what the tree shows. It has the public name of the
unit, taken from the list \texttt{\_\_all\_\_} of the module, and the
empty list of dependencies. It finds every dependent file: the package
that re-exports the function, the script, and the README as consumers
with their kinds, the file of the unit tests, and the evaluation file,
which it recognises by its name. The script and the README do not import
the module. The scan finds them through the name that the package
re-exports. No draft names a file that the record does not declare.

Apart from the pace, which the maintainer gives, a draft differs from
the record in two respects.

\begin{enumerate}
\def\labelenumi{\arabic{enumi}.}
\tightlist
\item
  The intent. No draft contains an intent item, and with it no link from
  a promise to a durable evaluation and no decision record. The two
  records hold 6 intent items. The static check refuses a draft with the
  error INTENT01. It also warns for every durable evaluation of the
  evaluation file that no intent item names it (COMPACT05).
\item
  The prose. The draft takes its summary from the first paragraph of the
  docstring of the module, which is \texttt{Fibonacci\ numbers.} and
  \texttt{Prime\ numbers.}. The summaries of the records state the
  public function and the first index. The draft leaves the note of
  every consumer empty. The records hold a note for each.
\end{enumerate}

The drafts have 58 \% of the lines of the records, comment lines
included. Everything the draft gets right can be recovered from the
source tree without reauthoring it. What the draft leaves empty is not
recovered by the static draft. This does not show that it is absent from
the tree. The docstring of the function \texttt{nth\_prime}, for
instance, states the first index, that no prime is skipped, and the
errors for a wrong index, which are the subjects of the intent items of
\texttt{primes}. A language model might propose intent items from such
text. The study did not try this. An agreement of draft and record is
partly circular (Section~\ref{sec-redraft}). The result matches the
division of labour that Fowler (2027, ch.~6) draws for decision records,
in which a machine collects the facts: ``The one thing you can't
automate is the business rationale''.

The two units are a simple case for the command. Each has one public
function, a list \texttt{\_\_all\_\_}, no import inside the package, and
test files with the conventional names. The comparison therefore shows
what a draft lacks in a simple case. It cannot show the differences that
only a larger unit has: public names beyond \texttt{\_\_all\_\_}, a
dependency that the record narrows to one name of a module, or a
consumer outside the repository. Table~\ref{tbl-records} summarises the
records of the two units and Table~\ref{tbl-size} their sizes.

\begin{longtable}[]{@{}
  >{\raggedright\arraybackslash}p{(\linewidth - 10\tabcolsep) * \real{0.1600}}
  >{\raggedleft\arraybackslash}p{(\linewidth - 10\tabcolsep) * \real{0.1500}}
  >{\raggedleft\arraybackslash}p{(\linewidth - 10\tabcolsep) * \real{0.1700}}
  >{\raggedleft\arraybackslash}p{(\linewidth - 10\tabcolsep) * \real{0.1500}}
  >{\raggedleft\arraybackslash}p{(\linewidth - 10\tabcolsep) * \real{0.1800}}
  >{\raggedleft\arraybackslash}p{(\linewidth - 10\tabcolsep) * \real{0.1900}}@{}}
\caption{The records of the two units. \emph{Intent items},
\emph{decision records}, and \emph{consumers} count the intent items of
the unit record, the decision records they cite, and the declared
consumers (Section~\ref{sec-deletion}). \emph{Durable evaluations}
counts them as pytest collects them, so each parametrised case counts
separately. \emph{Intent coverage} is the number of covered intent items
out of all intent items (Definition~\ref{def-coverage}). In the last row
a decision record that both units cite is counted
once.}\label{tbl-records}\tabularnewline
\toprule\noalign{}
\begin{minipage}[b]{\linewidth}\raggedright
unit
\end{minipage} & \begin{minipage}[b]{\linewidth}\raggedleft
intent items
\end{minipage} & \begin{minipage}[b]{\linewidth}\raggedleft
decision records
\end{minipage} & \begin{minipage}[b]{\linewidth}\raggedleft
consumers
\end{minipage} & \begin{minipage}[b]{\linewidth}\raggedleft
durable evaluations
\end{minipage} & \begin{minipage}[b]{\linewidth}\raggedleft
intent coverage
\end{minipage} \\
\midrule\noalign{}
\endfirsthead
\toprule\noalign{}
\begin{minipage}[b]{\linewidth}\raggedright
unit
\end{minipage} & \begin{minipage}[b]{\linewidth}\raggedleft
intent items
\end{minipage} & \begin{minipage}[b]{\linewidth}\raggedleft
decision records
\end{minipage} & \begin{minipage}[b]{\linewidth}\raggedleft
consumers
\end{minipage} & \begin{minipage}[b]{\linewidth}\raggedleft
durable evaluations
\end{minipage} & \begin{minipage}[b]{\linewidth}\raggedleft
intent coverage
\end{minipage} \\
\midrule\noalign{}
\endhead
\bottomrule\noalign{}
\endlastfoot
\texttt{fibonacci} & 3 & 2 & 3 & 10 & 3/3 \\
\texttt{primes} & 3 & 2 & 3 & 10 & 3/3 \\
both & 6 & 2 & 6 & 20 & 6/6 \\
\end{longtable}

The specification (Definition~\ref{def-size}) is larger than the
implementation in both units, by a ratio of 4.4 for \texttt{fibonacci}
and 3.3 for \texttt{primes}. The unit record alone is longer than the
module in both units. The evaluation file is longer than the module in
both units as well. With each decision record counted once, the
specification of the two units is 3.3 times as large as their
implementation (Table~\ref{tbl-size}). For modules of a few dozen lines
the ratio is high by construction: a record has a fixed part, and the
module has little code. \texttt{babregen\ measures} counts a decision
record for every unit that cites it, so its own total row states 323
lines and a ratio of 3.8 (Example~\ref{exm-tut-measures}).

\begin{longtable}[]{@{}
  >{\raggedright\arraybackslash}p{(\linewidth - 12\tabcolsep) * \real{0.1400}}
  >{\raggedleft\arraybackslash}p{(\linewidth - 12\tabcolsep) * \real{0.1600}}
  >{\raggedleft\arraybackslash}p{(\linewidth - 12\tabcolsep) * \real{0.1300}}
  >{\raggedleft\arraybackslash}p{(\linewidth - 12\tabcolsep) * \real{0.1600}}
  >{\raggedleft\arraybackslash}p{(\linewidth - 12\tabcolsep) * \real{0.1600}}
  >{\raggedleft\arraybackslash}p{(\linewidth - 12\tabcolsep) * \real{0.1700}}
  >{\raggedleft\arraybackslash}p{(\linewidth - 12\tabcolsep) * \real{0.0800}}@{}}
\caption{Lines of implementation and of specification per unit, as
\protect\texttt{babregen\ measures} prints them. All columns except
\emph{ratio} count lines. \emph{Specification size} is the sum of the
lines of the unit record, of the decision records its intent items cite
(each counted once) and of the evaluation files. \emph{Ratio} is
specification lines per line of implementation. In the last row a
decision record that both units cite is counted
once.}\label{tbl-size}\tabularnewline
\toprule\noalign{}
\begin{minipage}[b]{\linewidth}\raggedright
unit
\end{minipage} & \begin{minipage}[b]{\linewidth}\raggedleft
implementation
\end{minipage} & \begin{minipage}[b]{\linewidth}\raggedleft
unit record
\end{minipage} & \begin{minipage}[b]{\linewidth}\raggedleft
decision records
\end{minipage} & \begin{minipage}[b]{\linewidth}\raggedleft
evaluation files
\end{minipage} & \begin{minipage}[b]{\linewidth}\raggedleft
specification size
\end{minipage} & \begin{minipage}[b]{\linewidth}\raggedleft
ratio
\end{minipage} \\
\midrule\noalign{}
\endfirsthead
\toprule\noalign{}
\begin{minipage}[b]{\linewidth}\raggedright
unit
\end{minipage} & \begin{minipage}[b]{\linewidth}\raggedleft
implementation
\end{minipage} & \begin{minipage}[b]{\linewidth}\raggedleft
unit record
\end{minipage} & \begin{minipage}[b]{\linewidth}\raggedleft
decision records
\end{minipage} & \begin{minipage}[b]{\linewidth}\raggedleft
evaluation files
\end{minipage} & \begin{minipage}[b]{\linewidth}\raggedleft
specification size
\end{minipage} & \begin{minipage}[b]{\linewidth}\raggedleft
ratio
\end{minipage} \\
\midrule\noalign{}
\endhead
\bottomrule\noalign{}
\endlastfoot
\texttt{fibonacci} & 36 & 69 & 43 & 46 & 158 & 4.4 \\
\texttt{primes} & 50 & 68 & 43 & 54 & 165 & 3.3 \\
both & 86 & 137 & 43 & 100 & 280 & 3.3 \\
\end{longtable}

\subsection{RQ2: strength of the evaluations}\label{sec-rq2}

On the clone the static check reports 0 errors and 0 warnings, with the
archive of the study in the tree. All 20 durable evaluations of the two
units pass. Every intent item names at least one durable evaluation that
exists, so the intent coverage (Definition~\ref{def-coverage}) of both
units is complete (6/6 in total, Table~\ref{tbl-records}). It says that
every promise has a judge, not that the judge is strict.

The mutation score says how strict the judges are
(Definition~\ref{def-mutation-score}). It is read per unit.

\texttt{babregen\ mutate} ran on the two units at the tag, before the
first model run, and wrote one mutation report per unit
(Table~\ref{tbl-mutation}). The durable evaluations killed 11 of the 11
mutants of \texttt{fibonacci} and 28 of the 31 mutants of
\texttt{primes}. Of the killed mutants of \texttt{primes}, 7 ran into
the time limit of 10 seconds: they turned the search for the next prime
into a loop that does not return. The result is the one that was known
before the freeze (Section~\ref{sec-design}).

\begin{longtable}[]{@{}
  >{\raggedright\arraybackslash}p{(\linewidth - 12\tabcolsep) * \real{0.1400}}
  >{\raggedleft\arraybackslash}p{(\linewidth - 12\tabcolsep) * \real{0.1600}}
  >{\raggedleft\arraybackslash}p{(\linewidth - 12\tabcolsep) * \real{0.1200}}
  >{\raggedleft\arraybackslash}p{(\linewidth - 12\tabcolsep) * \real{0.1200}}
  >{\raggedleft\arraybackslash}p{(\linewidth - 12\tabcolsep) * \real{0.1200}}
  >{\raggedleft\arraybackslash}p{(\linewidth - 12\tabcolsep) * \real{0.1600}}
  >{\raggedleft\arraybackslash}p{(\linewidth - 12\tabcolsep) * \real{0.1800}}@{}}
\caption{Mutation analysis per unit at the tag, read from the mutation
reports of the archive. \emph{Score} is the mutation score
(Definition~\ref{def-mutation-score}). \emph{Time limit} counts the
killed mutants whose run exceeded its time limit. The scores are not
pooled across units.}\label{tbl-mutation}\tabularnewline
\toprule\noalign{}
\begin{minipage}[b]{\linewidth}\raggedright
unit
\end{minipage} & \begin{minipage}[b]{\linewidth}\raggedleft
evaluations
\end{minipage} & \begin{minipage}[b]{\linewidth}\raggedleft
mutants
\end{minipage} & \begin{minipage}[b]{\linewidth}\raggedleft
killed
\end{minipage} & \begin{minipage}[b]{\linewidth}\raggedleft
score
\end{minipage} & \begin{minipage}[b]{\linewidth}\raggedleft
time limit
\end{minipage} & \begin{minipage}[b]{\linewidth}\raggedleft
survivors
\end{minipage} \\
\midrule\noalign{}
\endfirsthead
\toprule\noalign{}
\begin{minipage}[b]{\linewidth}\raggedright
unit
\end{minipage} & \begin{minipage}[b]{\linewidth}\raggedleft
evaluations
\end{minipage} & \begin{minipage}[b]{\linewidth}\raggedleft
mutants
\end{minipage} & \begin{minipage}[b]{\linewidth}\raggedleft
killed
\end{minipage} & \begin{minipage}[b]{\linewidth}\raggedleft
score
\end{minipage} & \begin{minipage}[b]{\linewidth}\raggedleft
time limit
\end{minipage} & \begin{minipage}[b]{\linewidth}\raggedleft
survivors
\end{minipage} \\
\midrule\noalign{}
\endhead
\bottomrule\noalign{}
\endlastfoot
\texttt{fibonacci} & 10 & 11 & 11 & 100 \% & 0 & 0 \\
\texttt{primes} & 10 & 31 & 28 & 90 \% & 7 & 3 \\
\end{longtable}

Table~\ref{tbl-survivors} lists the 3 survivors as the report names
them. All 3 lie in the private helper \texttt{\_is\_prime} of
\texttt{primes}, on two lines that return \texttt{False}. Each survivor
is a change that the durable evaluations did not notice. A rebuild may
differ from the original in the same way and still be accepted, unless
the mutant is equivalent. What a survivor means is a verdict of a person
(Section~\ref{sec-gate}).

\begin{longtable}[]{@{}
  >{\raggedright\arraybackslash}p{(\linewidth - 10\tabcolsep) * \real{0.1200}}
  >{\raggedright\arraybackslash}p{(\linewidth - 10\tabcolsep) * \real{0.1000}}
  >{\raggedleft\arraybackslash}p{(\linewidth - 10\tabcolsep) * \real{0.0800}}
  >{\raggedright\arraybackslash}p{(\linewidth - 10\tabcolsep) * \real{0.1400}}
  >{\raggedright\arraybackslash}p{(\linewidth - 10\tabcolsep) * \real{0.2800}}
  >{\raggedright\arraybackslash}p{(\linewidth - 10\tabcolsep) * \real{0.2800}}@{}}
\caption{The surviving mutants, read from the mutation report of the
archive: the line of the module, the mutation operator, the original
code, and the mutated code.}\label{tbl-survivors}\tabularnewline
\toprule\noalign{}
\begin{minipage}[b]{\linewidth}\raggedright
unit
\end{minipage} & \begin{minipage}[b]{\linewidth}\raggedright
mutant
\end{minipage} & \begin{minipage}[b]{\linewidth}\raggedleft
line
\end{minipage} & \begin{minipage}[b]{\linewidth}\raggedright
operator
\end{minipage} & \begin{minipage}[b]{\linewidth}\raggedright
original
\end{minipage} & \begin{minipage}[b]{\linewidth}\raggedright
mutated
\end{minipage} \\
\midrule\noalign{}
\endfirsthead
\toprule\noalign{}
\begin{minipage}[b]{\linewidth}\raggedright
unit
\end{minipage} & \begin{minipage}[b]{\linewidth}\raggedright
mutant
\end{minipage} & \begin{minipage}[b]{\linewidth}\raggedleft
line
\end{minipage} & \begin{minipage}[b]{\linewidth}\raggedright
operator
\end{minipage} & \begin{minipage}[b]{\linewidth}\raggedright
original
\end{minipage} & \begin{minipage}[b]{\linewidth}\raggedright
mutated
\end{minipage} \\
\midrule\noalign{}
\endhead
\bottomrule\noalign{}
\endlastfoot
\texttt{primes} & M003 & 9 & return & \texttt{return\ False} &
\texttt{return\ None} \\
\texttt{primes} & M004 & 9 & constant & \texttt{False} &
\texttt{True} \\
\texttt{primes} & M010 & 13 & return & \texttt{return\ False} &
\texttt{return\ None} \\
\end{longtable}

\protect\phantomsection\label{gate-results}
\emph{The gate.} The gate was opened on the 3 survivors with
\texttt{babregen\ gate}. Two agents wrote the sealed proposals, and
\texttt{gate\ seal} drew 1 of the 3 survivors for a planted proposal.
The file \texttt{seal.toml} was committed and pushed with the mutation
reports, in the commit \texttt{f1a5884}, before the blind page was
built.

\par\ifdim\dimexpr\pagegoal-\pagetotal\relax<12\baselineskip\newpage\fi

\begin{longtable}[]{@{}
  >{\raggedright\arraybackslash}p{(\linewidth - 12\tabcolsep) * \real{0.1100}}
  >{\raggedright\arraybackslash}p{(\linewidth - 12\tabcolsep) * \real{0.1400}}
  >{\raggedright\arraybackslash}p{(\linewidth - 12\tabcolsep) * \real{0.1600}}
  >{\raggedright\arraybackslash}p{(\linewidth - 12\tabcolsep) * \real{0.1500}}
  >{\raggedright\arraybackslash}p{(\linewidth - 12\tabcolsep) * \real{0.0900}}
  >{\raggedright\arraybackslash}p{(\linewidth - 12\tabcolsep) * \real{0.1400}}
  >{\raggedright\arraybackslash}p{(\linewidth - 12\tabcolsep) * \real{0.2100}}@{}}
\caption{The ratings of the maintainer at the gate, read from the rating
files and the verdicts of the archive. \emph{Shown proposal} is the
verdict of the proposal on the second page. \emph{Drawn} marks the
survivor whose shown proposal was written for a verdict assigned at
random. The verdict is the third rating of a drawn survivor and the
second rating of every other.}\label{tbl-gate}\tabularnewline
\toprule\noalign{}
\begin{minipage}[b]{\linewidth}\raggedright
mutant
\end{minipage} & \begin{minipage}[b]{\linewidth}\raggedright
blind rating
\end{minipage} & \begin{minipage}[b]{\linewidth}\raggedright
shown proposal
\end{minipage} & \begin{minipage}[b]{\linewidth}\raggedright
second rating
\end{minipage} & \begin{minipage}[b]{\linewidth}\raggedright
drawn
\end{minipage} & \begin{minipage}[b]{\linewidth}\raggedright
third rating
\end{minipage} & \begin{minipage}[b]{\linewidth}\raggedright
verdict
\end{minipage} \\
\midrule\noalign{}
\endfirsthead
\toprule\noalign{}
\begin{minipage}[b]{\linewidth}\raggedright
mutant
\end{minipage} & \begin{minipage}[b]{\linewidth}\raggedright
blind rating
\end{minipage} & \begin{minipage}[b]{\linewidth}\raggedright
shown proposal
\end{minipage} & \begin{minipage}[b]{\linewidth}\raggedright
second rating
\end{minipage} & \begin{minipage}[b]{\linewidth}\raggedright
drawn
\end{minipage} & \begin{minipage}[b]{\linewidth}\raggedright
third rating
\end{minipage} & \begin{minipage}[b]{\linewidth}\raggedright
verdict
\end{minipage} \\
\midrule\noalign{}
\endhead
\bottomrule\noalign{}
\endlastfoot
M003 & equivalent & equivalent & equivalent & no & -- & equivalent \\
M004 & equivalent & output & equivalent & yes & equivalent &
equivalent \\
M010 & equivalent & equivalent & equivalent & no & -- & equivalent \\
\end{longtable}

The verdict of the maintainer is \emph{equivalent} for all 3 survivors.
Of the 3 second ratings, 2 name the verdict of the shown proposal. 2 of
them had named it in the blind rating already, and none changed to it.
The remaining second rating does not follow the shown proposal. The
drawn mutant is M004. Its second page showed a proposal for the verdict
\emph{output}, written for a verdict that was assigned at random. The
first agent had proposed \emph{equivalent}. The maintainer rated the
mutant \emph{equivalent} in the blind rating and kept this rating with
the planted proposal and after the reveal.

By these verdicts, none of the 3 survivors is a gap, so by the judgement
of the rater none shows a missing evaluation of the public contract.
With all 3 survivors rated equivalent, the durable evaluations of
\texttt{primes} kill 28 of the 28 mutants that the rater does not hold
to be equivalent.

The reasons of the verdicts, in the words of the rater, are these.

\begin{itemize}
\tightlist
\item
  M003: For nth-prime(), all promised return values are not affected.
\item
  M004: nth\_prime first calls \_is\_prime with 2 and then increases the
  candidate, so the mutated k \textless{} 2 branch is unreachable
  through the public interface.
\item
  M010: nth\_prime() uses the modified return value only in if
  \_is\_prime(candidate). Both return values (False as well as None)
  will return in a false condition, so that the public result is
  equivalent for both variants. Two unit tests detect M004, because they
  call the private helper. They do not judge a rebuild.
\end{itemize}

The proposal that was planted for M004 argued for output. By the wording
of the page this verdict is about a changed text, and M004 changes none,
so refusing the proposal says little.

Every rating is a judgement of one person, the maintainer of mathregen,
of its records, and of the tool. In the current study, there is no
second rater and no measure of agreement. The numbers of the gate are
counts of three survivors and of one drawn proposal, and there is no
rate. \texttt{babregen\ gate\ check} verifies the archive of the session
without a problem: the sealed files by their salted hashes, the chain of
the rating files, the order of the commits, and the rule of the plant
and of the verdicts.

\subsection{RQ3: regeneration with two local models}\label{sec-rq3}

This is the deletion test proper and the only part of the study in which
a language model writes a module. For each run the module of a unit is
deleted in a worktree, the generator receives the bundle of the unit,
and the acceptance rule judges what it writes
(Definition~\ref{def-regeneration}). The results are read from the file
\texttt{run.toml} of every run directory.

\emph{Controls.} The 7 controls were run at the tag before the first
model run, and all came out as they must. For both units the module of
the tag was accepted with all durable evaluations passed, and the module
that holds only its docstring was rejected. Of the three rebuilds kept
in babregen-examples, the Fibonacci numbers by fast doubling and the
primes by a sieve were accepted. The closed formula was rejected with 5
of 10 durable evaluations passed, although its unit tests and its
dependents pass. The driver repeated the positive control before the
first session of every run, with the counts of the positive control of
its unit.

\emph{Setup and canaries.} On both machines the manifest hash of the
model folder was the value of the protocol record, and the 26 tests of
the tag passed in the environment of the clone. The scoring environment
was the same on both machines: Python 3.14.2, pytest 9.2.0, and one hash
of the installed packages in every run record and in both mutation
reports. Each canary ran before the first run of its generator. The run
directory held the reply and the session log and nothing else, and the
only remote address of the driver's sockets was the model server on the
local machine.

\emph{Runs.} Table~\ref{tbl-regen-runs} has one row per run.
\emph{Attempts} gives, for each attempt in order, the passed durable
evaluations of its rebuild out of those that ran, or why the reply was
not scored: \emph{no module} is a reply from which the driver could not
take a complete module. The dependents are not in the table: both
dependents of the unit passed against the final rebuild of every run.
\emph{Unit tests} gives the unit tests of the original module that the
final rebuild passes, with the number of errors where there are any.
They are a second opinion and take no part in the acceptance.
\emph{Time} is the time of the model sessions of the run, the gap report
included. \emph{Input} and \emph{output} are the tokens of all sessions
of the run.

\begin{longtable}[]{@{}
  >{\raggedright\arraybackslash}p{(\linewidth - 16\tabcolsep) * \real{0.0600}}
  >{\raggedright\arraybackslash}p{(\linewidth - 16\tabcolsep) * \real{0.1300}}
  >{\raggedleft\arraybackslash}p{(\linewidth - 16\tabcolsep) * \real{0.1000}}
  >{\raggedright\arraybackslash}p{(\linewidth - 16\tabcolsep) * \real{0.2500}}
  >{\raggedleft\arraybackslash}p{(\linewidth - 16\tabcolsep) * \real{0.1400}}
  >{\raggedleft\arraybackslash}p{(\linewidth - 16\tabcolsep) * \real{0.0800}}
  >{\raggedleft\arraybackslash}p{(\linewidth - 16\tabcolsep) * \real{0.0800}}
  >{\raggedleft\arraybackslash}p{(\linewidth - 16\tabcolsep) * \real{0.0800}}
  >{\raggedright\arraybackslash}p{(\linewidth - 16\tabcolsep) * \real{0.0800}}@{}}
\caption{The regeneration runs of the study, one row per run, read from
the \protect\texttt{run.toml} of each run directory. The columns are
explained in the text.}\label{tbl-regen-runs}\tabularnewline
\toprule\noalign{}
\begin{minipage}[b]{\linewidth}\raggedright
gen.
\end{minipage} & \begin{minipage}[b]{\linewidth}\raggedright
unit
\end{minipage} & \begin{minipage}[b]{\linewidth}\raggedleft
seed
\end{minipage} & \begin{minipage}[b]{\linewidth}\raggedright
attempts
\end{minipage} & \begin{minipage}[b]{\linewidth}\raggedleft
unit tests
\end{minipage} & \begin{minipage}[b]{\linewidth}\raggedleft
time (s)
\end{minipage} & \begin{minipage}[b]{\linewidth}\raggedleft
input
\end{minipage} & \begin{minipage}[b]{\linewidth}\raggedleft
output
\end{minipage} & \begin{minipage}[b]{\linewidth}\raggedright
accepted
\end{minipage} \\
\midrule\noalign{}
\endfirsthead
\toprule\noalign{}
\begin{minipage}[b]{\linewidth}\raggedright
gen.
\end{minipage} & \begin{minipage}[b]{\linewidth}\raggedright
unit
\end{minipage} & \begin{minipage}[b]{\linewidth}\raggedleft
seed
\end{minipage} & \begin{minipage}[b]{\linewidth}\raggedright
attempts
\end{minipage} & \begin{minipage}[b]{\linewidth}\raggedleft
unit tests
\end{minipage} & \begin{minipage}[b]{\linewidth}\raggedleft
time (s)
\end{minipage} & \begin{minipage}[b]{\linewidth}\raggedleft
input
\end{minipage} & \begin{minipage}[b]{\linewidth}\raggedleft
output
\end{minipage} & \begin{minipage}[b]{\linewidth}\raggedright
accepted
\end{minipage} \\
\midrule\noalign{}
\endhead
\bottomrule\noalign{}
\endlastfoot
(c) & \texttt{fibonacci} & 1 & no module, no module, 9/10 & 2/2 & 18.6 &
8,896 & 1,815 & no \\
(c) & \texttt{fibonacci} & 2 & 9/10, 10/10 & 2/2 & 18.7 & 7,083 & 1,842
& yes \\
(c) & \texttt{fibonacci} & 3 & 9/10, 10/10 & 2/2 & 19.5 & 7,097 & 1,902
& yes \\
(c) & \texttt{fibonacci} & 4 & 9/10, 10/10 & 2/2 & 16.0 & 7,132 & 1,587
& yes \\
(c) & \texttt{fibonacci} & 5 & no module, 9/10, 10/10 & 2/2 & 22.0 &
9,254 & 2,183 & yes \\
(c) & \texttt{fibonacci} & 1~(2nd) & 9/10, 10/10 & 2/2 & 18.1 & 7,094 &
1,759 & yes \\
(c) & \texttt{primes} & 1 & no module, 9/10, 10/10 & 0/0 (1 error) &
22.3 & 9,232 & 2,091 & yes \\
(c) & \texttt{primes} & 2 & 9/10, 9/10, 10/10 & 0/0 (1 error) & 23.4 &
9,579 & 2,255 & yes \\
(c) & \texttt{primes} & 3 & 9/10, 10/10 & 0/0 (1 error) & 17.5 & 7,075 &
1,727 & yes \\
(c) & \texttt{primes} & 4 & no module, 9/10, 10/10 & 0/0 (1 error) &
20.8 & 9,163 & 1,997 & yes \\
(c) & \texttt{primes} & 5 & 9/10, 10/10 & 0/0 (1 error) & 19.9 & 6,951 &
1,996 & yes \\
(e) & \texttt{fibonacci} & 1 & 10/10 & 2/2 & 55.1 & 4,719 & 2,364 &
yes \\
(e) & \texttt{fibonacci} & 2 & 10/10 & 2/2 & 34.5 & 4,611 & 1,538 &
yes \\
(e) & \texttt{fibonacci} & 3 & 10/10 & 2/2 & 28.8 & 4,611 & 1,505 &
yes \\
(e) & \texttt{fibonacci} & 4 & 10/10 & 2/2 & 52.8 & 4,720 & 2,505 &
yes \\
(e) & \texttt{fibonacci} & 5 & 10/10 & 2/2 & 31.9 & 4,611 & 1,661 &
yes \\
(e) & \texttt{fibonacci} & 1~(2nd) & 10/10 & 2/2 & 41.7 & 4,611 & 2,194
& yes \\
(e) & \texttt{primes} & 1 & 10/10 & 0/0 (1 error) & 37.8 & 4,617 & 1,509
& yes \\
(e) & \texttt{primes} & 2 & 10/10 & 4/4 & 43.5 & 4,676 & 2,016 & yes \\
(e) & \texttt{primes} & 3 & 10/10 & 4/4 & 44.6 & 4,670 & 2,078 & yes \\
(e) & \texttt{primes} & 4 & 10/10 & 4/4 & 39.1 & 4,697 & 1,829 & yes \\
(e) & \texttt{primes} & 5 & 10/10 & 0/0 (1 error) & 32.5 & 4,617 & 1,699
& yes \\
\end{longtable}

Every cell ended in its first try: no run was aborted. In 5 of 5 runs of
\texttt{primes} with generator (c) and in 2 of 5 with generator (e) the
unit tests end in an error. The file of the unit tests imports the
private helper \texttt{\_is\_prime}, and the final rebuild of these runs
has no function of that name, so pytest collects none of its tests. All
of these runs were accepted: the unit tests are no criterion of the
acceptance rule (Definition~\ref{def-acceptance},
Example~\ref{exm-evaluation}). In the other runs of \texttt{primes} the
generator wrote a helper with this name, and the unit tests pass.

\par\ifdim\dimexpr\pagegoal-\pagetotal\relax<15\baselineskip\newpage\fi

\begin{longtable}[]{@{}
  >{\raggedright\arraybackslash}p{(\linewidth - 18\tabcolsep) * \real{0.0600}}
  >{\raggedright\arraybackslash}p{(\linewidth - 18\tabcolsep) * \real{0.1300}}
  >{\raggedleft\arraybackslash}p{(\linewidth - 18\tabcolsep) * \real{0.1000}}
  >{\raggedleft\arraybackslash}p{(\linewidth - 18\tabcolsep) * \real{0.1000}}
  >{\raggedleft\arraybackslash}p{(\linewidth - 18\tabcolsep) * \real{0.1000}}
  >{\raggedright\arraybackslash}p{(\linewidth - 18\tabcolsep) * \real{0.1600}}
  >{\raggedleft\arraybackslash}p{(\linewidth - 18\tabcolsep) * \real{0.1000}}
  >{\raggedleft\arraybackslash}p{(\linewidth - 18\tabcolsep) * \real{0.0900}}
  >{\raggedleft\arraybackslash}p{(\linewidth - 18\tabcolsep) * \real{0.0900}}
  >{\raggedleft\arraybackslash}p{(\linewidth - 18\tabcolsep) * \real{0.0700}}@{}}
\caption{The outcomes per generator and unit. \emph{First attempt},
\emph{after repair}, and \emph{rejected} count the runs with the seeds 1
to 5 by their final verdict. \emph{Second run} is the outcome of the
second run of seed 1. The other columns count over all runs of the row:
the \emph{attempts}, the attempts with \emph{no module}, the scored
attempts with \emph{one failing} durable evaluation, and the attempts
that passed \emph{all} of them. The re-scoring reproduced every
verdict.}\label{tbl-regen-summary}\tabularnewline
\toprule\noalign{}
\begin{minipage}[b]{\linewidth}\raggedright
gen.
\end{minipage} & \begin{minipage}[b]{\linewidth}\raggedright
unit
\end{minipage} & \begin{minipage}[b]{\linewidth}\raggedleft
first attempt
\end{minipage} & \begin{minipage}[b]{\linewidth}\raggedleft
after repair
\end{minipage} & \begin{minipage}[b]{\linewidth}\raggedleft
rejected
\end{minipage} & \begin{minipage}[b]{\linewidth}\raggedright
second run
\end{minipage} & \begin{minipage}[b]{\linewidth}\raggedleft
attempts
\end{minipage} & \begin{minipage}[b]{\linewidth}\raggedleft
no module
\end{minipage} & \begin{minipage}[b]{\linewidth}\raggedleft
one failing
\end{minipage} & \begin{minipage}[b]{\linewidth}\raggedleft
all
\end{minipage} \\
\midrule\noalign{}
\endfirsthead
\toprule\noalign{}
\begin{minipage}[b]{\linewidth}\raggedright
gen.
\end{minipage} & \begin{minipage}[b]{\linewidth}\raggedright
unit
\end{minipage} & \begin{minipage}[b]{\linewidth}\raggedleft
first attempt
\end{minipage} & \begin{minipage}[b]{\linewidth}\raggedleft
after repair
\end{minipage} & \begin{minipage}[b]{\linewidth}\raggedleft
rejected
\end{minipage} & \begin{minipage}[b]{\linewidth}\raggedright
second run
\end{minipage} & \begin{minipage}[b]{\linewidth}\raggedleft
attempts
\end{minipage} & \begin{minipage}[b]{\linewidth}\raggedleft
no module
\end{minipage} & \begin{minipage}[b]{\linewidth}\raggedleft
one failing
\end{minipage} & \begin{minipage}[b]{\linewidth}\raggedleft
all
\end{minipage} \\
\midrule\noalign{}
\endhead
\bottomrule\noalign{}
\endlastfoot
(c) & \texttt{fibonacci} & 0 & 4 & 1 & after repair & 14 & 3 & 6 & 5 \\
(c) & \texttt{primes} & 0 & 5 & 0 & -- & 13 & 2 & 6 & 5 \\
(e) & \texttt{fibonacci} & 5 & 0 & 0 & first attempt & 6 & 0 & 0 & 6 \\
(e) & \texttt{primes} & 5 & 0 & 0 & -- & 5 & 0 & 0 & 5 \\
\end{longtable}

A repair can follow a failing evaluation or a reply without a module. No
run was made without this feedback, so the table does not show its
effect. The one failing evaluation has the same name in every such
attempt and in both units,
\texttt{test\_a\_bool\_index\_raises\_type\_error}: it asserts that an
index of the type \texttt{bool} raises a \texttt{TypeError}. The intent
items FIB-3 and PRI-3 name it. In every accepted run of generator (c) a
repair passed it. The seconds of Table~\ref{tbl-regen-runs} do not
compare the two models, because the generators ran on two machines.

\emph{Repeated seed.} With each generator the second run of
\texttt{fibonacci} with seed 1 received the same first prompt as the
first run. With both generators the first reply differed from that of
the first run. A seed therefore does not fix the reply on these servers,
and the five runs of a cell are five samples. With generator (c) the two
runs also differ in their verdict: the first run of the cell is the
rejected run of the study, and the second run was accepted.

\emph{Reproduction.} The 22 final rebuilds were scored again at the tag,
each on the machine of its run. Every recorded verdict was reproduced,
with the same counts of the durable evaluations, the unit tests, and the
dependents. The reproduction of generator (c) ran directly after its
last run, while the runs of generator (e) on the other machine were
still going.

\subsection{RQ4: what the rejected run shows}\label{sec-rq4}

\protect\phantomsection\label{rejected-run}
The acceptance rule rejected one run: \texttt{fibonacci} with generator
(c) and seed 1, the run directory \texttt{fibonacci-c-seed1-try1}. The
archive of the runs holds these facts about it.

\begin{itemize}
\tightlist
\item
  Each reply of the first two attempts held a complete module, but the
  reply closed it with three backticks where the prompt asks for four.
  The driver therefore took no module from these replies and scored
  nothing. The third attempt, the last, wrote a module. No prompt of
  this run reported a failing evaluation.
\item
  This final rebuild passes 9 of the 10 durable evaluations. The one
  that fails is \texttt{test\_a\_bool\_index\_raises\_type\_error}: the
  rebuild does not raise a \texttt{TypeError} for an index of the type
  \texttt{bool}.
\item
  The failing evaluation is named by intent item FIB-3, whose invariant
  says that an index that is not an \texttt{int} raises
  \texttt{TypeError} and that this includes \texttt{bool}. The item is
  in the bundle that the generator received.
\item
  The final rebuild passes the 2 unit tests of the original and both
  dependents. The durable evaluation alone rejects it.
\item
  The generator wrote a gap report for the run, and the re-scoring
  reproduced the rejection (Section~\ref{sec-rq3}).
\end{itemize}

In every other run of generator (c), also in the second run of the same
cell, a later attempt passed the failing evaluation
(Section~\ref{sec-rq3}).

These facts say which promise the final rebuild broke. They do not say
why. Whether the records state the promise unclearly, whether the
generator did not do what the records say, or whether the driver is to
blame is a judgement, and the protocol leaves it to the coder
(Section~\ref{sec-failure}). The page of the coding does not show the
replies of the first two attempts, so the coder did not know that they
held a module.

\emph{The coding.} On the first page, which holds no proposal, the
maintainer put the run \texttt{fibonacci-c-seed1-try1} into the category
\emph{generator error}, with the confidence medium. To the question
whether the gap report of the generator names the dominant cause he
answered \emph{no}. On the second page, which shows the checked proposal
of an agent, his final category is \emph{generator error}, with the
confidence medium. It is the category of his own answer. The proposal of
the agent names the same category, \emph{generator error}, with the
confidence high, and the second agent that checked the proposal agrees.
The own answer was thus equal to the proposal before the maintainer saw
it, and the final answer is the own answer. The own answer took 563
seconds from the opening of the run on the page to the saving, and the
final answer 154 seconds (Table~\ref{tbl-coding}).

\begin{longtable}[]{@{}
  >{\raggedright\arraybackslash}p{(\linewidth - 10\tabcolsep) * \real{0.1900}}
  >{\raggedright\arraybackslash}p{(\linewidth - 10\tabcolsep) * \real{0.1700}}
  >{\raggedright\arraybackslash}p{(\linewidth - 10\tabcolsep) * \real{0.1100}}
  >{\raggedright\arraybackslash}p{(\linewidth - 10\tabcolsep) * \real{0.1400}}
  >{\raggedleft\arraybackslash}p{(\linewidth - 10\tabcolsep) * \real{0.0900}}
  >{\raggedright\arraybackslash}p{(\linewidth - 10\tabcolsep) * \real{0.3000}}@{}}
\caption{The coding of the rejected run
\protect\texttt{fibonacci-c-seed1-try1}. The own and the final answer of
the maintainer are read from the answers files of the archive, with the
seconds from the opening of the run to the saving and with the reason in
the words of the coder. The rows of the proposal and of its check are
copied from the README of the archive. The first page held no
proposal.}\label{tbl-coding}\tabularnewline
\toprule\noalign{}
\begin{minipage}[b]{\linewidth}\raggedright
answer
\end{minipage} & \begin{minipage}[b]{\linewidth}\raggedright
category
\end{minipage} & \begin{minipage}[b]{\linewidth}\raggedright
confidence
\end{minipage} & \begin{minipage}[b]{\linewidth}\raggedright
gap report names the cause
\end{minipage} & \begin{minipage}[b]{\linewidth}\raggedleft
seconds
\end{minipage} & \begin{minipage}[b]{\linewidth}\raggedright
reason of the coder
\end{minipage} \\
\midrule\noalign{}
\endfirsthead
\toprule\noalign{}
\begin{minipage}[b]{\linewidth}\raggedright
answer
\end{minipage} & \begin{minipage}[b]{\linewidth}\raggedright
category
\end{minipage} & \begin{minipage}[b]{\linewidth}\raggedright
confidence
\end{minipage} & \begin{minipage}[b]{\linewidth}\raggedright
gap report names the cause
\end{minipage} & \begin{minipage}[b]{\linewidth}\raggedleft
seconds
\end{minipage} & \begin{minipage}[b]{\linewidth}\raggedright
reason of the coder
\end{minipage} \\
\midrule\noalign{}
\endhead
\bottomrule\noalign{}
\endlastfoot
own answer & generator error & medium & no & 563 & The specification was
known by the generator, but the generator does not implement it
correctly. The generator did not consider that bool is a subclass of int
in Python, so the test failed. \\
proposal of the agent & generator error & high & no & -- & -- \\
check of the second agent & agrees: generator error & -- & -- & -- &
-- \\
final answer & generator error & medium & -- & 154 & Although line 44 in
the test script uses bool, they are in Python a subclass of int, so the
test fails. \\
\end{longtable}

The category is a judgement of one person, the maintainer of mathregen,
of its records, and of the tool. In the current study, there is no
second rater and no measure of agreement. The study reports the own and
the final answer of one run and no rate. One coded run shows how the two
stages work on a real rejection. It says nothing about how often a coder
follows a proposal. \texttt{babregen\ coding\ check} verifies the
archive of the session without a problem: both pages, the chain of the
two answers files, the order of the commits, and every quotation of the
proposal.

\section{Discussion}\label{sec-discussion}

\subsection{Answers to the research questions}\label{sec-disc-answers}

\emph{RQ1.} The tool drafts what the tree holds: the public name of a
unit, its dependencies inside the package, its test files with their
roles, and its consumers inside the repository
(Table~\ref{tbl-scaffold}). The rest of a record is authored or approved
by the maintainer: the intent with its evaluations and decisions, the
pace, and the prose. The two units are a simple case for the command.
The comparison counts fields and rates nothing.

\emph{RQ2.} The durable evaluations kill 11 of the 11 mutants of
\texttt{fibonacci} and 28 of the 31 mutants of \texttt{primes}
(Table~\ref{tbl-mutation}). The 3 survivors lie in the private helper of
\texttt{primes}. What they mean is the verdict of the maintainer at the
gate: \emph{equivalent} for all 3 survivors (Table~\ref{tbl-gate}). By
this verdict none of the three survivors shows a missing evaluation of
the public contract. The verdict is the judgement of one person. It
rests on one reading of the vocabulary, the one that asks what a caller
of the public function can observe.

\emph{RQ3.} Under this protocol the acceptance rule accepted, with
generator (c), 4 of the 5 runs of \texttt{fibonacci} and 5 of the 5 runs
of \texttt{primes}, and with generator (e), 5 of the 5 runs of
\texttt{fibonacci} and 5 of the 5 runs of \texttt{primes}
(Table~\ref{tbl-regen-summary}). The controls gave the expected verdict
for every file, and the re-scoring reproduced every recorded verdict
(Section~\ref{sec-rq3}). The verdicts are therefore statements about the
rebuilds and, where attempts were lost, about the format of the replies.
An accepted run shows that the bundle, for generator (c) together with
the failures shown in a repair, sufficed for this generator, within two
limits: behaviour that no evaluation guards was free to change
(Section~\ref{sec-undecided}), and a rebuild from memory is not excluded
(Section~\ref{sec-threats}). Besides the records, the bundle holds the
public signatures and docstrings and the source of the durable
evaluations. The study does not show what each part contributed. For
these units the dependents are the import of the package and one
docstring example. The script and the README are not run. Every run of
generator (e) was accepted at its first attempt, every run of generator
(c) but one after a repair.

\emph{RQ4.} One run was rejected. Its final rebuild fails one durable
evaluation, the one for an index of the type \texttt{bool}, and it
passes the unit tests and the dependents (Section~\ref{sec-rq4}). It
violated FIB-3. Two earlier replies with a wrong format had used up two
of the three attempts, so no prompt of the run reported the broken
promise. The durable evaluation therefore did what it is there for: it
kept a rebuild out that everything else would have let pass. Why the run
failed is the category of the coder. He coded the rejected run as
\emph{generator error} on both pages: first without a proposal, then
with the checked proposal of an agent (Table~\ref{tbl-coding}). By this
category the records say what the evaluation demands, and the rebuild
does not do it.

\subsection{What acceptance leaves undecided}\label{sec-undecided}

An accepted rebuild can differ from the original. Three kinds of
difference have to be told apart. The closed formula breaks promises
that the records state (FIB-2, FIB-3) and is rejected. The sieve takes
freedoms that the records grant: the architecture file lists the
algorithm, the private helpers, and the wording of error messages as
open choices. It is accepted. The third kind is neither promised nor
granted. The original functions accept an index whose type is a subclass
of \texttt{int} other than \texttt{bool}, such as a member of an
\texttt{IntEnum}. The sieve and the fast doubling refuse it with a
\texttt{TypeError}. Both are accepted. FIB-3 and PRI-3 say that an index
that is not an \texttt{int} is refused and that this includes
\texttt{bool}. Whether this refuses other subclasses can be read in two
ways, and no evaluation tries one. Fernandez (2026) states the lesson as
a rule: ``A regenerator turns an unstated type rule into a coin flip''.
A reviewer of a draft found the difference after the freeze. The final
rebuild of every run accepts a subclass as the original does, so no
verdict on a run depends on it. The maintainer decides that the choice
is free. It then belongs among the open choices of the next version of
the records. The records of the study are not changed.

The same comparison, made after the freeze, shows a loss in the runs.
The bundle does not hold the list \texttt{\_\_all\_\_} of a module, and
no final rebuild has one. In one accepted run of \texttt{primes}
(generator (c), seed 3) a star import therefore also exports a helper.
Neither the rule nor the static check notices it.

\subsection{Consequences of the human steps and of the draft
comparison}\label{sec-disc-consequences}

Regenerative coding claims that the knowledge about a module survives
the deletion of its code. The fourth contribution concerns what the
maintainer authors or approves before a rebuild. The third concerns what
a person judges around the rebuilds. Without the draft comparison it is
open whether the records hold more than a static tool recovers from the
code. If they did not, a tool could extract them as long as the code
exists. And nobody would have to maintain them. Without the two human
steps it is open what an acceptance or a rejection says about the
records. An acceptance is as strong as the durable evaluations. A
mutation score does not say whether a surviving mutant is a gap in them.
A rejection has several possible causes, and only some of them lie in
the records. The order of the human steps has a reason of its own. The
person answers before a proposal is shown, because a proposal of an
agent invites automation bias (Section~\ref{sec-bg-human}). The files
that the two commands write turn a judgement into evidence that a reader
can check.

\subsubsection{The place in the study}\label{the-place-in-the-study}

In Figure~\ref{fig-workflow} no arrow leaves the gate or the coding. The
first rating was saved after the last run had ended. No category changes
a verdict of the acceptance rule. The tool acts on neither result: what
follows is a decision of the maintainer. The draft comparison has no box
in the figure. It concerns the content of the first box. No draft
entered a bundle.

\subsubsection{The consequence of each
answer}\label{the-consequence-of-each-answer}

Table~\ref{tbl-consequences} lists what the maintainer can do with each
answer after the study. The answers say where a change would be due: in
the specification, in the generator or the driver, or nowhere. With
three survivors and one rejected run, most rows of the table have the
count zero. Section~\ref{sec-hypothetical} applies the table to invented
answers that fill it.

\begin{longtable}[]{@{}
  >{\raggedright\arraybackslash}p{(\linewidth - 8\tabcolsep) * \real{0.1000}}
  >{\raggedright\arraybackslash}p{(\linewidth - 8\tabcolsep) * \real{0.2000}}
  >{\raggedright\arraybackslash}p{(\linewidth - 8\tabcolsep) * \real{0.2800}}
  >{\raggedright\arraybackslash}p{(\linewidth - 8\tabcolsep) * \real{0.3400}}
  >{\raggedleft\arraybackslash}p{(\linewidth - 8\tabcolsep) * \real{0.0800}}@{}}
\caption{The answers of the two human steps and their consequences.
\emph{Says} is what the answer states about the specification.
\emph{Consequence} is a change that the maintainer can make after the
study. \emph{Count} is the number of survivors with the verdict and of
rejected runs with the category in this
study.}\label{tbl-consequences}\tabularnewline
\toprule\noalign{}
\begin{minipage}[b]{\linewidth}\raggedright
step
\end{minipage} & \begin{minipage}[b]{\linewidth}\raggedright
answer
\end{minipage} & \begin{minipage}[b]{\linewidth}\raggedright
says
\end{minipage} & \begin{minipage}[b]{\linewidth}\raggedright
consequence
\end{minipage} & \begin{minipage}[b]{\linewidth}\raggedleft
count
\end{minipage} \\
\midrule\noalign{}
\endfirsthead
\toprule\noalign{}
\begin{minipage}[b]{\linewidth}\raggedright
step
\end{minipage} & \begin{minipage}[b]{\linewidth}\raggedright
answer
\end{minipage} & \begin{minipage}[b]{\linewidth}\raggedright
says
\end{minipage} & \begin{minipage}[b]{\linewidth}\raggedright
consequence
\end{minipage} & \begin{minipage}[b]{\linewidth}\raggedleft
count
\end{minipage} \\
\midrule\noalign{}
\endhead
\bottomrule\noalign{}
\endlastfoot
gate & equivalent & No evaluation through the public function can detect
the mutant. & none. The mutation score stays as it is. & 3 \\
gate & gap & A promise is not guarded. An accepted rebuild may break it.
& add a durable evaluation for the promise. & 0 \\
gate & output & No intent item pins the behaviour. & check whether the
records leave it open. If they are silent, the maintainer decides and
records the choice. & 0 \\
coding & missing knowledge & The records are silent. & add the intent
item, or remove the demand from the evaluation. & 0 \\
coding & ambiguous knowledge & The records allow two readings. & reword
the intent item so that one reading remains. & 0 \\
coding & byte-level pinning & An evaluation demands more than the
records. & loosen the evaluation, or state the text in an intent item. &
0 \\
coding & generator error & The records say it. & none in the records.
Use more attempts or another generator. & 1 \\
coding & harness error & The run says nothing about the records. &
repair the driver or the judge. & 0 \\
\end{longtable}

A change of the intent, the contract, or the consumers has a price. It
changes the bundle. The static check then reports that the newest
archived run of the unit judged a different bundle (PROV08). The
verdicts of the study are evidence about the records of the tag. An
improved specification needs new runs.

\subsubsection{What followed in this
study}\label{what-followed-in-this-study}

At the gate all three survivors were rated \emph{equivalent}. No durable
evaluation is added. The accepted rebuilds of \texttt{primes} lose no
support from the survivors, because a mutant that cannot change the
behaviour of the public function marks no promise that a rebuild could
break unnoticed. Had a survivor been a gap, the accepted runs would not
have shown that the promise is kept. The planted proposal has no
consequence for the verdict. It shows whether the rating followed a
proposal, which here it did not. The rejected run was coded
\emph{generator error}. No intent item is added or reworded, because the
records state the broken promise (FIB-3). The run remains in the results
as a failure under the acceptance rule. By its category, held with
medium confidence, it does not count as missing knowledge. Had the
category been \emph{missing knowledge}, the same rejection would have
counted against the records. Both consequences rest on the judgement of
one rater and inherit its limits. With the other reading of the
vocabulary a verdict, and with it the consequence, could differ.

The draft comparison has three consequences for a maintainer. First, it
divides the labour. The facts of the source tree are left to the tool,
and the attention of the maintainer goes to the intent, the reasons, and
the notes. Second, it qualifies the cost of the records. The
specification of both units is larger than the code
(Table~\ref{tbl-size}). A part of it need not be typed. Third, it
justifies a guard. A draft holds no intent, and the static check refuses
it (INTENT01), so a draft as the tool writes it does not pass the check.
The comparison has no consequence for the runs. It does not decide
whether a language model could draft the intent.

The rejected run connects the two contributions. An intent item such as
FIB-3 is what a draft lacks, and the coding attributes the failure to
the generator and not to this item.

\subsection{Two ways to use the tool}\label{sec-disc-usage}

The study starts from a small package whose records were written for it.
Two other uses are plausible: an existing package that has grown over
years, and a package that is still being written. The tool supports each
of them in part, and the study shows neither.

\begin{itemize}
\tightlist
\item
  The first use is an existing package. The tool works unit by unit. The
  command \texttt{new-unit} drafts a record for a module that has none.
  The static check accepts a tree in which only some modules have a
  record: on a copy of mathregen without the record of
  \texttt{fibonacci} it counts one unit and reports nothing about the
  module that lost its record. The tool is recorded in this way itself.
  Release 0.2.0 keeps 18 unit records for 50 Python files. A bundle, a
  mutation report, and a run each belong to one unit, so the tool offers
  no step that rebuilds a package as a whole. A maintainer of an old
  package can therefore record and rebuild one module at a time. The
  order of the modules is advice and no result, because the
  documentation of the tool gives none and the study has no old package.
  A reasonable order follows from what makes a unit hard to rebuild
  (Section~\ref{sec-conclusion}): a module without state, without
  imports from its own package, and with consumers inside its repository
  comes first, and the pace of a unit says how much evidence a change of
  it needs (Section~\ref{sec-pace}). Whether the records of an aged
  module can be written at all, when the people who knew its reasons
  have left, is the open question behind the first of the next steps.
\item
  The second use is a package that is being written, also one whose code
  a language model writes. The worked example shows this use on a small
  scale: a third function arrives, and the check refuses its module
  until the module has a place in the architecture
  (Section~\ref{sec-tut-third}). Two kinds of repetition are kept apart
  here. The static check is deterministic and fast. The tool places it
  in the guard that runs before every push, and mathregen does so. With
  every new unit it compares all records with the tree again, so a new
  module cannot use an earlier unit unnoticed: an import that the
  architecture does not allow and a consumer that no record declares are
  findings. The mutation analysis, the scoring, and a regeneration are
  slow and stay outside the guard. They are due again when a unit
  changes. After a change of a record the check reports that the newest
  archived run of the unit judged a different bundle (PROV08,
  Section~\ref{sec-redraft}), which marks the archived verdict as one
  about an older specification. A repeated check thus renews no verdict.
  The study does not show a model that writes a new unit: its two models
  rebuild modules whose records exist.
\end{itemize}

\subsection{What a small demonstrator can and cannot
show}\label{sec-disc-limits}

The object of the study is small. Every step of the procedure has
something to do on it. Its use is that of a demonstrator. Every step
that needs no model can be repeated on a laptop. The price is a ceiling.
The protocol record expected few or no rejected runs. The runs confirm
it: the acceptance rule rejected one run. With almost every run
accepted, the study cannot rank the generators or the units, and an
accepted run says little about the knowledge in the records. The package
has no state, no input or output, no dependency, and no consumer outside
its repository. The aged and tacit knowledge that motivates the deletion
test is not in it. The results are therefore statements about the
procedure: every machine step ran, every verdict was reproduced, and
each of the two human steps leaves an archive that a reader can check.
They are no statements about how well records preserve the knowledge of
a mature module.

The human steps have the same character. Their outcome is as uniform as
it can be: the same verdict for every survivor in every rating, also
against the planted proposal, and the same category in the own answer,
in the proposal, and in the final answer. Such an outcome shows no
influence of a proposal.

\subsection{Lessons for regeneration drivers}\label{sec-disc-lessons}

Every accepted run of generator (c) reached the rebuild that passed
every durable evaluation in a repair. In each of them a repair that
received a summary of the failures passed, in one run the second such
repair. The attempts that gave no module (Table~\ref{tbl-regen-summary})
closed it with the wrong number of backticks. Such an attempt is used up
like any other, so the number of attempts is a budget for both kinds of
reply. On these servers a driver cannot replay a run from its seed. It
has to archive every prompt and every reply, as \texttt{babregen\ local}
does in the session log of a run.

\subsection{Threats to validity}\label{sec-threats}

\subsubsection{Construct validity}\label{construct-validity}

Acceptance stands for the claim that the records hold the knowledge a
module needs. The acceptance rule guards the obvious shortcuts. A
rebuild must pass every durable evaluation and every dependent. Mutation
analysis grades the durable evaluations. Two risks remain. An accepted
run does not show that the intent is complete
(Section~\ref{sec-undecided}). And for a textbook function an accepted
run may owe little to the records. The architecture file leaves the
algorithm open. A model is likely to know either function without a
bundle. The records add the contract: the first index, the exact result,
and the refusal of an index that is no \texttt{int}.

The mutation score shows that the durable evaluations would notice a
change, not that they test the right thing (Fowler 2027, ch.~5). The
verdict on a survivor is a judgement of one person and no measurement.
Without the verdicts the score mixes gaps with equivalent mutants
(Definition~\ref{def-mutation-score}). The sample of mutants keeps a
bias: each operator changes one node, so a fault that spans several
statements is not in it. No operator removes a function body. The
mutation tool is part of babregen and is not cross-checked with a second
tool. A mutation score depends on the operators of its tool.

The category of a rejected run is a judgement as well. The five
categories are fixed in the tool, and no second person applied them. One
rejected run shows how the coding works and nothing about the
distribution of causes. A place of a record that no rebuild reached was
not tested, so the absence of a failure is no evidence that a record is
complete. The comparison of drafts and records is mechanical
(Section~\ref{sec-redraft}). It shows which fields a draft holds, not
how much work the missing fields are, and no person judged whether a
draft would have been a good start. Specification size counts lines. A
line count is a proxy for what the maintainer must author or approve and
keep current, not a measure of effort. The durable evaluations are
durable within one language only. Each is a pytest function, which
survives a rewrite of a module in Python but not a rewrite in another
language (Section~\ref{sec-evaluation}). Pace is recorded but not
exercised. Both units are fast. The tool does not run the verification
that a class requires (Section~\ref{sec-pace}).

\subsubsection{Internal validity}\label{internal-validity}

One person holds every human role. The maintainer of the package also
maintains its records, its durable evaluations, and the tool. He
approved the protocol, he rates the survivors of his own evaluations. He
also codes the rejection of a rebuild of his own module. The protocol
states this as a gap and not as a design. The two commands take his
judgement before they show him a proposal, while an agent operates them.
The archive supports the order of the commitments
(Section~\ref{sec-gate}).

The blind rating of the gate is blind to the proposals of the agents. It
is not blind to what the rater knew before. It was not given without
help. The mutation analysis had been run in a trial before the freeze.
The maintainer was told how many mutants survived and on which lines.

A rejected run may reflect a limitation of the specification or the
generator, or a defect of the harness. The positive control shows that
the durable evaluations and the dependents of a unit can be passed at
all. The reproduction shows that a second scoring, soon after the first
and on the same machine, gives the same verdict. The negative controls
are an empty module and a module with the closed formula. They show that
the judge can reject. Whether it rejects a rebuild that is wrong in
another detail is what the mutation analysis probes.

The two generators differ in more than their weights. They ran on two
machines, with two context lengths and two output budgets. One of them
has a reasoning mode that the protocol switches off. A difference
between their counts is therefore a difference between two setups. The
scoring environment was the same on both machines, by the hash of the
installed packages that every run records.

A model may have seen the deleted modules during its training, or
thousands of modules like them. A rebuild from memory would then count
as a rebuild from the records. The study does not exclude this. For two
textbook functions it is likely for the algorithm. It matters less for
the contract of the two modules, which is their own.

The confinement of a generator rests on the driver, which offers no
tools and records the remote addresses of its own sockets. The driver
does not observe the model server, which runs on the same machine and
could itself reach another machine. The protocol is an internal
pre-specification. It is a record of babregen-examples in a pushed tag.
Every run records the commit of the tag as its base, so a change of the
protocol after the freeze would show in the archive. One condition of
the record was met late: the hashes of the model files were computed
after the tag and before the first run. The protocol is not registered
with a third party.

\subsubsection{External validity}\label{external-validity}

The case is one young package that was written as an example. It has one
maintainer and two modules of a few dozen lines with pure functions
(Section~\ref{sec-disc-limits}). The guidelines of Runeson and Höst
(2009) exclude studies on toy programs, so the term case study applies
in a limited sense only. The results do not extend to other packages, to
units with state or with dependencies, to older code whose lessons were
never written down, or to teams. The answer to RQ3 holds for two
generators with one set of sampling options, for one prompt, on two
machines, and for a protocol in which the generator writes a whole
module in one reply and may repair it twice
(Table~\ref{tbl-generators}).

Fowler (2026c, slide 42) names three places where regeneration is the
wrong move for now: when the hard part is the intent, when evidence must
be assurance, and when nobody is allowed to fail. The study does not
reach the second and the third. Nothing is cut over. Every rebuild lives
in a worktree and is archived as evidence. The measures and the verdicts
are evidence, not assurance.

\subsubsection{Conclusion validity}\label{conclusion-validity}

The sample is small: 22 runs, five per unit and generator and two second
runs. The runs of a unit are not independent. They share one bundle, so
a gap in the bundle can be a common cause of their failures. Even
versions written separately from one carefully checked specification did
not fail independently (Knight and Leveson 1986). The counts describe
these runs. They are not estimates for other units, seeds, or prompts.
No test compares two of them. The human part is smaller still
(Section~\ref{sec-disc-limits}). The reproduction shows that the
verdicts are reproducible. It does not show that they are right: a
defect that the judge has at the tag is the same in both scorings.

A verdict holds for the bundle and the module of its day. The static
check reports a run whose hashes no longer match the current bundle or
module (Example~\ref{exm-provenance}). The warning answers half of the
open question of Fowler (2026c, slide 43), whether the retained
representation ``hold{[}s{]} what the next replacement needs, and do we
notice when it stops''. No check answers the first half. The study has
not been replicated by another group.

\section{Conclusion and future work}\label{sec-conclusion}

This report introduces \emph{regenerative coding}. It maps the seven
primitives of Fowler (2027) to the records, the check IDs, and the
commands of babregen 0.2.0 (Table~\ref{tbl-primitives}) and carries out
a deletion test on the two modules of the package mathregen. The command
\texttt{new-unit} drafts the facts of the source tree and no intent
item, although the docstrings already express a part of the intent. The
intent and the prose are authored or approved by the maintainer (RQ1).
The durable evaluations killed 11 of the 11 mutants of
\texttt{fibonacci} and 28 of the 31 mutants of \texttt{primes} (RQ2).
Two local generators rebuilt the modules from their bundles: generator
(c) in 4 of 5 runs of \texttt{fibonacci} and 5 of 5 runs of
\texttt{primes}, generator (e) in 5 of 5 runs of \texttt{fibonacci} and
5 of 5 runs of \texttt{primes}, with the seeds 1 to 5 (RQ3). The one
rejected run fails one durable evaluation and passes the unit tests and
the dependents. Two of its three attempts were lost to the format of the
reply (RQ4). Every verdict was reproduced from the archive. The
acceptance rule thus rejected one rebuild that violated a documented
input contract. A check after the freeze found that one accepted
rebuild, \texttt{primes} with generator (c) and seed 3, exposed an
additional name through a star import (Section~\ref{sec-undecided}).

At the gate the verdict of the maintainer is \emph{equivalent} for all 3
survivors, also for the one survivor whose shown proposal was planted.
He coded the rejected run as \emph{generator error} on both pages: first
without a proposal, then with the checked proposal of an agent. Both are
the judgement of one rater, who is the maintainer of the package, of its
records, and of the tool, so the study reports counts of three and of
one and no rate.

The study is a demonstrator (Section~\ref{sec-disc-limits}). It shows
that the procedure, from the records to the reproduced verdict, runs on
a package that a reader can clone. Its benefits are a person who answers
for the knowledge, a check that compares the records with the code, a
rule that asks no model and does not compare a rebuild with the old
code, failures that a person can trace, and open evidence
(Section~\ref{sec-bg-benefits}).

Five steps are next.

\begin{enumerate}
\def\labelenumi{\arabic{enumi}.}
\tightlist
\item
  First, the procedure is applied to units that are harder to rebuild:
  modules with state, with dependencies inside their package, and with
  consumers outside their repository, in an older repository with
  another maintainer. The package spanregen, the second case in
  babregen-examples, is a first step in this direction. It was added
  after the study with release 0.4.0 of the tool and is not an object of
  this study. It stores sets of integers as spans and has four units
  that depend on each other: a value type for a closed interval of
  integers, the normal form of a set, the set algebra, and a text
  notation. In a first run with the smaller local generator, two of its
  four units were accepted and two were rejected. A maintainer of an
  older repository meets the records first as text to be typed.
  Section~\ref{sec-hmi} describes how the records are written and what
  the tool does about typical input mistakes. A comparison of a bundle
  that holds only the contract and the evaluations with one that also
  holds the intent and the reasons, on unfamiliar requirements and with
  an independently written evaluation, can show what the records
  contribute. An older repository also raises the question of where to
  begin. A table can order its units by two properties: the pace that
  the maintainer declares (Definition~\ref{def-pace}) and the degree to
  which a unit reads or writes state outside itself. A unit of fast pace
  without state is a first candidate. A unit that writes to a shared
  database is a late one. So far this order is advice that the study
  does not test (Section~\ref{sec-disc-usage}). A unit that calls an
  external service needs a replacement of that service while
  \texttt{babregen\ score} runs. The replacement has to follow the
  contract of the service, so that no rebuild is accepted against an
  interface that does not exist.
\item
  Second, the cost of the records is weighed against their use. The
  study counts lines (Table~\ref{tbl-size}) and no hours, so it cannot
  say when the records pay off. A follow-up study measures the time to
  write and to maintain the records and the time that a later rebuild
  saves, for example after a dependency was replaced. Until then, two
  questions can guide the choice of a unit. Is the loss of the knowledge
  about the unit likely and costly? Is its implementation expected to be
  replaced often? A script that has run unchanged for years answers both
  with no, and its records would not repay their cost.
\item
  Third, the human steps get a second rater and more material. With at
  least 15 survivors more than one proposal is drawn, and with two
  raters an agreement can be stated. Only then can a study say something
  about the effect of a proposal on a rating. For a team this becomes a
  rule of two keys: two maintainers rate a survivor or code a rejected
  run blind and independently. And an answer enters the counts only when
  both give it. A disagreement between generator error and missing
  knowledge is a finding of its own, because a rater who also maintains
  the records has a reason to prefer the generator error. With enough
  cases a chance-corrected measure of agreement such as kappa is
  reported with the counts.
\item
  Fourth, the tool takes up what the two sessions showed: an explanation
  of the verdicts where the rater reads it, a vocabulary that says whose
  observation counts as behaviour, a wording of the verdict output that
  separates what the records leave open from what they are silent about,
  and a check that notices a stage taken in twice.
\item
  Last, but not least, babregen's user interface will be improved. Three
  interfaces are available in babregen: (i) an interactive Quarto
  document (Allaire et al. 2026), (ii) a terminal-based procedure, and
  (iii) an HTML-based interface. All of them are subject of further
  research, because generating and maintaining the knowledge in a
  user-friendly manner is a central element of the regenerative coding
  approach.
\end{enumerate}

\section*{Data and code availability}\label{data-and-code-availability}
\addcontentsline{toc}{section}{Data and code availability}

The tool is the package babregen (Bartz-Beielstein 2026b),
\url{https://gitlab.git.nrw/thomas.bartz-beielstein/babregen}, release
0.2.0, which is on PyPI and is documented at
\url{https://advm1.gm.fh-koeln.de/~bartz/babregen/} (Bartz-Beielstein
2026a). The object of the study, the package mathregen with its records
and its durable evaluations, is the folder \texttt{mathregen/} of the
repository babregen-examples,
\url{https://gitlab.git.nrw/thomas.bartz-beielstein/babregen-examples}.
The freeze of the study is the tag \texttt{mathregen-study-v1} of
babregen-examples, which points at the commit \texttt{eae4473}. The
archive of the study is on the branch \texttt{archive/study-v1}, under
\texttt{mathregen/regen/runs/study/}. The source of this report is the
Quarto document \texttt{index.qmd} of the repository bart26x-regen.
Rendering it runs every example with babregen 0.2.0 against a clone of
that branch. A reader needs read access only. The controls, the mutation
analysis, the runs, and the reproduction run in a clone of
babregen-examples and write below \texttt{regen/runs/}, which needs no
commit. The gate and the coding need commits, which a reader makes on a
branch of the clone or in a fork. The gate asks for a push of the sealed
key only if the branch has an upstream.\footnote{A branch is a line of
  work in a copy of a repository. It has an upstream when it is linked
  to a branch on a server, so that the changes made on one's own machine
  can be sent there, which is called a push. A branch that a reader
  creates in a copy has no upstream until the reader links it to one.}
Without one, the blind page records that the key was not pushed.

Release 0.3.1 of babregen, published on 30 November 2026 after the
study, provides a browser interface, the command \texttt{babregen\ ui}.
It shows the state of every unit, runs commands from pages on the
machine of the maintainer, and serves the pages of the two human steps.
\texttt{gate\ assign} and \texttt{gate\ seal} stay in the terminal. The
commands of release 0.2.0, their options, and the checks are unchanged,
and the command \texttt{example} has one more kind. No byte of a review
page changes: the code of release 0.3.1 rebuilds the five pages of the
study byte for byte. A review directory is verified with the release
that started it, so this archive is checked with release 0.2.0. The
interface has forms for the four commands that write a record. They use
the functions of these commands, and the argument of
Section~\ref{sec-hmi} stands: the prose of an invariant and the names in
the tree are typed as before. The interface is described in a guide,
\url{https://advm1.gm.fh-koeln.de/~bartz/babregen/docs/ui.html}, and in
the reference of the commands,
\url{https://advm1.gm.fh-koeln.de/~bartz/babregen/docs/commands.html\#ui}.
Its design record is ADR 0002 in the repository of babregen
(Bartz-Beielstein 2026b). Release 0.3.2 and release 0.4.0, both of 1
December 2026, change the interface only. Release 0.5.0 of the same day
adds a report of the results of the two human steps
(Section~\ref{sec-hypo-report}).

\section*{Use of AI tools}\label{use-of-ai-tools}
\addcontentsline{toc}{section}{Use of AI tools}

Transparency notice: parts of this paper were researched, drafted, and
checked with the support of artificial intelligence. This information is
provided voluntarily for complete transparency, as no legal requirement
mandates its disclosure.

\ifPDFTeX\else\catcode"0107=\active \protected\def^^^^0107{\'c}\fi

\section*{References}\label{references}
\addcontentsline{toc}{section}{References}

\protect\phantomsection\label{refs}

\clearpage
\appendix

\section*{Appendix}\label{appendix}

\section{Notation}\label{sec-notation}

The symbols of the measures (Section~\ref{sec-measures}) and of the
regeneration runs (Section~\ref{sec-regeneration}) are listed in
Table~\ref{tbl-symbols}. Each symbol has one meaning throughout the
report. The lower-case \(c\) and \(s\) are measures, the calligraphic
\(\mathcal{C}\) is a set, and the upper-case \(S\) is a size, so the
pairs \(c\), \(\mathcal{C}\) and \(s\), \(S\) name different quantities.

\begin{longtable}[]{@{}
  >{\raggedright\arraybackslash}p{(\linewidth - 6\tabcolsep) * \real{0.1600}}
  >{\raggedright\arraybackslash}p{(\linewidth - 6\tabcolsep) * \real{0.5200}}
  >{\raggedright\arraybackslash}p{(\linewidth - 6\tabcolsep) * \real{0.1700}}
  >{\raggedright\arraybackslash}p{(\linewidth - 6\tabcolsep) * \real{0.1500}}@{}}
\caption{Symbols of the report, ordered Latin lower case, Latin upper
case, calligraphic, Greek, operators and relations. Every symbol is a
function of the unit \(u\) except \(n\), \(u\) and the
operators.}\label{tbl-symbols}\tabularnewline
\toprule\noalign{}
\begin{minipage}[b]{\linewidth}\raggedright
Symbol
\end{minipage} & \begin{minipage}[b]{\linewidth}\raggedright
Meaning
\end{minipage} & \begin{minipage}[b]{\linewidth}\raggedright
Type
\end{minipage} & \begin{minipage}[b]{\linewidth}\raggedright
Defined in
\end{minipage} \\
\midrule\noalign{}
\endfirsthead
\toprule\noalign{}
\begin{minipage}[b]{\linewidth}\raggedright
Symbol
\end{minipage} & \begin{minipage}[b]{\linewidth}\raggedright
Meaning
\end{minipage} & \begin{minipage}[b]{\linewidth}\raggedright
Type
\end{minipage} & \begin{minipage}[b]{\linewidth}\raggedright
Defined in
\end{minipage} \\
\midrule\noalign{}
\endhead
\bottomrule\noalign{}
\endlastfoot
\(c(u)\) & intent coverage of \(u\), the share of covered intent items &
share in \([0, 1]\) & Definition~\ref{def-coverage},
Equation~\ref{eq-coverage} \\
\(\ell_{\mathrm{code}}(u)\) & number of lines of the implementation of
\(u\) & count & Definition~\ref{def-size} \\
\(\ell_{\mathrm{dec}}(u)\) & number of lines of the decision records
that the intent items of \(u\) cite, each counted once & count &
Definition~\ref{def-size} \\
\(\ell_{\mathrm{eval}}(u)\) & number of lines of the evaluation files of
\(u\) & count & Definition~\ref{def-size} \\
\(\ell_{\mathrm{unit}}(u)\) & number of lines of the unit record of
\(u\) & count & Definition~\ref{def-size} \\
\(n\) & number of an attempt in a run. The repair of attempt \(n\) uses
the seed of the run plus \(n\) & positive integer &
Section~\ref{sec-regeneration} \\
\(s(u)\) & mutation score of \(u\), the share of killed mutants & share
in \([0, 1]\) & Definition~\ref{def-mutation-score},
Equation~\ref{eq-mutation-score} \\
\(u\) & a unit of the study & unit & Definition~\ref{def-coverage} \\
\(S(u)\) & specification size of \(u\) in lines & count &
Definition~\ref{def-size}, Equation~\ref{eq-size} \\
\(\mathcal{C}(u)\) & covered intent items of \(u\), which name at least
one durable evaluation and whose named evaluations are all defined in an
evaluation file of \(u\) & set,
\(\mathcal{C}(u) \subseteq \mathcal{I}(u)\) &
Definition~\ref{def-coverage} \\
\(\mathcal{I}(u)\) & intent items of the unit record of \(u\) & set &
Definition~\ref{def-coverage} \\
\(\mathcal{K}(u)\) & mutants of \(u\) that the durable evaluations of
\(u\) kill & set, \(\mathcal{K}(u) \subseteq \mathcal{M}(u)\) &
Definition~\ref{def-mutation-score} \\
\(\mathcal{M}(u)\) & mutants that the mutation analysis generates from
the module of \(u\) & set & Definition~\ref{def-mutation-score} \\
\(\rho(u)\) & ratio of the specification size of \(u\) to the lines of
its implementation & ratio & Definition~\ref{def-size},
Equation~\ref{eq-size} \\
\(\lvert \cdot \rvert\) & number of elements of a set & operator &
Equation~\ref{eq-coverage} \\
\(\subseteq\) & subset or equal & relation &
Definition~\ref{def-coverage} \\
\end{longtable}

Table~\ref{tbl-acronyms} lists the acronyms of the report and the
identifier schemes of the records. Each is written out where it is first
used.

\begingroup\small

\begin{longtable}[]{@{}
  >{\raggedright\arraybackslash}p{(\linewidth - 4\tabcolsep) * \real{0.2400}}
  >{\raggedright\arraybackslash}p{(\linewidth - 4\tabcolsep) * \real{0.5600}}
  >{\raggedright\arraybackslash}p{(\linewidth - 4\tabcolsep) * \real{0.2000}}@{}}
\caption{Acronyms and identifier schemes of the report, in alphabetical
order.}\label{tbl-acronyms}\tabularnewline
\toprule\noalign{}
\begin{minipage}[b]{\linewidth}\raggedright
Acronym
\end{minipage} & \begin{minipage}[b]{\linewidth}\raggedright
Meaning
\end{minipage} & \begin{minipage}[b]{\linewidth}\raggedright
Introduced in
\end{minipage} \\
\midrule\noalign{}
\endfirsthead
\toprule\noalign{}
\begin{minipage}[b]{\linewidth}\raggedright
Acronym
\end{minipage} & \begin{minipage}[b]{\linewidth}\raggedright
Meaning
\end{minipage} & \begin{minipage}[b]{\linewidth}\raggedright
Introduced in
\end{minipage} \\
\midrule\noalign{}
\endhead
\bottomrule\noalign{}
\endlastfoot
AI & artificial intelligence & Section~\ref{sec-bg-regen} \\
COMPACT, COMPILE, DELETE, EVAL, INTENT, PACE, PROV & families of check
IDs, named after the primitives compaction, compilation, deletion,
evaluation, intent, pace, and provenance & Section~\ref{sec-records} \\
FIB & prefix of the intent items of the unit \texttt{fibonacci} &
Section~\ref{sec-records} \\
ID & identifier & Section~\ref{sec-intro} \\
LLM & large language model & Section~\ref{sec-intro} \\
PRI & prefix of the intent items of the unit \texttt{primes} &
Section~\ref{sec-records} \\
PyPI & Python Package Index & Section~\ref{sec-intro} \\
RQ & research question & Section~\ref{sec-intro} \\
SHA-256 & Secure Hash Algorithm with a 256-bit digest &
Section~\ref{sec-evaluation} \\
TOML & Tom's Obvious, Minimal Language, the format of the records &
Section~\ref{sec-records} \\
\end{longtable}

\endgroup

\section{Tool reference}\label{sec-tool-reference}

This appendix holds what Section~\ref{sec-approach} leaves out: the
commands of babregen, the conventions of their output, the files of the
two human steps, the table of check IDs, and one longer listing.

\subsection{Commands and their output}\label{commands-and-their-output}

\begin{longtable}[]{@{}
  >{\raggedright\arraybackslash}p{(\linewidth - 2\tabcolsep) * \real{0.2000}}
  >{\raggedright\arraybackslash}p{(\linewidth - 2\tabcolsep) * \real{0.8000}}@{}}
\caption{The commands of babregen with the descriptions that its help
prints. Table~\ref{tbl-primitives} relates the first nine to the
primitives, except \protect\texttt{new-unit}. The last two are the
commands of the two human steps.}\label{tbl-commands}\tabularnewline
\toprule\noalign{}
\begin{minipage}[b]{\linewidth}\raggedright
command
\end{minipage} & \begin{minipage}[b]{\linewidth}\raggedright
what it does
\end{minipage} \\
\midrule\noalign{}
\endfirsthead
\toprule\noalign{}
\begin{minipage}[b]{\linewidth}\raggedright
command
\end{minipage} & \begin{minipage}[b]{\linewidth}\raggedright
what it does
\end{minipage} \\
\midrule\noalign{}
\endhead
\bottomrule\noalign{}
\endlastfoot
\texttt{check} & validate regen/ against the source tree \\
\texttt{deletion-test} & blast radius and knowledge gaps of one unit \\
\texttt{context} & write the regeneration bundle of one unit \\
\texttt{compact} & list compaction candidates \\
\texttt{measures} & pace, intent coverage and specification size of
every unit \\
\texttt{mutate} & mutation score of one unit: run its evaluations
against mutants \\
\texttt{score} & judge a candidate module for one unit: evaluations,
unit tests, dependents \\
\texttt{new-unit} & draft regen/units/\textless unit\textgreater.toml
for a module from the tree \\
\texttt{local} & regenerate a unit with a local model in short sessions
(non-deterministic) \\
\texttt{schema} & print the JSON Schema of a record kind (for
editors) \\
\texttt{explain} & explain a check ID: condition, a wrong and a right
fragment, the usual fix \\
\texttt{example} & print one complete, valid, commented record of a
kind \\
\texttt{new-decision} & create regen/decisions/ID.toml from the options
(never overwrites) \\
\texttt{add-intent} & append one {[}{[}intent{]}{]} block to the record
of a unit \\
\texttt{add-consumer} & append one {[}{[}consumers{]}{]} block to the
record of a unit \\
\texttt{coding} & code why rejected runs failed, on an offline page:
start, check, page, intake \\
\texttt{gate} & rate surviving mutants blind, in three stages with
offline pages \\
\end{longtable}

Every command reads the records of the repository in which it is called,
except \texttt{explain}, \texttt{example}, and \texttt{schema}, which
need no records. It exits with code 0 when it succeeds and with 1 when
its answer is negative, for instance when the static check reports an
error, \texttt{babregen\ score} rejects a rebuild, or a mutant survives.
A user error, such as a record with an unknown key or a value of the
wrong type, or an unknown unit, ends a command with code 2. A finding of
the static check is one line of the form
\texttt{path:line:\ level\ ID\ message}, without the line number where a
check has none. The last line counts the units, the errors, and the
warnings. A warning fails the check only with the option
\texttt{-\/-strict}. Every step of \texttt{coding} and \texttt{gate}
uses the same three codes: 0 for success, 1 when content that was
checked fails, such as a quotation or a chain, and 2 for an error of the
input, such as a step that is taken out of its order or finds a changed
file.

\subsection{The files of the two human
steps}\label{the-files-of-the-two-human-steps}

\texttt{babregen\ gate} and \texttt{babregen\ coding} work in a
\emph{review directory} below \texttt{regen/runs/} that holds no run
record. Table~\ref{tbl-review-files} lists the files of the two review
directories of the study. Every data file is TOML and starts with
\texttt{format\ =\ 1}. The briefs for the agents are not listed. The
tool reads each data file strictly: an unknown key, a value of the wrong
type, or a word outside its vocabulary is an error. Whether a rating
followed a proposal is computed and stored nowhere.

\begin{longtable}[]{@{}
  >{\raggedright\arraybackslash}p{(\linewidth - 4\tabcolsep) * \real{0.3100}}
  >{\raggedright\arraybackslash}p{(\linewidth - 4\tabcolsep) * \real{0.2500}}
  >{\raggedright\arraybackslash}p{(\linewidth - 4\tabcolsep) * \real{0.4400}}@{}}
\caption{The files of the two review directories of the study, relative
to
\protect\texttt{regen/runs/study/}.}\label{tbl-review-files}\tabularnewline
\toprule\noalign{}
\begin{minipage}[b]{\linewidth}\raggedright
file
\end{minipage} & \begin{minipage}[b]{\linewidth}\raggedright
written by
\end{minipage} & \begin{minipage}[b]{\linewidth}\raggedright
holds
\end{minipage} \\
\midrule\noalign{}
\endfirsthead
\toprule\noalign{}
\begin{minipage}[b]{\linewidth}\raggedright
file
\end{minipage} & \begin{minipage}[b]{\linewidth}\raggedright
written by
\end{minipage} & \begin{minipage}[b]{\linewidth}\raggedright
holds
\end{minipage} \\
\midrule\noalign{}
\endhead
\bottomrule\noalign{}
\endlastfoot
\texttt{gate/gate.toml} & \texttt{gate\ start} & the survivors with
their diffs and excerpts, the mutation reports with their hashes, the
rule of the draw \\
\texttt{gate/sealed/} & the two agents, \texttt{gate\ assign},
\texttt{gate\ seal} & the proposals, the assigned verdicts, and the key.
It is kept out of every commit until the last intake \\
\texttt{gate/seal.toml} & \texttt{gate\ seal} & the salted hashes of the
sealed files, the number of survivors, and the number drawn \\
\texttt{gate/gate-\textless{}stage\textgreater{}.html} &
\texttt{gate\ page} & the page of a stage: blind, second, third \\
\texttt{gate/rating-\textless{}stage\textgreater{}.toml} & the page;
\texttt{gate\ intake} copies the bytes & the ratings of a stage with
their reasons and times, the hash of the page, of the seal, and of the
rating file before \\
\texttt{gate/verdicts.toml} & the last \texttt{gate\ intake} & the
verdict of every survivor and whether it was drawn \\
\texttt{coding/coding.toml} & \texttt{coding\ start} & the coded runs
with the hashes of their run records \\
\texttt{coding/proposals/} & an agent, and a second agent that checks &
the proposal for a run with its quotations \\
\texttt{coding/coding-\textless{}stage\textgreater{}.html} &
\texttt{coding\ page} & the page of a stage: own, final \\
\texttt{coding/answers-\textless{}stage\textgreater{}.toml} & the page;
\texttt{coding\ intake} copies the bytes & the category, the confidence,
and the reason of a stage with their times; the final answers name the
hash of the own answers \\
\end{longtable}

\subsection{Check IDs}\label{check-ids}

Table~\ref{tbl-checks} lists every check ID with its level and its
condition, as the table of checks of babregen states them.

\begin{longtable}[]{@{}
  >{\raggedright\arraybackslash}p{(\linewidth - 4\tabcolsep) * \real{0.1400}}
  >{\raggedright\arraybackslash}p{(\linewidth - 4\tabcolsep) * \real{0.1000}}
  >{\raggedright\arraybackslash}p{(\linewidth - 4\tabcolsep) * \real{0.7600}}@{}}
\caption{The check IDs of babregen, grouped by family, with their levels
and conditions. The rows are read from the table of checks of the tool
while this document renders.}\label{tbl-checks}\tabularnewline
\toprule\noalign{}
\begin{minipage}[b]{\linewidth}\raggedright
check ID
\end{minipage} & \begin{minipage}[b]{\linewidth}\raggedright
level
\end{minipage} & \begin{minipage}[b]{\linewidth}\raggedright
condition
\end{minipage} \\
\midrule\noalign{}
\endfirsthead
\toprule\noalign{}
\begin{minipage}[b]{\linewidth}\raggedright
check ID
\end{minipage} & \begin{minipage}[b]{\linewidth}\raggedright
level
\end{minipage} & \begin{minipage}[b]{\linewidth}\raggedright
condition
\end{minipage} \\
\midrule\noalign{}
\endhead
\bottomrule\noalign{}
\endlastfoot
INTENT01 & error & a unit record declares no intent item \\
INTENT02 & error & an intent item names no evaluation \\
COMPILE01 & error & a Python file of the package does not parse \\
COMPILE02 & error & a subpackage imports a subpackage that
\texttt{{[}allowed{]}} does not permit \\
COMPILE03 & error & a subpackage of the package is missing from
\texttt{{[}allowed{]}} \\
COMPILE04 & error & an import cycle between subpackages is not listed in
\texttt{known\_cycles} \\
COMPILE05 & error & the module file of a unit does not exist \\
COMPILE06 & error & a name in \texttt{public} is not defined at the top
level of the module \\
COMPILE07 & error & the module imports a package module that is not in
\texttt{depends\_on} \\
COMPILE08 & error & a \texttt{depends\_on} entry names no module of the
package, or a name its module does not define \\
EVAL01 & error & an evaluation file named in \texttt{evaluations} does
not exist \\
EVAL02 & error & an evaluation file does not parse \\
EVAL03 & error & an evaluation named by an intent item is not defined in
the unit's evaluation files \\
EVAL04 & error & an evaluation file is not marked
\texttt{pytest.mark.regen}: no module-level \texttt{pytestmark} is, or
contains, the marker, and not every test carries it as a decorator (or
the file has no test) \\
EVAL05 & error & an evaluation file imports a name of the unit that is
not in \texttt{public} \\
PROV01 & error & an intent item cites a decision that does not exist \\
PROV02 & error & \texttt{architecture.toml} cites a decision that does
not exist \\
PROV03 & error & \texttt{file:} evidence does not exist in the
repository \\
PROV04 & error & \texttt{commit:} evidence is not a commit of the
repository (not reported when \texttt{git} is missing) \\
PROV05 & error & evidence has a prefix other than \texttt{file:},
\texttt{commit:}, \texttt{KB:}, \texttt{url:} \\
PROV06 & warning & an active decision is past \texttt{review\_after} \\
PROV07 & warning & an intent item cites a decision whose status is not
\texttt{active} \\
PROV08 & warning & the newest archived run of the unit judged a
different bundle (\texttt{bundle\_sha256}) \\
PROV09 & warning & the newest archived run of the unit replaced a
different module (\texttt{module\_sha256}) \\
PROV10 & warning & a \texttt{run.toml} under \texttt{regen/runs/} does
not parse \\
PACE01 & error & a unit record declares no \texttt{pace} \\
PACE02 & error & a unit's \texttt{pace} is not \texttt{fast},
\texttt{medium} or \texttt{slow} \\
PACE03 & error & \texttt{{[}pace{]}} has no policy for a pace that a
unit declares \\
PACE04 & error & a unit lists in \texttt{depends\_on} a unit (or a name
of it) whose pace is faster than its own \\
DELETE01 & error & a file depends on the unit but is not declared
(consumer, \texttt{unit\_tests} or \texttt{evaluations}) \\
DELETE02 & error & a declared consumer of kind \texttt{code},
\texttt{test} or \texttt{doc} does not exist \\
DELETE03 & error & a consumer outside \texttt{unit\_tests} and
\texttt{evaluations} uses a name of the unit that is not in
\texttt{public} \\
COMPACT01 & warning & \texttt{{[}allowed{]}} lists a subpackage that
does not exist \\
COMPACT02 & warning & an allowed edge of \texttt{{[}allowed{]}} is not
used by any import \\
COMPACT03 & warning & a pair in \texttt{known\_cycles} is no longer a
cycle \\
COMPACT04 & warning & an entry of \texttt{depends\_on} is not imported
by the module \\
COMPACT05 & warning & a test function in an evaluation file is named by
no intent item \\
COMPACT06 & warning & an active decision is cited by no intent item and
not by \texttt{architecture.toml} \\
COMPACT07 & warning & a declared consumer of kind \texttt{code},
\texttt{test} or \texttt{doc} no longer references the unit \\
\end{longtable}

\subsection{\texorpdfstring{One intent item in the bundle of
\texttt{primes}}{One intent item in the bundle of primes}}\label{sec-listing-bundle}

The excerpt follows the intent item PRI-2 of Example~\ref{exm-intent}
through the bundle of \texttt{primes} that Example~\ref{exm-bundle}
builds: from the intent through the decision record that it cites and
the public contract to the durable evaluation that checks the item.
\texttt{{[}...{]}} marks omitted lines. Long lines are wrapped.

\begin{verbatim}
# Regeneration bundle: mathregen.primes

Write `src/mathregen/primes.py` from scratch. You do not get the current implementation.
    Keep every public name and signature of the contract below, satisfy every intent item,
    and import nothing outside `depends_on`.

[...]

## Intent
### PRI-2

- scope: nth_prime
- invariant: Every result is a prime, the results increase strictly with the index, and no
    prime lies between `nth_prime(n)` and `nth_prime(n + 1)`.
- reason: A table of known values checks only the indices it lists. An implementation with a
    fault in its search, for example a sieve with a wrong bound, can pass the table and skip
    a prime or return a composite number elsewhere.
- checked by: test_every_result_is_prime_and_no_prime_is_skipped
[...]

## Provenance
### integers-only: Arguments and results are Python ints, nothing else

- status: active, date 2026-11-27, approved by bartzbeielstein
- driver: Both functions map an index to an exact integer. Python treats a bool as an int
    and converts between int and float in many places without a warning.
[...]

## Public contract (signatures and docstrings, no bodies)
def nth_prime(n: int) -> int:
    """Return the n-th prime number, counted from ``nth_prime(1) == 2``.
[...]

## Durable evaluations
def test_every_result_is_prime_and_no_prime_is_skipped():
    last = 1
    for n in range(1, 80):
        prime = nth_prime(n)
        assert prime > last
        assert not has_divisor(prime)
        assert all(has_divisor(k) for k in range(last + 1, prime))
        last = prime
[...]
\end{verbatim}

\section{A worked example: babregen step by step}\label{sec-tutorial}

\begin{tcolorbox}[enhanced jigsaw, arc=.35mm, bottomrule=.15mm, bottomtitle=1mm, breakable, colback=white, colbacktitle=quarto-callout-note-color!10!white, colframe=quarto-callout-note-color-frame, coltitle=black, left=2mm, leftrule=.75mm, opacityback=0, opacitybacktitle=0.6, rightrule=.15mm, title=\textcolor{quarto-callout-note-color}{\faInfo}\hspace{0.5em}{Running the example interactively}, titlerule=0mm, toprule=.15mm, toptitle=1mm]

This appendix is a part of the report and does not run on its own,
because its cells use values of the sections before it. The same
walkthrough can be obtained as a single Quarto file:
\texttt{docs/tutorials/step-by-step.qmd} in the repository of babregen,
\url{https://gitlab.git.nrw/thomas.bartz-beielstein/babregen}. Note:
This report uses release 0.2.0 of the tool, whereas the current version
of babregen is higher (version 0.5.0 or higher).

To run the Quarto file, clone the repository, create the environment
with \texttt{uv\ sync} at its root, and then either render the page with
\texttt{uv\ run\ quarto\ render\ docs/tutorials/step-by-step.qmd} or
open the file in an editor with the Quarto extension, for example VS
Code.

An interactive run needs the kernel of \texttt{.venv} and
\texttt{docs/tutorials/} as its working directory, because the cells
import \texttt{\_mathregen\_demo.py}. The cells share one throw-away git
repository, so they are run once and in order, from the first cell on,
and the kernel is restarted before a second pass. The note ``How to run
this page'' at the top of that file gives the details.

\end{tcolorbox}

This appendix introduces babregen on the package of the study, which is
small enough to read in full. The package, \texttt{mathregen}, has two
functions: one returns the n-th Fibonacci number, the other the n-th
prime. We build it in the order in which a developer would work. At
every step we run the real \texttt{babregen} command and show what it
prints. The terms are those of Section~\ref{sec-approach}, and the seven
primitives are those of Table~\ref{tbl-primitives}. As in
Section~\ref{sec-evaluation}, a module file that is offered in place of
the module of a unit is a \emph{rebuild}. The output and the file names
of the tool say candidate.

Every output block is captured from a real run when this document is
rendered. The cells copy the files of \texttt{mathregen/} into a
throw-away git repository, one stage at a time, and call
\texttt{babregen}, \texttt{pytest}, and \texttt{git} there.
\texttt{\textless{}repo\textgreater{}} in the output is the root of the
throw-away repository and \texttt{\textless{}tmp\textgreater{}} its
parent directory. The one exception is the model run in
Section~\ref{sec-tut-local}, which is a run of the study and is read
from its archive. The package and its records are those of the tag of
the study (Section~\ref{sec-object}). The Quickstart of babregen takes a
reader through the same commands in a terminal.

The examples of Section~\ref{sec-approach} run on the same package,
mostly on its unit \texttt{primes}. This appendix starts with the unit
\texttt{fibonacci}. Where Section~\ref{sec-approach} has shown a step,
the appendix refers to it and does not repeat it.

\subsection{The package skeleton}\label{sec-tut-skeleton}

\texttt{mathregen} is an ordinary package with a \texttt{src/} layout,
managed with \href{https://docs.astral.sh/uv/}{uv}. babregen is one
development dependency. It needs Python 3.13 or later and has no
dependencies of its own.

\begin{Shaded}
\begin{Highlighting}[]
\ExtensionTok{uv}\NormalTok{ init }\AttributeTok{{-}{-}lib}\NormalTok{ mathregen}
\BuiltInTok{cd}\NormalTok{ mathregen}
\ExtensionTok{uv}\NormalTok{ add }\AttributeTok{{-}{-}dev}\NormalTok{ pytest ruff babregen}
\end{Highlighting}
\end{Shaded}

The result is this \texttt{pyproject.toml}. Besides the development
dependency, two entries have to do with babregen: the marker
\texttt{regen}, which Section~\ref{sec-tut-evaluations} uses, and the
line that keeps the linter out of the archive of the runs.

\begin{verbatim}
# pyproject.toml
[project]
name = "mathregen"
version = "0.1.0"
description = "Math functions kept regenerable with babregen: Fibonacci numbers and
    primes"
readme = "README.md"
requires-python = ">=3.13"
# Standard library only (regen/architecture.toml, conventions).
dependencies = []

[build-system]
requires = ["uv_build>=0.9.18"]
build-backend = "uv_build"

[dependency-groups]
dev = [
    "pytest>=9.0.2",
    "ruff>=0.3.0",
    "babregen>=0.2.0",
]

[tool.pytest.ini_options]
testpaths = ["tests"]
markers = [
    "regen: durable evaluations of a regenerable unit (regen/units/*.toml); they must
    survive a rewrite of the implementation",
]

[tool.ruff]
line-length = 88
target-version = "py313"
# Archived candidates are evidence and are kept as the generator wrote them.
extend-exclude = ["regen/runs"]
\end{verbatim}

\begin{verbatim}
$ babregen --version
babregen 0.2.0
[exit 0]
\end{verbatim}

\subsection{Two functions and their unit tests}\label{sec-tut-functions}

Nothing in this step is special to babregen. We write the code and its
tests first. Each function gets a module of its own. babregen works on
modules: a regenerable unit is one module, so two modules can be deleted
and rebuilt independently of each other.

The unit tests are ordinary pytest files,
\texttt{tests/test\_fibonacci.py} and \texttt{tests/test\_primes.py},
one per module. No command of babregen generates them. No generator of
the study writes or changes them. They are part of the package mathregen
in babregen-examples. This worked example copies them from there, as it
does the two modules. One commit of the maintainer adds the modules,
their unit tests, and the records to that repository. A unit record
lists the files in \texttt{unit\_tests}
(Section~\ref{sec-tut-new-unit}). The acceptance rule runs them after
the durable evaluations.

\begin{verbatim}
.gitignore
Makefile
README.md
pyproject.toml
scripts/table.py
src/mathregen/__init__.py
src/mathregen/fibonacci.py
src/mathregen/primes.py
tests/test_fibonacci.py
tests/test_primes.py
\end{verbatim}

\begin{verbatim}
# src/mathregen/fibonacci.py
"""Fibonacci numbers."""

__all__ = ["fibonacci"]

def fibonacci(n: int) -> int:
    """Return the n-th Fibonacci number, counted from ``fibonacci(0) == 0``.

    The sequence is 0, 1, 1, 2, 3, 5, 8, ...: every value is the sum of the two before
    it. The result is exact for every index, because only integers are added.

    Args:
        n: Index of the number, an ``int`` that is not negative.

    Returns:
        int: The n-th Fibonacci number.

    Raises:
        TypeError: If `n` is not an ``int``, or is a ``bool``.
        ValueError: If `n` is negative.

    Examples:
        ```{python}
        from mathregen.fibonacci import fibonacci

        print([fibonacci(n) for n in range(10)])
        ```
    """
    if isinstance(n, bool) or not isinstance(n, int):
        raise TypeError(f"n must be an int, not {type(n).__name__}")
    if n < 0:
        raise ValueError(f"n must not be negative, got {n}")
    previous, current = 0, 1
    for _ in range(n):
        previous, current = current, previous + current
    return previous
\end{verbatim}

The module for the primes has a private helper, \texttt{\_is\_prime}.

\begin{verbatim}
# src/mathregen/primes.py, lines 1 to 16
"""Prime numbers."""

__all__ = ["nth_prime"]

def _is_prime(k: int) -> bool:
    """Whether `k` is prime, by trial division up to its square root."""
    if k < 2:
        return False
    divisor = 2
    while divisor * divisor <= k:
        if k % divisor == 0:
            return False
        divisor += 1
    return True
# src/mathregen/primes.py, lines 38 to 46
        ```
    """
    if isinstance(n, bool) or not isinstance(n, int):
        raise TypeError(f"n must be an int, not {type(n).__name__}")
    if n < 1:
        raise ValueError(f"n must be at least 1, got {n}")
    found = 0
    candidate = 1
    while found < n:
\end{verbatim}

Both functions state three things that a reader could not guess from the
name:

\begin{longtable}[]{@{}
  >{\raggedright\arraybackslash}p{(\linewidth - 4\tabcolsep) * \real{0.3333}}
  >{\raggedright\arraybackslash}p{(\linewidth - 4\tabcolsep) * \real{0.3333}}
  >{\raggedright\arraybackslash}p{(\linewidth - 4\tabcolsep) * \real{0.3333}}@{}}
\caption{The contract of the two
functions.}\label{tbl-tut-contract}\tabularnewline
\toprule\noalign{}
\begin{minipage}[b]{\linewidth}\raggedright
\end{minipage} & \begin{minipage}[b]{\linewidth}\raggedright
\texttt{fibonacci(n)}
\end{minipage} & \begin{minipage}[b]{\linewidth}\raggedright
\texttt{nth\_prime(n)}
\end{minipage} \\
\midrule\noalign{}
\endfirsthead
\toprule\noalign{}
\begin{minipage}[b]{\linewidth}\raggedright
\end{minipage} & \begin{minipage}[b]{\linewidth}\raggedright
\texttt{fibonacci(n)}
\end{minipage} & \begin{minipage}[b]{\linewidth}\raggedright
\texttt{nth\_prime(n)}
\end{minipage} \\
\midrule\noalign{}
\endhead
\bottomrule\noalign{}
\endlastfoot
first index & 0, with \texttt{fibonacci(0)\ ==\ 0} & 1, with
\texttt{nth\_prime(1)\ ==\ 2} \\
below the first index & \texttt{ValueError} & \texttt{ValueError} \\
not an \texttt{int}, or a \texttt{bool} & \texttt{TypeError} &
\texttt{TypeError} \\
\end{longtable}

The package re-exports both names. A script uses them:

\begin{verbatim}
# src/mathregen/__init__.py
"""Math functions, kept regenerable with babregen."""

from mathregen.fibonacci import fibonacci
from mathregen.primes import nth_prime

__all__ = ["fibonacci", "nth_prime"]
$ python scripts/table.py
 n  fibonacci(n)  nth_prime(n)
 1             1             2
 2             1             3
 3             2             5
 4             3             7
 5             5            11
 6             8            13
 7            13            17
 8            21            19
 9            34            23
10            55            29
[exit 0]
\end{verbatim}

The unit tests pass. The tests of \texttt{primes} call the private
helper \texttt{\_is\_prime}. This detail matters later: such a test is
coupled to one implementation.

\begin{verbatim}
$ pytest -q
......                                                                   [100%]
6 passed in 0.00s
[exit 0]
\end{verbatim}

\subsection{The architecture}\label{sec-tut-architecture}

The file \texttt{regen/architecture.toml} is the \emph{compilation
target} (Definition~\ref{def-compilation}): the shape that every module
of the package must fit into, whoever writes it. There is exactly one.

It names the package, lists for every module which other modules it may
import, states the conventions that all code follows, and leaves some
choices open on purpose.

\begin{verbatim}
# regen/architecture.toml
# Compilation target of mathregen: the shape every regenerated module must fit into.
# `babregen check` enforces [allowed] on the whole src/ tree.

package = "mathregen"
source_dir = "src"

# Provenance of this file: the protocol under which a unit is regenerated.
decisions = []

# Where consumers are searched: Python files below scan_dirs, text of doc_globs.
scan_dirs = ["src", "tests", "scripts"]
doc_globs = ["README.md"]

conventions = [
    "Python >= 3.13, standard library only: [project].dependencies stays empty",
    "one module per function; the package __init__ re-exports the public names",
    "arguments and results are Python ints; no floating-point arithmetic",
    "Google-style docstring with a live ```{python} Examples block for every public
    function",
]

open_choices = [
    "the algorithm (iteration or fast doubling; trial division or a sieve)",
    "private helpers and their names",
    "the wording of error messages",
]

known_cycles = []

[pace]
fast = "the durable evaluations (make regen-eval) and a clean babregen check"
slow = "the durable evaluations, a clean babregen check and a decision record"

# Module -> modules of the package it may import. The two function modules import
# nothing; only the package __init__ imports them.
[allowed]
"__init__" = ["fibonacci", "primes"]
fibonacci = []
primes = []
\end{verbatim}

(The final file names two decision records in \texttt{decisions}. They
are written in Section~\ref{sec-tut-decisions}, so the list is empty
here.)

Three parts deserve a remark. \texttt{{[}allowed{]}} says that the two
function modules import nothing from the package and that only
\texttt{\_\_init\_\_} imports them. \texttt{open\_choices} tells a
future author, human or model, what is deliberately not fixed: the
algorithm is free. \texttt{{[}pace{]}} states the verification that a
change requires. A unit declares one of the paces, and the policy of
that pace applies to it (Section~\ref{sec-pace}).

\begin{example}[The first
check]\protect\hypertarget{exm-tut-first-check}{}\label{exm-tut-first-check}

~

\end{example}

\begin{verbatim}
$ babregen check
regen check: 0 units, 0 errors, 0 warnings
[exit 0]
\end{verbatim}

\texttt{babregen\ check} reads \texttt{regen/} and the source tree and
prints one line per finding and a summary. It finds
\texttt{regen/architecture.toml} by searching upwards from the current
directory. There are no units yet, and the imports in \texttt{src/}
respect \texttt{{[}allowed{]}}.

\subsection{A unit record, drafted by the tool}\label{sec-tut-new-unit}

A \emph{unit} (Definition~\ref{def-record}) is the part of the program
that a \emph{unit record}
\texttt{regen/units/\textless{}stem\textgreater{}.toml} describes. It is
implemented by one module, which a regeneration may replace. The record
states what the module promises (intent), which tests judge it
(evaluations) and who depends on it (consumers). The file
\texttt{fibonacci.toml} describes the module
\texttt{mathregen.fibonacci}; the name of the file is the only link. A
unit is identified by its record, not by its current implementation: the
module is what is deleted and rebuilt, the record is what survives.

\texttt{babregen\ new-unit} drafts a record from what the tree can tell.

\begin{example}[The draft for
\texttt{fibonacci}]\protect\hypertarget{exm-tut-new-unit}{}\label{exm-tut-new-unit}

~

\end{example}

\begin{verbatim}
$ babregen new-unit fibonacci --pace fast
wrote regen/units/fibonacci.toml
next: write the intent items and evaluations, then run babregen check
still open (babregen check will report):
regen/units/fibonacci.toml:4: error INTENT01 declares no intent item (a unit without
    intent cannot be regenerated)
[exit 0]
# regen/units/fibonacci.toml
# Drafted by `babregen new-unit` on 2026-12-02. Review every field;
# `babregen check` fails with INTENT01 until the first [[intent]] item is written.

summary = "Fibonacci numbers."
pace = "fast"
public = ["fibonacci"]
depends_on = []
evaluations = []  # e.g. tests/test_regen_eval_fibonacci.py, marked pytest.mark.regen
unit_tests = ["tests/test_fibonacci.py"]

# Intent: what must stay true for any implementation, and why. Ask "what would a
# competent stranger get wrong?" and turn every incident that touched the module into
# an item with an evaluation and a decision record. `babregen add-intent` appends one.
#
# [[intent]]
# id = "FIB-1"
# scope = ""  # where the requirement applies
# invariant = ""  # what must remain true
# reason = ""  # why the invariant matters
# evaluations = []  # test functions in the evaluation files above
# decisions = []  # ids of regen/decisions/<id>.toml

[[consumers]]
ref = "scripts/table.py"
kind = "code"
note = ""

[[consumers]]
ref = "src/mathregen/__init__.py"
kind = "code"
note = ""

[[consumers]]
ref = "README.md"
kind = "doc"
note = ""

# Add what the tree cannot show as kind = "external": downstream files
# (ref = "<repo>:<path>"), scheduled runs and people (a free-text ref).
# `babregen add-consumer` appends a consumer; say in its note what it relies on.
\end{verbatim}

The tool filled in what it could read: the summary from the docstring,
the public name from \texttt{\_\_all\_\_}, the unit test, and three
consumers. The file \texttt{tests/test\_regen\_eval\_fibonacci.py} would
be listed under \texttt{evaluations}. It does not exist yet. It found
\texttt{scripts/table.py} and \texttt{README.md} although neither
imports \texttt{mathregen.fibonacci} directly: both use the name that
the package re-exports, and the scan follows re-exports.

What the tool does not draft is left open, and the draft says so. It
fails the check on purpose, so that a draft does not pass as the tool
writes it:

\begin{verbatim}
$ babregen check
regen/units/fibonacci.toml:4: error INTENT01 declares no intent item (a unit without
    intent cannot be regenerated)
regen check: 1 units, 1 errors, 0 warnings
[exit 1]
\end{verbatim}

\texttt{INTENT01} is the first finding of this tutorial. A finding line
is \texttt{\{path\}:\{line\}:\ \{level\}\ \{id\}\ \{message\}}
(Section~\ref{sec-tool-reference}). The prefix of the ID is the
primitive it belongs to, here intent. The exit code is 1: the check
fails.

\subsection{Evaluations and intent}\label{sec-tut-evaluations}

A \emph{durable evaluation} (Definition~\ref{def-evaluation}) is a test
that uses only the public names of a unit and says nothing about how the
module works. It stays valid when the module is deleted and written
again. Evaluation files carry the pytest marker \texttt{regen}.

The unit tests of Section~\ref{sec-tut-functions} travel with the
implementation and may be rewritten with it. The durable evaluations are
the opposite: they belong to the knowledge that stays. They are the
judge of every future implementation. The question that tells the two
apart is that of Fowler (2027, ch.~3): would this assertion still be
meaningful if we replaced the implementation tomorrow? The records of a
unit and its durable evaluations together are its \emph{specification}.

\begin{verbatim}
# tests/test_regen_eval_fibonacci.py
"""Durable evaluations of the unit mathregen.fibonacci (regen/units/fibonacci.toml).

They use the public function only and say nothing about how it computes, so they stay
valid when the module is deleted and written again.
"""

import pytest

from mathregen.fibonacci import fibonacci

pytestmark = pytest.mark.regen

def test_the_sequence_starts_at_index_zero_with_zero_and_one():
    assert fibonacci(0) == 0
    assert fibonacci(1) == 1
    assert [fibonacci(n) for n in range(11)] == [0, 1, 1, 2, 3, 5, 8, 13, 21, 34, 55]

def test_every_value_is_the_sum_of_the_two_before_it():
    for n in range(2, 200):
        assert fibonacci(n) == fibonacci(n - 1) + fibonacci(n - 2)

def test_a_large_value_is_an_exact_int():
    assert fibonacci(100) == 354224848179261915075
    assert fibonacci(200) == 280571172992510140037611932413038677189525
    assert type(fibonacci(200)) is int

def test_a_negative_index_raises_value_error():
    for n in (-1, -2, -100):
        with pytest.raises(ValueError):
            fibonacci(n)

@pytest.mark.parametrize("n", [1.0, 2.5, "3", None, [1]])
def test_an_index_that_is_not_an_int_raises_type_error(n):
    with pytest.raises(TypeError):
        fibonacci(n)

def test_a_bool_index_raises_type_error():
    for n in (True, False):
        with pytest.raises(TypeError):
            fibonacci(n)
\end{verbatim}

An \emph{intent item} (Definition~\ref{def-intent}) states one
requirement in three parts: the scope (where it applies), the invariant
(what must remain true) and the reason (why it matters). It names the
evaluations that check it and the decisions behind it.

A good intent item answers the question ``what would a competent
stranger get wrong?'' For \texttt{fibonacci} there are three such
things: the first index, the temptation of the closed formula, and the
fact that Python treats \texttt{True} as the integer 1. We replace the
draft by the finished record.

\begin{verbatim}
# regen/units/fibonacci.toml
summary = """
The n-th Fibonacci number, `fibonacci(n)`, for an index counted from 0. The module is
one public function without state."""
pace = "fast"
public = ["fibonacci"]
depends_on = []
evaluations = ["tests/test_regen_eval_fibonacci.py"]
unit_tests = ["tests/test_fibonacci.py"]

[[intent]]
id = "FIB-1"
scope = "fibonacci"
invariant = """
`fibonacci(0) == 0`, `fibonacci(1) == 1`, and `fibonacci(n) == fibonacci(n - 1) +
fibonacci(n - 2)` for every n >= 2."""
reason = """
Two conventions are in use: the sequence 1, 1, 2, ... counted from index 1, and the
sequence 0, 1, 1, 2, ... counted from index 0. An implementation that picks the other
one is off by one index for every caller."""
evaluations = [
    "test_the_sequence_starts_at_index_zero_with_zero_and_one",
    "test_every_value_is_the_sum_of_the_two_before_it",
]
decisions = ["indexing-conventions"]

[[intent]]
id = "FIB-2"
scope = "fibonacci"
invariant = """
The result is an `int` and is exact for every index, however large:
`fibonacci(100) == 354224848179261915075`."""
reason = """
The closed formula with the golden ratio is the obvious shortcut, and in
    floating-point
arithmetic it gives wrong last digits from about n = 70 on. Callers use the values as
exact integers."""
evaluations = ["test_a_large_value_is_an_exact_int"]
decisions = ["integers-only"]

[[intent]]
id = "FIB-3"
scope = "fibonacci"
invariant = """
A negative index raises `ValueError`. An index that is not an `int` raises
    `TypeError`;
this includes `bool`, and a `float` with an integral value such as `3.0`."""
reason = """
A silent result for `fibonacci(True)` or `fibonacci(3.0)` hides a mistake of the
caller, and Python accepts both wherever an `int` is expected unless the function
refuses them."""
evaluations = [
    "test_a_negative_index_raises_value_error",
    "test_an_index_that_is_not_an_int_raises_type_error",
    "test_a_bool_index_raises_type_error",
]
decisions = ["integers-only"]

[[consumers]]
ref = "src/mathregen/__init__.py"
kind = "code"
note = "re-exports fibonacci"

[[consumers]]
ref = "scripts/table.py"
kind = "code"
note = "prints the first values"

[[consumers]]
ref = "README.md"
kind = "doc"
note = "states the index convention and shows fibonacci(10)"
\end{verbatim}

The evaluation file is now listed under \texttt{evaluations}. Every
intent item names the test functions that check it. The evaluation file
needs no entry under \texttt{consumers}: declaring it as an evaluation
is enough.

\begin{verbatim}
$ babregen check
regen/units/fibonacci.toml:24: error PROV01 intent FIB-1: unknown decision
    'indexing-conventions'
regen/units/fibonacci.toml:37: error PROV01 intent FIB-2: unknown decision
    'integers-only'
regen/units/fibonacci.toml:54: error PROV01 intent FIB-3: unknown decision
    'integers-only'
regen check: 1 units, 3 errors, 0 warnings
[exit 1]
\end{verbatim}

\texttt{INTENT01} is resolved. The new findings are of the family
\texttt{PROV}, provenance: each intent item cites a decision that does
not exist yet.

\subsection{Decisions}\label{sec-tut-decisions}

A \emph{decision record}
\texttt{regen/decisions/\textless{}id\textgreater{}.toml} is a
provenance record (Definition~\ref{def-provenance}). It states what
drove a decision, what was decided, who approved it, which alternatives
were rejected and why, the evidence, and the assumptions the decision
rests on. It may carry a review date (\texttt{review\_after}), after
which the check reports it as due for review.

The reason of an intent item says why an invariant matters. The decision
record says why this invariant was chosen and not another one. Without
it, the next author sees a rule and cannot tell whether it is a
considered choice or an accident.

\begin{verbatim}
# regen/decisions/indexing-conventions.toml
title = "Fibonacci numbers count from 0, primes count from 1"
date = 2026-11-27
status = "active"
driver = """
Both sequences have two conventions for the first index, and the package has to state
one of them for each function before the first caller relies on it."""
decision = """
`fibonacci` counts from 0 with fibonacci(0) == 0, the convention of the OEIS (A000045)
and of most programming texts. `nth_prime` counts from 1 with nth_prime(1) == 2, the
convention p_1 = 2 of number theory (OEIS A000040). Each function follows the
convention of its own literature; the README states both."""
approved_by = "bartzbeielstein"
evidence = ["file:README.md", "url:https://oeis.org/A000045",
    "url:https://oeis.org/A000040"]
assumes = ["callers read the README or the docstring before they index"]

[[alternatives]]
option = "count both from 0"
rejected_because = "a zeroth prime exists in no text; every table of primes would be
    shifted by one"

[[alternatives]]
option = "count both from 1"
rejected_because = "fibonacci(1) == fibonacci(2) == 1 leaves no place for 0, and the
    identities of the sequence are stated for F(0) = 0"
\end{verbatim}

The second record, \texttt{integers-only}, explains why a \texttt{bool}
and a \texttt{float} are refused. The other two records are protocols,
and the architecture file cites both. \texttt{regen-protocol} fixes how
a unit is regenerated with a local model in the tutorial of the tool.
\texttt{study-protocol} is the protocol of the study of this report
(Section~\ref{sec-design}).

We add the records of the second unit, \texttt{primes}, which were
written the same way, and the full architecture file. We also add the
directory \texttt{candidates/} with three rebuilds kept in
babregen-examples. Section~\ref{sec-tut-regeneration} scores them. The
protocol record of the study cites one of them as evidence.

\begin{verbatim}
$ babregen check
regen check: 2 units, 0 errors, 0 warnings
[exit 0]
\end{verbatim}

Two units, no finding. The evaluations pass as well; \texttt{-m\ regen}
selects them by their marker.

\begin{verbatim}
$ pytest -q -m regen
....................                                                     [100%]
20 passed, 6 deselected in 0.01s
[exit 0]
\end{verbatim}

This is the state that is worth a commit.

\begin{verbatim}
$ git add -A && git commit -m 'mathregen with its regen records'
093c17b mathregen with its regen records
\end{verbatim}

\subsection{What the check catches}\label{sec-tut-faults}

A passing check is not informative on its own.
Example~\ref{exm-compilation} plants a forbidden import and shows the
four findings that the one line causes. Here we plant a fault that
changes no line of the package. The check parses files and runs no code
of the package. It reports where the records and the code disagree, with
an error for what must be fixed and a warning for what asks for
attention.

\begin{example}[An undeclared
script]\protect\hypertarget{exm-tut-faults}{}\label{exm-tut-faults}

A colleague adds a script that uses \texttt{nth\_prime}.

\end{example}

\begin{verbatim}
$ babregen check
regen/units/primes.toml:1: error DELETE01 scripts/first_primes.py depends on the unit
    but is not a declared consumer
regen check: 2 units, 1 errors, 0 warnings
[exit 1]
\end{verbatim}

\texttt{DELETE01} protects the deletion test. If \texttt{primes} were
regenerated today, nobody would know that
\texttt{scripts/first\_primes.py} has to keep working. The fix is to
declare the script.

\begin{verbatim}
$ babregen check
regen check: 2 units, 0 errors, 0 warnings
[exit 0]
\end{verbatim}

(The script was only an illustration. The sandbox returns to the commit
of Section~\ref{sec-tut-decisions}.)

A clean check shows that the records agree with the code. It cannot see
what was never written down, so it does not show that the records hold
enough knowledge to write a unit again, or that the evaluations would
notice a wrong module. The next three sections test both.

\subsection{The deletion test}\label{sec-tut-deletion}

The \emph{deletion test} of a unit has four components
(Definition~\ref{def-deletion}): the blast radius, the knowledge outside
the implementation, the rebuild, and the judgement.
\texttt{babregen\ deletion-test} checks the first two and deletes
nothing.

\begin{example}[The deletion test of
\texttt{fibonacci}]\protect\hypertarget{exm-tut-deletion-test}{}\label{exm-tut-deletion-test}

~

\end{example}

\begin{verbatim}
$ babregen deletion-test fibonacci
# Deletion test: mathregen.fibonacci

File `src/mathregen/fibonacci.py`, pace fast.

## Blast radius

| consumer | kind | declared | note |
|---|---|---|---|
| `README.md` | doc | yes | states the index convention and shows fibonacci(10) |
| `scripts/table.py` | code | yes | prints the first values |
| `src/mathregen/__init__.py` | code | yes | re-exports fibonacci |
| `tests/test_fibonacci.py` | test | yes |  |
| `tests/test_regen_eval_fibonacci.py` | test | yes | evaluation |

## Knowledge outside the implementation

| intent | evaluations | decisions |
|---|---|---|
| FIB-1 | 2 | indexing-conventions |
| FIB-2 | 1 | integers-only |
| FIB-3 | 3 | integers-only |

Verdict: regenerable. 3 intent items, all evaluated; 5 dependents, all declared.
[exit 0]
\end{verbatim}

The blast radius lists the files that use the unit, as far as the scan
sees them: the three declared consumers and the unit's own tests. The
second table is the knowledge outside the implementation. The verdict is
the exit code: 0 for regenerable.

This verdict is a precondition for an attempt, not its result.
Components 3 and 4 are the subject of
Section~\ref{sec-tut-regeneration}: \texttt{babregen\ score} judges a
given rebuild, and \texttt{babregen\ local} deletes, rebuilds and
judges.

The \emph{bundle} of a unit (Definition~\ref{def-bundle}) is the
document that an author receives in order to write the module again.
\texttt{babregen\ context} assembles it from the records as a single
Markdown file.

\begin{example}[The bundle of
\texttt{fibonacci}]\protect\hypertarget{exm-tut-bundle}{}\label{exm-tut-bundle}

~

\end{example}

\begin{verbatim}
# Regeneration bundle: mathregen.fibonacci
## Purpose
## Architecture (compilation target)
## Intent
### FIB-1
### FIB-2
### FIB-3
## Provenance
### indexing-conventions: Fibonacci numbers count from 0, primes count from 1
### integers-only: Arguments and results are Python ints, nothing else
## Public contract (signatures and docstrings, no bodies)
## Durable evaluations
## Consumers (must keep working)
\end{verbatim}

The headings show the structure: purpose, architecture, intent,
provenance, the public contract, the evaluations and the consumers. The
section with the contract is the part of the module that survives:

\begin{verbatim}
## Public contract (signatures and docstrings, no bodies)

```python
"""Fibonacci numbers."""

def fibonacci(n: int) -> int:
    """Return the n-th Fibonacci number, counted from ``fibonacci(0) == 0``.

    The sequence is 0, 1, 1, 2, 3, 5, 8, ...: every value is the sum of the two before
    it. The result is exact for every index, because only integers are added.

    Args:
        n: Index of the number, an ``int`` that is not negative.

    Returns:
        int: The n-th Fibonacci number.

    Raises:
        TypeError: If `n` is not an ``int``, or is a ``bool``.
        ValueError: If `n` is negative.

    Examples:
        ```{python}
        from mathregen.fibonacci import fibonacci

        print([fibonacci(n) for n in range(10)])
        ```
    """
    ...
```
\end{verbatim}

The body is replaced by \texttt{...}. Example~\ref{exm-bundle} searches
both bundles for every code line of the function bodies and finds none.

Everything that we wrote in the last three sections ends up in this one
document. It is what a model sees in the first session of
Section~\ref{sec-tut-local}, after a fixed instruction.

\subsection{How good are the evaluations?}\label{sec-tut-measures}

\begin{example}[The measures of both
units]\protect\hypertarget{exm-tut-measures}{}\label{exm-tut-measures}

~

\end{example}

\begin{verbatim}
$ babregen measures
| unit | pace | coverage | decisions | consumers | code | record | decision lines |
    evaluation lines | spec | ratio |
|---|---|---:|---:|---:|---:|---:|---:|---:|---:|---:|
| mathregen.fibonacci | fast | 3/3 | 2 | 3 | 36 | 69 | 43 | 46 | 158 | 4.4 |
| mathregen.primes | fast | 3/3 | 2 | 3 | 50 | 68 | 43 | 54 | 165 | 3.3 |
| total | - | 6/6 | 4 | 6 | 86 | 137 | 86 | 100 | 323 | 3.8 |
[exit 0]
\end{verbatim}

\texttt{coverage} is the \emph{intent coverage}: the share of the intent
items that name an evaluation and whose named evaluations all exist, 3
of 3 for each unit. It is not code coverage. It counts the recorded
promises that have a judge, not the lines that ran. The columns from
\texttt{code} on are line counts. Unit record, cited decisions and
evaluation files add up to the \emph{specification size}, and the last
column is its ratio to the code: for two functions of a dozen lines the
specification is more than three times as long as the implementation
(Table~\ref{tbl-size}). That is the price of being able to delete the
code.

Intent coverage says that every recorded promise has a judge. It does
not say that all necessary promises are recorded. It also does not say
that the evaluations are sharp. For the last point, babregen plants
faults.

\texttt{babregen\ mutate} (Definition~\ref{def-mutation-analysis}) makes
small changes to the module, one at a time (a \emph{mutant}), with a
fixed set of operators, and runs the durable evaluations on each. A
mutant is \emph{killed} if an evaluation fails or ends in an error, or
if the run exceeds its time limit, and \emph{survives} otherwise. The
\emph{mutation score} is the share of killed mutants among all mutants.
Every survivor then needs a verdict by a person: \emph{gap},
\emph{output}, or \emph{equivalent} (Table~\ref{tbl-verdicts}).

\begin{example}[All mutants of \texttt{fibonacci} are
killed]\protect\hypertarget{exm-tut-mutate}{}\label{exm-tut-mutate}

~

\end{example}

\begin{verbatim}
$ babregen mutate fibonacci --timeout 10
M001 killed (5 failed)
M002 killed (1 failed)
M003 killed (4 failed)
M004 killed (1 failed)
M005 killed (2 failed)
M006 killed (2 failed)
M007 killed (1 failed)
M008 killed (2 failed)
M009 killed (2 failed)
M010 killed (3 failed)
M011 killed (3 failed)
# Mutation score: mathregen.fibonacci

10 evaluations in tests/test_regen_eval_fibonacci.py, 11 mutants.

Base commit `093c17b7a89359804e74e0f07958a53602da232d`, working tree clean; Python
    3.14.2, pytest 9.2.0.

Mutation score: 11/11 (100 %).
[exit 0]
\end{verbatim}

To see what a survivor looks like, suppose we had forgotten one
evaluation, the one for \texttt{bool}.

\begin{example}[A forgotten evaluation lets two mutants
survive]\protect\hypertarget{exm-tut-survivor}{}\label{exm-tut-survivor}

~

\end{example}

\begin{verbatim}
$ babregen mutate fibonacci --timeout 10
M002 SURVIVED (survived)
M004 SURVIVED (survived)
# Mutation score: mathregen.fibonacci

9 evaluations in tests/test_regen_eval_fibonacci.py, 11 mutants.

Base commit `093c17b7a89359804e74e0f07958a53602da232d`, working tree dirty (the
    mutants were made from it); Python 3.14.2, pytest 9.2.0.

## Surviving mutants

| mutant | line | operator | original | mutated |
|---|---:|---|---|---|
| M002 | 29 | boolean | `isinstance(n, bool) or not isinstance(n, int)` |
    `isinstance(n, bool) and (not isinstance(n, int))` |
| M004 | 30 | raise | `raise TypeError(f'n must be an int, not {type(n).__name__}')` |
    `pass` |

Mutation score: 9/11 (82 %).
[exit 1]
\end{verbatim}

Two mutants of \texttt{fibonacci} survive on this copy, and the exit
code is 1. Its mutant \texttt{M004} is the instructive one: the whole
\texttt{raise\ TypeError} is replaced by \texttt{pass}, and the five
evaluations with two floats, a string, \texttt{None} and a list still
pass. The reason is that Python raises a \texttt{TypeError} by itself a
few lines later, when it compares a string with 0 or calls
\texttt{range} with a \texttt{float}. Only a \texttt{bool} gets through.
The explicit type check exists for the \texttt{bool} alone. Only the
\texttt{bool} evaluation proves that it is there. By construction an
evaluation is missing here, which is what the verdict gap means. The
remedy is the evaluation that we restore now.

The unit \texttt{primes} keeps three survivors with its full set of
durable evaluations (Table~\ref{tbl-survivors}). What they mean is the
question of the gate of the study (Section~\ref{sec-rq2}). This appendix
does not discuss them.

\subsection{Regeneration}\label{sec-tut-regeneration}

We now do what the records were written for: replace a module by a new
one that was written without looking at the old one.

\texttt{babregen\ score\ UNIT\ FILE} judges a rebuild by the acceptance
rule (Definition~\ref{def-acceptance}). It puts the rebuild in the place
of the original in a scratch copy of the last commit and runs the
evaluations of the unit, its unit tests, and its dependents.
\texttt{score} calls no model. Its copy is no sandbox: a rebuild runs
with the rights of the user. It judges against the last commit, so the
project must lie in a git repository.

\subsubsection{A rebuild kept in the
repository}\label{a-rebuild-kept-in-the-repository}

The first rebuild computes the Fibonacci numbers by fast doubling, an
algorithm that has nothing in common with the loop of the original. The
architecture left the algorithm open.

\begin{verbatim}
"""Fibonacci numbers."""

__all__ = ["fibonacci"]

def _pair(n: int) -> tuple[int, int]:
    """Return (F(n), F(n + 1)) by fast doubling."""
    if n == 0:
        return 0, 1
    a, b = _pair(n // 2)
    c = a * (2 * b - a)
    d = a * a + b * b
    return (d, c + d) if n % 2 else (c, d)

def fibonacci(n: int) -> int:
    """Return the n-th Fibonacci number, counted from ``fibonacci(0) == 0``.

    Args:
        n: Index of the number, an ``int`` that is not negative.

    Returns:
        int: The n-th Fibonacci number.

    Raises:
        TypeError: If `n` is not an ``int``, or is a ``bool``.
        ValueError: If `n` is negative.
    """
    if type(n) is not int:
        raise TypeError(f"n must be an int, not {type(n).__name__}")
    if n < 0:
        raise ValueError(f"n must not be negative, got {n}")
    return _pair(n)[0]
\end{verbatim}

\begin{example}[A different algorithm is
accepted]\protect\hypertarget{exm-tut-score-accepted}{}\label{exm-tut-score-accepted}

~

\end{example}

\begin{verbatim}
$ babregen score fibonacci candidates/fibonacci_doubling.py --out
    ../runs/fibonacci_doubling
... (42 lines before)
run directory: <tmp>/runs/fibonacci_doubling
[exit 0]
accepted = true   (rule accept/1)
evaluations 10 passed, 0 failed, 0 errors
unit tests  2 passed, 0 failed, 0 errors
dependents  2 passed, 0 failed
\end{verbatim}

The ten evaluations pass. The two dependents are the import of the
package and the docstring example of the original, run against the new
module. The exit code is 0. \texttt{scripts/table.py} and
\texttt{README.md} are declared consumers too, but they lie outside
\texttt{src/}, so the judge does not run them: they are part of the
blast radius that a person checks. The module refuses an index whose
type is a subclass of \texttt{int}, which the original accepts
(Section~\ref{sec-undecided}).

The second rebuild takes the shortcut that intent item \texttt{FIB-2}
warns about.

\begin{verbatim}
"""Fibonacci numbers."""

__all__ = ["fibonacci"]

PHI = (1 + 5**0.5) / 2

def fibonacci(n: int) -> int:
    """Return the n-th Fibonacci number by the closed formula with the golden ratio."""
    if n < 0:
        raise ValueError(f"n must not be negative, got {n}")
    return round(PHI**n / 5**0.5)
\end{verbatim}

\begin{example}[The closed formula is
rejected]\protect\hypertarget{exm-tut-score-rejected}{}\label{exm-tut-score-rejected}

~

\end{example}

\begin{verbatim}
$ babregen score fibonacci candidates/fibonacci_binet.py --out ../runs/fibonacci_binet
... (42 lines before)
run directory: <tmp>/runs/fibonacci_binet
[exit 1]
accepted = false   (rule accept/1)
evaluations 5 passed, 5 failed, 0 errors
unit tests  2 passed, 0 failed, 0 errors
dependents  2 passed, 0 failed
\end{verbatim}

The rebuild is rejected with exit code 1. Note what did not catch it:
the unit tests pass and the dependents pass, because they look at the
first ten values only. The durable evaluations for exactness and for the
argument type are what stand between this module and the package.

The repository holds a third rebuild, a sieve for \texttt{primes}.
Example~\ref{exm-evaluation} scores it: it is accepted although its unit
tests end in an error. The three files are among the controls of the
study (Section~\ref{sec-controls}).

\subsubsection{A rebuild written by a model}\label{sec-tut-local}

\texttt{babregen\ local} does the same with a language model as the
author (Definition~\ref{def-regeneration}). A driver sends the bundle to
a model that a local server such as LM Studio serves on this machine,
takes the module from the reply, and judges it with the same rule. If
evaluations fail, the failures go back to the model for a repair, until
\texttt{-\/-attempts} replies have been used.

The command wants to know what produced a run. Four options are
required: the model, the SHA-256 of its files, which the command records
and cannot verify, the context length it was loaded with, and
\texttt{-\/-protocol}, the id of an active decision record that fixes
the procedure before the first session. For the runs of the study this
record is \texttt{study-protocol}.

A model run cannot be repeated while this document renders, so this
subsection reads one run of the study from its archive: the run of
\texttt{fibonacci} with generator (c) and seed 2
(Table~\ref{tbl-regen-runs}). This is its call, as the log of the
archive records it:

\begin{verbatim}
$ babregen local mathregen.fibonacci --model qwen/qwen3-coder-30b --model-sha256
    6b189cafac4813766f88107e601855a267017432f11d319bb0cfde94a65d7683 --reasoning none
    --context-length 32768 --template chatml --attempts 3 --temperature 0.7 --top-p
    0.8 --top-k 20 --score-timeout 120 --protocol study-protocol --seed 2 --host
    p040025 --out regen/runs/study/fibonacci-c-seed2-try1
\end{verbatim}

The run leaves a directory under \texttt{regen/runs/}:

\begin{verbatim}
bundle.md
dependents.txt
evaluations-1.txt
evaluations-2.txt
fibonacci-1.py
fibonacci-2.py
report.md
run.toml
sessions.jsonl
unit_tests.txt
\end{verbatim}

\begin{example}[A run of
\texttt{fibonacci}]\protect\hypertarget{exm-tut-local-run}{}\label{exm-tut-local-run}

\texttt{run.toml} is the record of the run. These are its verdict and
its attempts:

\end{example}

\begin{verbatim}
accepted = true   (rule accept/1)
model    = qwen/qwen3-coder-30b
base     = eae4473ca986
attempt 1: 9 passed, 1 failed, 0 errors
attempt 2: 10 passed, 0 failed, 0 errors
session 1: generate   310 tokens   3.0 s
session 2: repair     362 tokens   4.0 s
session 3: report    1170 tokens  11.7 s
\end{verbatim}

The rebuild was accepted on the second attempt. The first attempt failed
one evaluation:

\begin{verbatim}
# regen/runs/study/fibonacci-c-seed2-try1/evaluations-1.txt
E           Failed: DID NOT RAISE TypeError

tests/test_regen_eval_fibonacci.py:45: Failed
1 failed, 9 passed in <time>
babregen: failing evaluations
FAILED tests/test_regen_eval_fibonacci.py::test_a_bool_index_raises_type_error
    tests/test_regen_eval_fibonacci.py:45: Failed: DID NOT RAISE TypeError
\end{verbatim}

It is the evaluation for an index of the type \texttt{bool}, which
intent item \texttt{FIB-3} names. The driver handed the failure back,
and the repair changed the type check:

\begin{verbatim}
--- fibonacci-1.py
+++ fibonacci-2.py
@@ -26,3 +26,6 @@
     """
-    if not isinstance(n, int):
+    # BUNDLE-CONFLICT: The bundle requires that bool values raise TypeError,
+    # but the current implementation only checks for isinstance(n, int).
+    # This is a strict interpretation of the intent FIB-3.
+    if not isinstance(n, int) or isinstance(n, bool):
         raise TypeError("Index must be an int")
\end{verbatim}

After its work the model is asked for a gap report: a list of the places
where the bundle was ambiguous, silent or contradictory. This is the
first item of the report of the run:

\begin{verbatim}
# regen/runs/study/fibonacci-c-seed2-try1/report.md
1. **Bundle Section: FIB-3**
   - **Problem**: The intent states that "A negative index raises `ValueError`. An
    index that is not an `int` raises `TypeError`; this includes `bool`, and a `float`
    with an integral value such as `3.0`." However, the specification does not define
    a clear distinction between what constitutes an "int" versus a "float with
    integral value" in terms of the function's behavior.
   - **Module Choice**: The implementation strictly checks `isinstance(n, int)` and
    explicitly excludes booleans with `isinstance(n, bool)`. This choice is more
    restrictive than the wording of the intent implies, which could allow floats like
    `3.0` to pass through if interpreted as a valid integer representation.
\end{verbatim}

A gap report is raw material that a person has to read, and where it
points at a real ambiguity, the fix belongs in the record, not in the
code. Section~\ref{sec-rq3} reports all runs of the study, and
Section~\ref{sec-rq4} the one run that was rejected.

\begin{tcolorbox}[enhanced jigsaw, arc=.35mm, bottomrule=.15mm, bottomtitle=1mm, breakable, colback=white, colbacktitle=quarto-callout-warning-color!10!white, colframe=quarto-callout-warning-color-frame, coltitle=black, left=2mm, leftrule=.75mm, opacityback=0, opacitybacktitle=0.6, rightrule=.15mm, title=\textcolor{quarto-callout-warning-color}{\faExclamationTriangle}\hspace{0.5em}{What the archive can and cannot do}, titlerule=0mm, toprule=.15mm, toptitle=1mm]

\texttt{score} and \texttt{local} judge against a commit, so the project
must lie in a git repository. A subfolder of one is enough. The runs of
the study were made in \texttt{mathregen/} of a clean clone of
babregen-examples at the tag of the study, and the \texttt{{[}base{]}}
of each \texttt{run.toml} names the commit of that tag. An archived
rebuild can be scored again in place with \texttt{babregen\ score} and
the option \texttt{-\/-base}.

No rebuild was put into \texttt{src/}. An accepted rebuild is a result
to review, not a merge.

\end{tcolorbox}

\texttt{babregen\ check} reads the archive too. Each run records the
SHA-256 of the bundle it was given and of the module it replaced. When
the bundle or the module changes later, the check reports that the
newest completed run judged something else
(Example~\ref{exm-provenance}). It does not report older runs.

\subsection{Compaction and the guard}\label{sec-tut-compaction}

Knowledge goes stale. A consumer stops using a unit, an allowed import
is never used, a decision is cited by nothing. The check reports such
things as warnings of the family \texttt{COMPACT}. The command
\texttt{babregen\ compact} lists only them.

\begin{example}[Two stale
entries]\protect\hypertarget{exm-tut-compact}{}\label{exm-tut-compact}

We allow \texttt{fibonacci} an import that it does not use. We also
rewrite the README so that it no longer mentions the primes.

\end{example}

\begin{verbatim}
$ babregen compact
regen/architecture.toml:37: warning COMPACT02 allowed edge fibonacci -> primes is
    unused
regen/units/primes.toml:66: warning COMPACT07 consumer README.md no longer references
    the unit
regen compact: 2 candidates
[exit 0]
\end{verbatim}

Warnings do not fail the check. \texttt{-\/-strict} makes them fail it:

\begin{verbatim}
$ babregen check --strict
... (2 lines before)
regen check: 2 units, 0 errors, 2 warnings
[exit 1]
\end{verbatim}

The last step is to make the check a habit. \texttt{mathregen} runs it
in \texttt{make\ guard}, with \texttt{-\/-strict}, next to the linter.
The check takes a fraction of a second, so it belongs before every push.
\texttt{mutate}, \texttt{score} and \texttt{local} do not: they run
tests or call a model, and they are for the moments when a unit is
examined or replaced.

\begin{verbatim}
# Makefile, lines 17 to 18
regen-check:  ## Check regen/ against the source tree
    $(RUN) babregen check --strict
# Makefile, lines 37 to 37
guard: sync regen-check lint  ## Compliance checks before a push: regen spec and lint
\end{verbatim}

\subsection{A third function}\label{sec-tut-third}

The package will grow. Adding a function repeats the steps of this
appendix in small. We start one, a module for factorials, and stop after
the first step to see what the check says.

\begin{example}[A new module without a place in the
architecture]\protect\hypertarget{exm-tut-third}{}\label{exm-tut-third}

~

\end{example}

\begin{verbatim}
$ babregen check
src/mathregen/factorial.py: error COMPILE03 subpackage 'factorial' is not listed in
    [allowed] of architecture.toml
regen check: 2 units, 1 errors, 0 warnings
[exit 1]
\end{verbatim}

\texttt{COMPILE03}: the module is not part of the architecture. From
here the list is:

\begin{enumerate}
\def\labelenumi{\arabic{enumi}.}
\tightlist
\item
  Add \texttt{factorial\ =\ {[}{]}} to \texttt{{[}allowed{]}} in
  \texttt{regen/architecture.toml}, and \texttt{"factorial"} to the list
  of \texttt{"\_\_init\_\_"} if the package re-exports the name
  (Section~\ref{sec-tut-architecture}).
\item
  Run \texttt{babregen\ new-unit\ factorial\ -\/-pace\ fast}
  (Section~\ref{sec-tut-new-unit}). A module that other units import
  would declare \texttt{slow}; the policy of that pace asks for a
  decision record with every change, and a slow unit may not depend on a
  fast one.
\item
  Write \texttt{tests/test\_regen\_eval\_factorial.py} with the marker
  \texttt{regen}, and the intent items that name its tests
  (Section~\ref{sec-tut-evaluations}). The questions are the same as
  before: what is \texttt{factorial(0)}, and what happens for a negative
  number, a \texttt{bool}, a \texttt{float}?
\item
  Cite a decision. \texttt{integers-only} already covers the argument
  type, so no new record is needed unless a new choice is made
  (Section~\ref{sec-tut-decisions}).
\item
  Run \texttt{babregen\ check},
  \texttt{babregen\ deletion-test\ factorial} and
  \texttt{babregen\ mutate\ factorial}, and commit
  (Section~\ref{sec-tut-deletion}, Section~\ref{sec-tut-measures}).
\end{enumerate}

No existing record changes except two lines in the architecture. That is
what the split into one module per function buys.

\subsection{Summary}\label{sec-tut-summary}

\begin{longtable}[]{@{}
  >{\raggedright\arraybackslash}p{(\linewidth - 4\tabcolsep) * \real{0.3333}}
  >{\raggedright\arraybackslash}p{(\linewidth - 4\tabcolsep) * \real{0.3333}}
  >{\raggedright\arraybackslash}p{(\linewidth - 4\tabcolsep) * \real{0.3333}}@{}}
\caption{The steps of the worked example and the primitives that each
step serves. Table~\ref{tbl-primitives} names only the commands that
show a record.}\label{tbl-tut-summary}\tabularnewline
\toprule\noalign{}
\begin{minipage}[b]{\linewidth}\raggedright
step
\end{minipage} & \begin{minipage}[b]{\linewidth}\raggedright
command
\end{minipage} & \begin{minipage}[b]{\linewidth}\raggedright
primitive
\end{minipage} \\
\midrule\noalign{}
\endfirsthead
\toprule\noalign{}
\begin{minipage}[b]{\linewidth}\raggedright
step
\end{minipage} & \begin{minipage}[b]{\linewidth}\raggedright
command
\end{minipage} & \begin{minipage}[b]{\linewidth}\raggedright
primitive
\end{minipage} \\
\midrule\noalign{}
\endhead
\bottomrule\noalign{}
\endlastfoot
state the shape of the package & edit \texttt{regen/architecture.toml} &
compilation, pace \\
draft a unit record & \texttt{babregen\ new-unit\ MODULE\ -\/-pace\ P} &
deletion \\
write evaluations and intent & edit
\texttt{tests/test\_regen\_eval\_*.py}, \texttt{regen/units/*.toml} &
evaluation, intent \\
record the reasons & edit \texttt{regen/decisions/*.toml} &
provenance \\
check records against the tree &
\texttt{babregen\ check\ {[}-\/-strict{]}} & all \\
check the preconditions of a deletion &
\texttt{babregen\ deletion-test\ UNIT} & deletion \\
see what an author would receive & \texttt{babregen\ context\ UNIT} &
intent, provenance \\
measure intent coverage and specification size &
\texttt{babregen\ measures} & pace, intent \\
test the evaluations & \texttt{babregen\ mutate\ UNIT} & evaluation \\
judge a rebuild & \texttt{babregen\ score\ UNIT\ FILE} & evaluation,
deletion \\
regenerate with a local model & \texttt{babregen\ local\ UNIT\ ...} &
deletion, provenance \\
give every surviving mutant a verdict & \texttt{babregen\ gate\ ...} &
evaluation \\
say why a rejected run was rejected & \texttt{babregen\ coding\ ...} &
deletion \\
find stale knowledge & \texttt{babregen\ compact} & compaction \\
\end{longtable}

babregen also has commands that write a decision record, an intent item,
or a consumer (\texttt{new-decision}, \texttt{add-intent},
\texttt{add-consumer}, Section~\ref{sec-tool-reference}). The worked
example adds its records as files. The two human steps, \texttt{gate}
and \texttt{coding}, are not replayed here, because a person takes them.
Section~\ref{sec-gate} and Section~\ref{sec-failure} describe them, and
Section~\ref{sec-results} reports them for this package. The check IDs
are listed in Section~\ref{sec-tool-reference}. The schema of the three
kinds of TOML files and the options of \texttt{local} are in the
\href{https://advm1.gm.fh-koeln.de/~bartz/babregen/}{documentation of
babregen}.

\section{Writing the records}\label{sec-hmi}

The records of Section~\ref{sec-records} are authored or approved by the
maintainer. They are text files in TOML ({Preston-Werner et al.} 2021).
This appendix describes that interface in babregen 0.2.0: what is typed,
which input mistakes it invites, what the tool does about them, why the
format is TOML, and which help the tool and the editor give.

The numbers and the trials of this appendix are cells that run babregen
on mathregen, the package of the study, when the document is rendered. A
trial that changes a record works on a throw-away copy. The design
record of the features described here is ADR 0001 in the repository of
babregen (Bartz-Beielstein 2026b).

\subsection{The records in numbers}\label{sec-hmi-records}

\protect\phantomsection\label{hmi-records}
\begin{longtable}[]{@{}
  >{\raggedright\arraybackslash}p{(\linewidth - 2\tabcolsep) * \real{0.7200}}
  >{\raggedleft\arraybackslash}p{(\linewidth - 2\tabcolsep) * \real{0.2800}}@{}}
\caption{The records of mathregen in
numbers.}\label{tbl-hmi-records}\tabularnewline
\toprule\noalign{}
\begin{minipage}[b]{\linewidth}\raggedright
measure
\end{minipage} & \begin{minipage}[b]{\linewidth}\raggedleft
value
\end{minipage} \\
\midrule\noalign{}
\endfirsthead
\toprule\noalign{}
\begin{minipage}[b]{\linewidth}\raggedright
measure
\end{minipage} & \begin{minipage}[b]{\linewidth}\raggedleft
value
\end{minipage} \\
\midrule\noalign{}
\endhead
\bottomrule\noalign{}
\endlastfoot
unit records & 2 (137 lines) \\
decision records & 4 (232 lines) \\
intent items & 6 \\
names of durable evaluations typed into intent items & 12 \\
ids of decision records typed into intent items & 6 \\
declared consumers & 6 \\
share of prose in the unit records & 48 \% \\
share of prose in the decision records & 77 \% \\
multi-line strings & 22 \\
string values wrapped over several lines & 22 of 147 \\
backslashes at the end of a line & 0 \\
\end{longtable}

Table~\ref{tbl-hmi-records} counts what an author of the records has to
type. The numbers are small, because the package is. The share of prose
is the share of the characters of a record file that lie in string
values longer than 80 characters. Two properties decide what follows.
First, the records are to a large part prose, and the rest is mostly
names: every one of the 12 names of a durable evaluation stands in an
evaluation file and again in a unit record. Second, 22 string values are
paragraphs that are wrapped in the file and must be one line in the
bundle. TOML joins the lines of such a string only when each line ends
with a backslash ({Preston-Werner et al.} 2021). babregen asks for none:
its loader folds a prose field, so that a line break and the blanks
around it load as one blank.

\subsection{Input mistakes and what the tool
does}\label{sec-hmi-failures}

Table~\ref{tbl-hmi-failures} lists the input mistakes that such records
invite and what babregen does about each. The first four follow from the
table of check IDs (Section~\ref{sec-tool-reference}) and from the
loader of the records, which accepts no unknown key and no value of a
wrong type. The last three are frictions of writing that the maintainer
of the records named.

\begin{longtable}[]{@{}
  >{\raggedright\arraybackslash}p{(\linewidth - 4\tabcolsep) * \real{0.0600}}
  >{\raggedright\arraybackslash}p{(\linewidth - 4\tabcolsep) * \real{0.4600}}
  >{\raggedright\arraybackslash}p{(\linewidth - 4\tabcolsep) * \real{0.4800}}@{}}
\caption{Input mistakes and frictions of the records, and what babregen
does about them.}\label{tbl-hmi-failures}\tabularnewline
\toprule\noalign{}
\begin{minipage}[b]{\linewidth}\raggedright
\end{minipage} & \begin{minipage}[b]{\linewidth}\raggedright
mistake or friction
\end{minipage} & \begin{minipage}[b]{\linewidth}\raggedright
what babregen does
\end{minipage} \\
\midrule\noalign{}
\endfirsthead
\toprule\noalign{}
\begin{minipage}[b]{\linewidth}\raggedright
\end{minipage} & \begin{minipage}[b]{\linewidth}\raggedright
mistake or friction
\end{minipage} & \begin{minipage}[b]{\linewidth}\raggedright
what babregen does
\end{minipage} \\
\midrule\noalign{}
\endhead
\bottomrule\noalign{}
\endlastfoot
F1 & a name in a record does not match the tree: a durable evaluation, a
decision id, a consumer path, or a public name is misspelled or was
renamed & a finding of the static check at the line of the wrong name,
with a similar valid name for an evaluation, a decision id, or a public
name when one exists (Example~\ref{exm-hmi-suggestion}) \\
F2 & a name is typed twice, in the evaluation file and in the unit
record & nothing prevents it. A mismatch becomes F1. \texttt{add-intent}
checks the names before it writes \\
F3 & a paragraph of prose is wrapped over several lines & the loader
folds it. A blank line inside a prose field is a user error
(Example~\ref{exm-hmi-folded}) \\
F4 & a key is misspelled or unknown & a user error with file, line, and
key, and with a similar key when one exists
(Example~\ref{exm-hmi-key}) \\
F5 & no feedback while typing & schema files for the editor and an
editor task that reads the finding lines
(Section~\ref{sec-hmi-options}) \\
F6 & not knowing what to write: which keys exist, which are required,
how a good intent item reads & \texttt{example\ KIND} prints a complete
record, \texttt{explain\ ID} a wrong and a right fragment.
\texttt{new-unit} and \texttt{new-decision} write a draft \\
F7 & TOML syntax: \texttt{{[}{[}intent{]}{]}}, \texttt{"""}, quotes,
commas in lists & a parse error, mostly with line and column. The
commands that write a record append a complete block \\
\end{longtable}

\begin{example}[Wrapped prose without a
backslash]\protect\hypertarget{exm-hmi-folded}{}\label{exm-hmi-folded}

The cell prints the first two lines of the reason of intent item PRI-2
of the unit \texttt{primes} as the record holds them, the line of the
bundle that carries the reason, and the static check after a blank line
was put into the paragraph on a throw-away copy.

\end{example}

\begin{verbatim}
# regen/units/primes.toml, lines 33 and 34
A table of known values checks only the indices it lists. An implementation with a
fault in its search, for example a sieve with a wrong bound, can pass the table and
$ babregen context primes
...
- reason: A table of known values checks only the indices it lists ...
(a blank line is put between the two lines of the record)
$ babregen check
babregen: regen/units/primes.toml:32: key 'reason' in the [[intent]] item 'PRI-2'
    contains a blank line; a prose field holds one paragraph
[exit code 2]
\end{verbatim}

A wrapped paragraph loads as one line and reaches the bundle as one list
item. A forgotten backslash cannot break a paragraph, because no
backslash is asked for. The mistake that is left, a blank line inside a
paragraph, stops every command that reads the records with a user error
that names the file, the line, and the key.

\begin{example}[One letter, one
suggestion]\protect\hypertarget{exm-hmi-suggestion}{}\label{exm-hmi-suggestion}

The cell removes one letter from the name of a durable evaluation in the
same intent item.

\end{example}

\begin{verbatim}
$ babregen check
regen/units/primes.toml:36: error EVAL03 intent PRI-2: evaluation
    'test_every_result_is_prime_and_no_prime_is_skiped' not found; a similar name
    exists: 'test_every_result_is_prime_and_no_prime_is_skipped'
tests/test_regen_eval_primes.py:29: warning COMPACT05 mathregen.primes: evaluation
    'test_every_result_is_prime_and_no_prime_is_skipped' is not linked to any intent
    item
regen check: 2 units, 1 errors, 1 warnings
[exit code 1]
line of the wrong name in the record: 36
$ babregen check --format json   (the finding EVAL03)
{
 "path": "regen/units/primes.toml",
 "line": 36,
 "level": "error",
 "id": "EVAL03",
 "message": "intent PRI-2: evaluation 'test_every_result_is_prime_and_no_prime_is_skip
    ed' not found",
 "suggestion": "test_every_result_is_prime_and_no_prime_is_skipped"
}
\end{verbatim}

\texttt{EVAL03} stands at line 36, the line of the wrong name, and not
at line 27, where the intent item begins. It states the name that was
meant. The set of valid names is small and closed, here the 6 durable
evaluations of the unit, so a call of the standard library finds it. For
a name that has no close match, such as
\texttt{test\_the\_result\_is\_cached}, the tool suggests nothing. That
is right as well: such a name is a missing evaluation and no typo. The
one mistake still gives two findings, and \texttt{COMPACT05} is not
connected with the first. The JSON document carries the suggestion as a
field of its own, so a program does not have to read it from the
message.

\begin{example}[A misspelled
key]\protect\hypertarget{exm-hmi-key}{}\label{exm-hmi-key}

The cell writes the key \texttt{evaluations} of the unit record as
\texttt{evalutions}.

\end{example}

\begin{verbatim}
$ babregen check
babregen: regen/units/primes.toml:7: unknown key 'evalutions' in a unit record; a
    similar name exists: 'evaluations'
[exit code 2]
\end{verbatim}

The loader stops with a user error and names the file, the line, the
key, and the key that was meant.

\subsection{The format}\label{sec-hmi-format}

The format of the records was compared with three alternatives on the
criteria that the mistakes above suggest (Table~\ref{tbl-hmi-formats}).
The rows on silent and on loud errors are trials with \texttt{tomllib},
\texttt{json}, and PyYAML in the cell \texttt{hmi-formats}.
Configuration languages with a schema of their own, such as CUE, Pkl,
and KDL, were assessed from their documentation only. Each needs an
external tool, and none was pursued.

\begin{longtable}[]{@{}
  >{\raggedright\arraybackslash}p{(\linewidth - 8\tabcolsep) * \real{0.1800}}
  >{\raggedright\arraybackslash}p{(\linewidth - 8\tabcolsep) * \real{0.2400}}
  >{\raggedright\arraybackslash}p{(\linewidth - 8\tabcolsep) * \real{0.2200}}
  >{\raggedright\arraybackslash}p{(\linewidth - 8\tabcolsep) * \real{0.1600}}
  >{\raggedright\arraybackslash}p{(\linewidth - 8\tabcolsep) * \real{0.2000}}@{}}
\caption{The formats compared for the
records.}\label{tbl-hmi-formats}\tabularnewline
\toprule\noalign{}
\begin{minipage}[b]{\linewidth}\raggedright
criterion
\end{minipage} & \begin{minipage}[b]{\linewidth}\raggedright
TOML
\end{minipage} & \begin{minipage}[b]{\linewidth}\raggedright
YAML
\end{minipage} & \begin{minipage}[b]{\linewidth}\raggedright
JSON
\end{minipage} & \begin{minipage}[b]{\linewidth}\raggedright
Markdown with front matter
\end{minipage} \\
\midrule\noalign{}
\endfirsthead
\toprule\noalign{}
\begin{minipage}[b]{\linewidth}\raggedright
criterion
\end{minipage} & \begin{minipage}[b]{\linewidth}\raggedright
TOML
\end{minipage} & \begin{minipage}[b]{\linewidth}\raggedright
YAML
\end{minipage} & \begin{minipage}[b]{\linewidth}\raggedright
JSON
\end{minipage} & \begin{minipage}[b]{\linewidth}\raggedright
Markdown with front matter
\end{minipage} \\
\midrule\noalign{}
\endhead
\bottomrule\noalign{}
\endlastfoot
wrapped prose & \texttt{"""}. Plain TOML keeps the line breaks unless
each line ends with a backslash. The loader of babregen folds them &
folded block, no escape & one long line & prose is the body \\
list of records & \texttt{{[}{[}intent{]}{]}}, good at one level &
nested lists & possible, noisy & needs rules for headings \\
silent errors & none found in the trials & a doubled key; a list item
indented differently; \texttt{0.10} loads as 0.1, \texttt{yes} as true &
a doubled key & those of the front matter \\
loud errors & a doubled key, a missing quote or comma & wrong
indentation of a key & a missing comma, a comment & \\
comments & yes & yes & no & free text \\
parser in the standard library of Python & \texttt{tomllib}, reads only
& no & \texttt{json}, reads and writes & no \\
schema support in the editor & JSON Schema through an extension & JSON
Schema through an extension & built in & front matter only \\
\end{longtable}

\begin{longtable}[]{@{}
  >{\raggedright\arraybackslash}p{(\linewidth - 6\tabcolsep) * \real{0.4000}}
  >{\raggedleft\arraybackslash}p{(\linewidth - 6\tabcolsep) * \real{0.2000}}
  >{\raggedleft\arraybackslash}p{(\linewidth - 6\tabcolsep) * \real{0.2000}}
  >{\raggedleft\arraybackslash}p{(\linewidth - 6\tabcolsep) * \real{0.2000}}@{}}
\caption{The unit record of \protect\texttt{primes} in three formats.
The three files load to the same value once the prose is
folded.}\label{tbl-hmi-sizes}\tabularnewline
\toprule\noalign{}
\begin{minipage}[b]{\linewidth}\raggedright
format
\end{minipage} & \begin{minipage}[b]{\linewidth}\raggedleft
lines
\end{minipage} & \begin{minipage}[b]{\linewidth}\raggedleft
characters
\end{minipage} & \begin{minipage}[b]{\linewidth}\raggedleft
longest line
\end{minipage} \\
\midrule\noalign{}
\endfirsthead
\toprule\noalign{}
\begin{minipage}[b]{\linewidth}\raggedright
format
\end{minipage} & \begin{minipage}[b]{\linewidth}\raggedleft
lines
\end{minipage} & \begin{minipage}[b]{\linewidth}\raggedleft
characters
\end{minipage} & \begin{minipage}[b]{\linewidth}\raggedleft
longest line
\end{minipage} \\
\midrule\noalign{}
\endhead
\bottomrule\noalign{}
\endlastfoot
TOML & 68 & 2,225 & 88 \\
YAML, prose as folded blocks & 63 & 2,155 & 94 \\
JSON & 72 & 2,580 & 238 \\
\end{longtable}

JSON is not an option for records that hold so much prose: a paragraph
becomes one line, the longest of 238 characters
(Table~\ref{tbl-hmi-sizes}), and a reason cannot be commented. YAML is
the one serious competitor. Its folded block is what the records need, a
wrapped paragraph without an escape.

Four arguments speak for TOML.

\begin{enumerate}
\def\labelenumi{\arabic{enumi}.}
\tightlist
\item
  First, it is the only format of Table~\ref{tbl-hmi-formats} that is
  written for people and that the standard library of Python reads
  (Hukkinen and Jain 2022). Python packaging chose it three times, for
  the build requirements (Cannon et al. 2016), for the metadata of a
  script (Lev 2023), and for lock files (Cannon 2024). The last of these
  considered JSON and found it less readable for people without a
  sufficient gain for programs.
\item
  Second, its errors are loud where those of YAML are silent, as the
  trials show. The Kubernetes project states the same about indentation:
  a wrongly indented file can stay valid and describe another object
  than intended (Verma 2026).
\item
  Third, the records need one level of \texttt{{[}{[}...{]}{]}} tables
  only. The usual criticism of TOML concerns deeper nesting.
\item
  Fourth, the one controlled study of the usability of such formats that
  we found, with 215 participants and seven formats, observed ``no
  systematic performance advantage for any single format, aside from
  some minor differences in certain editing tasks'', where YAML did
  better in some, and concludes that better linters, formatters, and
  editor support may yield more than a new syntax (He et al. 2026). Its
  tasks did not include wrapped prose.
\end{enumerate}

Three arguments speak against it.

\begin{enumerate}
\def\labelenumi{\arabic{enumi}.}
\tightlist
\item
  First, plain TOML does not fold a wrapped paragraph. babregen folds in
  its loader, so a prose field means one paragraph only to babregen:
  another TOML parser reads the same file with its line breaks. The rule
  is one line of the schema, and it is the price of records without
  backslashes.
\item
  Second, language models mostly wrote TOML worse than JSON or YAML in
  the benchmark StructEval (Yang et al. 2025): GPT-4o reached 90.2
  points for TOML against 99.4 for JSON and 97.7 for YAML, and the six
  open models averaged 8.6 points for TOML. The score mixes the syntax
  with the content of the output, the models are older than the
  generators of this study, and the task is the writing of a new file,
  not the editing of a record.
\item
  Third, the standard library does not write TOML. The writer was left
  out on purpose (Hukkinen and Jain 2022). The commands of babregen that
  write a record therefore follow one rule: a command creates a file or
  appends one complete \texttt{{[}{[}...{]}{]}} block at the end of a
  record, and it never rewrites existing text. TOML appends the tables
  of an array in the order in which they stand in the file
  ({Preston-Werner et al.} 2021), and intent items and consumers are
  such arrays. A library that rewrites a file and keeps its comments
  exists (Eustace 2026) and would be the first runtime dependency of the
  tool. Tools for decision records, such as MADR (Kopp et al. 2018),
  keep their prose in Markdown.
\end{enumerate}

YAML would bring the folded block at the price of a runtime dependency
or a parser of our own and of a new class of silent errors. Markdown for
the decision records alone would put two formats into one directory and
would need a second loader for fields that the static check reads
(Section~\ref{sec-provenance}). We found no evidence on how the formats
differ in diffs and merges, and none on language models that edit
existing records.

\subsection{Help from the editor and from the
tool}\label{sec-hmi-options}

The frictions F5 to F7 are frictions of the editor.
Table~\ref{tbl-hmi-help} lists what the release holds against them and
against the mistakes F1 to F4.

\begin{longtable}[]{@{}
  >{\raggedright\arraybackslash}p{(\linewidth - 4\tabcolsep) * \real{0.3000}}
  >{\raggedright\arraybackslash}p{(\linewidth - 4\tabcolsep) * \real{0.5200}}
  >{\raggedright\arraybackslash}p{(\linewidth - 4\tabcolsep) * \real{0.1800}}@{}}
\caption{What babregen holds for the author of a
record.}\label{tbl-hmi-help}\tabularnewline
\toprule\noalign{}
\begin{minipage}[b]{\linewidth}\raggedright
help
\end{minipage} & \begin{minipage}[b]{\linewidth}\raggedright
what the author of a record gets
\end{minipage} & \begin{minipage}[b]{\linewidth}\raggedright
against
\end{minipage} \\
\midrule\noalign{}
\endfirsthead
\toprule\noalign{}
\begin{minipage}[b]{\linewidth}\raggedright
help
\end{minipage} & \begin{minipage}[b]{\linewidth}\raggedright
what the author of a record gets
\end{minipage} & \begin{minipage}[b]{\linewidth}\raggedright
against
\end{minipage} \\
\midrule\noalign{}
\endhead
\bottomrule\noalign{}
\endlastfoot
\texttt{schema\ KIND} and the three schema files in
\texttt{regen/schemas/} & a JSON Schema for each kind of record, derived
from the table of fields that the loader reads. With a TOML extension
(Tamás 2025) the editor marks an unknown key or a wrong type & F4, F5,
F7 \\
an editor task and snippets in the repository & the findings of the
static check as markers in the files, and a template for an intent item,
a consumer, and a decision record & F5, F6 \\
a similar name in a finding and in the error of an unknown key & the
name that was probably meant, at the line of the wrong name & F1, F4 \\
\texttt{explain\ ID} and \texttt{example\ KIND} & the condition of a
check with a wrong and a right fragment, and one complete, commented
record of a kind & F6 \\
\texttt{new-unit}, \texttt{new-decision}, \texttt{add-intent}, and
\texttt{add-consumer} & a draft or a complete block that the strict
loader has read before it is written. The names are checked first, and
\texttt{-\/-dry-run} shows the result & F1, F2, F6, F7 \\
\texttt{check\ -\/-format\ json} & the findings, their suggestions, and
a summary as one document for agents and pipelines & F5 \\
\end{longtable}

A test of the tool compares its three schema files with the output of
the command, so the files cannot go stale. The cell \texttt{hmi-help}
repeats the comparison for the copies that mathregen keeps and tries the
two patterns of the editor task on the finding lines of this appendix:
every line matches, and the summary line does not. The format of a
finding line is fixed (Section~\ref{sec-tool-reference}), so such a
pattern does not go stale either. A schema cannot know that a durable
evaluation exists. F1 and F2, which concern names in the tree, stay with
the static check. A similar name is stated and not asked: the guide to
the diagnostics of the Rust compiler asks for a statement, such as that
a similar name exists, and not for a question (The Rust Project
Developers 2026). The evidence that better messages help is mixed and
comes mostly from novices (Becker et al. 2019). \texttt{explain} and
\texttt{example} serve agents as well as people. A tool that teaches its
format on request may serve an agent better than a long instruction
file: in one evaluation, context files did not generally improve the
success rate of coding agents and raised the cost of inference by over
20 \% (Gloaguen et al. 2026).

Four things are not built. A fix that applies a suggestion is tempting
and mostly unsafe, because a misspelled name and a missing evaluation
look the same to the tool. The double typing of F2 could end if the link
from an intent item to its durable evaluations lived in the evaluation
file. A record would then no longer show its judges, and every bundle
would change. A language server would give findings while typing and
would be a second program with its own releases. A form or a graphical
interface does not help with what is hard about a record, the prose of
an invariant and the names in the tree. One risk of the schema files
should be named. The owner of the engine behind the common TOML
extension stepped down as its maintainer in December 2024 (Tamás 2024).
The schema files therefore use no key that is specific to one extension,
so that any validator of JSON Schema can read them.

This appendix has limits of its own. The frictions F5 to F7 are the
report of one maintainer. Errors that were corrected before a commit
left no trace, so the appendix gives no rate for any mistake, and it
cannot say whether the help of the tool lowers one. Completion and hover
inside the editor were not tried here. The benchmark on language models
covers the writing of new files and not the editing of records.

\section{A hypothetical example: from the answers to the next
step}\label{sec-hypothetical}

\begin{tcolorbox}[enhanced jigsaw, arc=.35mm, bottomrule=.15mm, bottomtitle=1mm, breakable, colback=white, colbacktitle=quarto-callout-warning-color!10!white, colframe=quarto-callout-warning-color-frame, coltitle=black, left=2mm, leftrule=.75mm, opacityback=0, opacitybacktitle=0.6, rightrule=.15mm, title=\textcolor{quarto-callout-warning-color}{\faExclamationTriangle}\hspace{0.5em}{Invented data}, titlerule=0mm, toprule=.15mm, toptitle=1mm]

Every name, number, and answer of the example in this appendix is
invented for illustration. No person rated a mutant or coded a run for
it, no command wrote it, and nothing of it is a result of the study. The
protocol record of the study states that no human rating is simulated,
and this appendix simulates none: it shows how answers of this kind
would be read.

\end{tcolorbox}

The study ends with three survivors, all rated \emph{equivalent}, and
one rejected run, coded \emph{generator error}
(Section~\ref{sec-results}). By Table~\ref{tbl-consequences} these
answers ask for no change of the records, so the study cannot show what
a maintainer does when an answer asks for one. This appendix shows it on
an invented package, \texttt{shopcart}, with the two units
\texttt{pricing} and \texttt{inventory} and the two generators (x) and
(y). Its gate has 12 survivors, and its coding has six rejected runs.

\begin{figure}[!htb]

\centering{

\includegraphics[width=6.5in,height=\textheight,keepaspectratio]{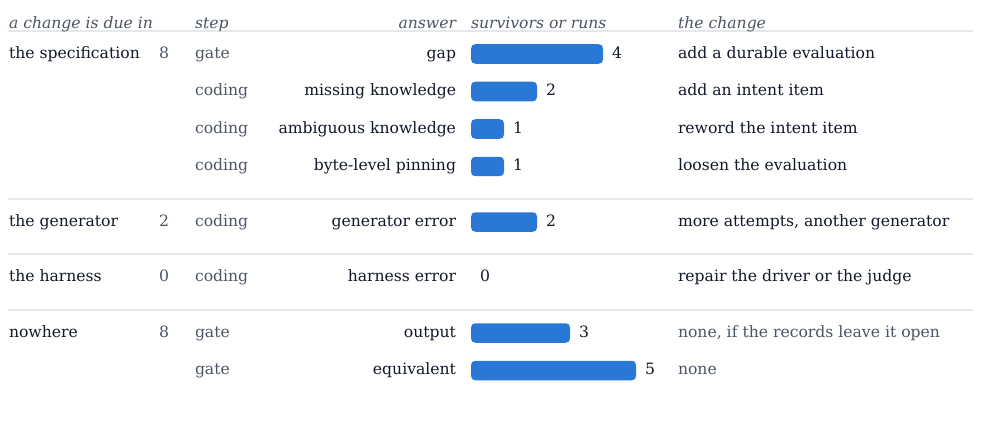}

}

\caption{\label{fig-answers}Hypothetical example, invented data: where
the answers of the two human steps point. One row per answer, grouped by
the place in which a change is due, with the number of survivors (gate)
or of rejected runs (coding) that received the answer, and the change
that Table~\ref{tbl-consequences} names for it. The number behind a
place is the sum of its rows.}

\end{figure}%

\begin{longtable}[]{@{}
  >{\raggedright\arraybackslash}p{(\linewidth - 14\tabcolsep) * \real{0.1100}}
  >{\raggedright\arraybackslash}p{(\linewidth - 14\tabcolsep) * \real{0.0900}}
  >{\raggedright\arraybackslash}p{(\linewidth - 14\tabcolsep) * \real{0.1200}}
  >{\raggedright\arraybackslash}p{(\linewidth - 14\tabcolsep) * \real{0.1200}}
  >{\raggedright\arraybackslash}p{(\linewidth - 14\tabcolsep) * \real{0.1200}}
  >{\raggedright\arraybackslash}p{(\linewidth - 14\tabcolsep) * \real{0.1100}}
  >{\raggedright\arraybackslash}p{(\linewidth - 14\tabcolsep) * \real{0.1200}}
  >{\raggedright\arraybackslash}p{(\linewidth - 14\tabcolsep) * \real{0.2100}}@{}}
\caption{Hypothetical example, invented data: the ratings of the 12
survivors in their order. \emph{Shown} is the verdict of the proposal on
the second page. \emph{Second rating} says whether the second rating
named that verdict already in the blind rating, changed to it, or did
not follow it. The survivor M021 of \protect\texttt{pricing} is the one
drawn survivor: its shown proposal was planted, and its third rating,
after the reveal, is its verdict.}\label{tbl-hypo-gate}\tabularnewline
\toprule\noalign{}
\begin{minipage}[b]{\linewidth}\raggedright
unit
\end{minipage} & \begin{minipage}[b]{\linewidth}\raggedright
mutant
\end{minipage} & \begin{minipage}[b]{\linewidth}\raggedright
blind
\end{minipage} & \begin{minipage}[b]{\linewidth}\raggedright
shown
\end{minipage} & \begin{minipage}[b]{\linewidth}\raggedright
second
\end{minipage} & \begin{minipage}[b]{\linewidth}\raggedright
third
\end{minipage} & \begin{minipage}[b]{\linewidth}\raggedright
verdict
\end{minipage} & \begin{minipage}[b]{\linewidth}\raggedright
second rating
\end{minipage} \\
\midrule\noalign{}
\endfirsthead
\toprule\noalign{}
\begin{minipage}[b]{\linewidth}\raggedright
unit
\end{minipage} & \begin{minipage}[b]{\linewidth}\raggedright
mutant
\end{minipage} & \begin{minipage}[b]{\linewidth}\raggedright
blind
\end{minipage} & \begin{minipage}[b]{\linewidth}\raggedright
shown
\end{minipage} & \begin{minipage}[b]{\linewidth}\raggedright
second
\end{minipage} & \begin{minipage}[b]{\linewidth}\raggedright
third
\end{minipage} & \begin{minipage}[b]{\linewidth}\raggedright
verdict
\end{minipage} & \begin{minipage}[b]{\linewidth}\raggedright
second rating
\end{minipage} \\
\midrule\noalign{}
\endhead
\bottomrule\noalign{}
\endlastfoot
\texttt{pricing} & M002 & gap & gap & gap & -- & gap & matched before \\
\texttt{pricing} & M005 & equivalent & equivalent & equivalent & -- &
equivalent & matched before \\
\texttt{pricing} & M007 & output & gap & gap & -- & gap & changed to
it \\
\texttt{pricing} & M011 & equivalent & equivalent & equivalent & -- &
equivalent & matched before \\
\texttt{pricing} & M014 & output & output & output & -- & output &
matched before \\
\texttt{pricing} & M019 & gap & gap & gap & -- & gap & matched before \\
\texttt{pricing} & M021 & equivalent & gap & gap & equivalent &
equivalent & changed to it \\
\texttt{inventory} & M003 & gap & output & gap & -- & gap & not
followed \\
\texttt{inventory} & M004 & equivalent & equivalent & equivalent & -- &
equivalent & matched before \\
\texttt{inventory} & M008 & output & output & output & -- & output &
matched before \\
\texttt{inventory} & M009 & equivalent & equivalent & equivalent & -- &
equivalent & matched before \\
\texttt{inventory} & M012 & output & equivalent & output & -- & output &
not followed \\
\end{longtable}

Figure~\ref{fig-answers} gives the first reading. Of the 18 answers,
eight ask for a change of the specification, two point at the generator,
and eight ask for no change. Three of these eight are output verdicts,
for which the maintainer first checks whether the records leave the
behaviour open. Table~\ref{tbl-hypo-gate} gives the second reading,
which the counts hide: how each verdict came about. Four cases show what
a maintainer does next.

\begin{enumerate}
\def\labelenumi{\arabic{enumi}.}
\tightlist
\item
  \emph{A gap.} The survivors M002, M007, and M019 of \texttt{pricing}
  and M003 of \texttt{inventory} are gaps. Each marks a promise that no
  durable evaluation guards, so the accepted rebuilds of the two units
  do not show that the promise is kept. The maintainer writes a durable
  evaluation for each promise, names it in the intent item, and runs
  \texttt{babregen\ mutate} again. The four mutants are then killed, or
  the new evaluation is not the one that was missing. The mutation score
  is read against the mutants that are not equivalent: the evaluations
  of \texttt{pricing} kill 33 of 37 before the change, since three of
  its seven survivors are equivalent.
\item
  \emph{Missing knowledge.} Both generators fail the same evaluation of
  \texttt{inventory} with seed 2, and the coder finds no place in the
  records that states the demand. Two generators that miss the same
  point are a hint, and the category is the judgement. The maintainer
  adds the intent item, or removes the demand from the evaluation if
  nobody relies on it. The bundle of \texttt{inventory} then changes,
  the static check reports PROV08 for the unit, and the runs of
  \texttt{inventory} are repeated with the same seeds.
\item
  \emph{Generator error.} The two runs of \texttt{pricing} that received
  this category come from generator (x), and generator (y) passes the
  same evaluations. The records state the broken promise, so they stay
  as they are. The maintainer allows more attempts or uses the other
  generator for this unit.
\item
  \emph{A rating that followed a plant.} The rater called M021
  \emph{equivalent} in the blind rating, changed to \emph{gap} on the
  second page, which showed a planted proposal, and returned to
  \emph{equivalent} after the reveal. The verdict is the third rating,
  so the plant did no harm to the result. It did show that a proposal
  moved this rater once. The second change, M007, followed an honest
  proposal and stands, and the reader of the archive can see that it
  rests on a proposal and not on the blind judgement.
\end{enumerate}

The answers can start more than these repairs. A new study can take the
gaps as its hypothesis: after the evaluations are added, the rebuilds
that were accepted before are scored again with
\texttt{babregen\ score}, and the number that now fail measures how much
the acceptance rule had let pass. The category \emph{missing knowledge}
gives a second design, a comparison of the runs of \texttt{inventory}
before and after the new intent item with the same generators and seeds.
With 12 survivors one proposal is drawn, and from 15 on a second one.
Two raters on such material give an agreement, and several plants give a
first count of followed proposals.

The limits are those of the study (Section~\ref{sec-threats}). One rater
gives counts and no rate, and one plant that was followed is an
observation. The tool acts on no answer. Every change above is a
decision of the maintainer. Also, a verdict belongs to the specification
that it judged.

\section{A report of the results in the tool}\label{sec-hypo-report}

Reading the result files by hand is the step that a user of the tool
would repeat. This report does it with a module of its own. Release
0.5.0 of babregen, published on 1 December 2026 after the study, has two
steps for it, \texttt{babregen\ gate\ report\ DIR} and
\texttt{babregen\ coding\ report\ DIR}. Both are deterministic, read the
manifest and the result files only, never a file below \texttt{sealed/}
or \texttt{proposals/}, and write nothing. A report prints the counts
per unit, the answers of every survivor or run in their order, and a
block that names for each answer the change that is due, with the
survivors or runs it concerns. It is no check: \texttt{gate\ check} and
\texttt{coding\ check} verify the files that a report trusts.

This report runs release 0.2.0 only, so the output below is a literal.
It was produced on 1 December 2026 with the code of release 0.5.0 on the
archive of the study, whose data are real, and the cell after it
compares every number of it with the archive.

\begingroup\fontsize{7.5}{9}\selectfont

\begin{Shaded}
\begin{Highlighting}[]
\NormalTok{gate regen/runs/study/gate: 3 survivors, 1 drawn, rater bartzbeielstein}
\NormalTok{a report is no check: \textasciigrave{}babregen gate check regen/runs/study/gate\textasciigrave{} verifies the files}

\NormalTok{verdicts per unit}
\NormalTok{unit                 mutants  killed  survivors  gap  output  equivalent}
\NormalTok{mathregen.fibonacci  11       11      0          0    0       0}
\NormalTok{mathregen.primes     31       28      3          0    0       3}

\NormalTok{ratings in their order}
\NormalTok{survivor               blind       shown       second      third       verdict     second rating}
\NormalTok{mathregen.primes M003  equivalent  equivalent  equivalent  {-}           equivalent  matched before}
\NormalTok{mathregen.primes M004  equivalent  output      equivalent  equivalent  equivalent  not followed, planted}
\NormalTok{mathregen.primes M010  equivalent  equivalent  equivalent  {-}           equivalent  matched before}

\NormalTok{what follows}
\NormalTok{gap (0): a promise is not guarded, and an accepted rebuild may break it}
\NormalTok{  change: add a durable evaluation for the promise, then run \textasciigrave{}babregen mutate\textasciigrave{} again}
\NormalTok{output (0): the behaviour is free}
\NormalTok{  change: nothing; if a consumer relies on the behaviour, add an intent item and an evaluation}
\NormalTok{equivalent (3): no evaluation through the public names can detect the mutant}
\NormalTok{  change: nothing; the mutation score stays as it is}
\NormalTok{  for: mathregen.primes M003, mathregen.primes M004, mathregen.primes M010}

\NormalTok{a change is due in: specification 0, generator 0, harness 0, nowhere 3}
\end{Highlighting}
\end{Shaded}

\endgroup

For the coding of the study the report ends with the line
\texttt{a\ change\ is\ due\ in:\ specification\ 0,\ generator\ 1,\ harness\ 0,\ nowhere\ 0}.

The two reports say for the study what
Section~\ref{sec-disc-consequences} says in prose: no answer asks for a
change of the records. On answers such as those of the example in
Section~\ref{sec-hypothetical}, the last block of a report would name
the four gaps with their change and the four runs whose category points
at the specification. The report of release 0.5.0 calls the behaviour
behind the verdict output free. This wording predates the distinction
between a difference that the records leave open and one on which they
are silent (Section~\ref{sec-undecided}, Table~\ref{tbl-consequences}).
The benefit is small and concrete. The maintainer reads one screen where
this report needed a module, and the mapping from an answer to a change
is stated in one place of the tool, next to the vocabulary that defines
the answers.

\end{document}

%% file: dirtree.tex
 \def\fileversion{0.32}
 \def\filedate{2012/12/11}
\message{`dirtree' v\fileversion, \filedate\space (jcc)}
\edef\DTAtCode{\the\catcode`\@}
\catcode`\@=11
\long\def\DT@loop#1\DT@repeat{%
  \def\DT@iterate{#1\relax\expandafter\DT@iterate\fi}%
  \DT@iterate
  \let\DT@iterate\relax
}
\let\DT@repeat=\fi
\expandafter\ifx\csname DT@fromsty\endcsname\relax
  \def\@namedef#1{\expandafter\def\csname #1\endcsname}
  \def\@nameuse#1{\csname #1\endcsname}
  \long\def\@gobble#1{}
\fi
\def\@nameedef#1{\expandafter\edef\csname #1\endcsname}
\newdimen\DT@offset \DT@offset=0.2em
\newdimen\DT@width \DT@width=1em
\newdimen\DT@sep \DT@sep=0.2em
\newdimen\DT@all
\DT@all=\DT@offset
\advance\DT@all \DT@width
\advance\DT@all \DT@sep
\newdimen\DT@rulewidth \DT@rulewidth=0.4pt
\newdimen\DT@dotwidth \DT@dotwidth=1.6pt
\newdimen\DTbaselineskip \DTbaselineskip=\baselineskip
\newcount\DT@counti
\newcount\DT@countii
\newcount\DT@countiii
\newcount\DT@countiv
\def\DTsetlength#1#2#3#4#5{%
  \DT@offset=#1\relax
  \DT@width=#2\relax
  \DT@sep=#3\relax
  \DT@all=\DT@offset
  \advance\DT@all by\DT@width
  \advance\DT@all by\DT@sep
  \DT@rulewidth=#4\relax
  \DT@dotwidth=#5\relax
}
\expandafter\ifx\csname DT@fromsty\endcsname\relax
  \def\DTstyle{\tt}
  \def\DTstylecomment{\rm}
\else
  \def\DTstyle{\ttfamily}
  \def\DTstylecomment{\rmfamily}
\fi
\def\DTcomment#1{%
  \kern\parindent\dotfill
  {\DTstylecomment{#1}}%
}
\newdimen\DT@indent
\newdimen\DT@parskip
\newdimen\DT@baselineskip
\def\dirtree#1{%
  \DT@indent=\parindent
  \parindent=\z@
  \DT@parskip=\parskip
  \parskip=\z@
  \DT@baselineskip=\baselineskip
  \baselineskip=\DTbaselineskip
  \let\DT@strut=\strut
  \def\strut{\vrule width\z@ height0.7\baselineskip depth0.3\baselineskip}%
  \DT@counti=\z@
  \let\next\DT@readarg
  \next#1\@nil
  \dimen\z@=\hsize
  \advance\dimen\z@ -\DT@offset
  \advance\dimen\z@ -\DT@width
  \setbox\z@=\hbox to\dimen\z@{%
    \hsize=\dimen\z@
    \vbox{\@nameuse{DT@body@1}}%
  }%
  \dimen\z@=\ht\z@
  \advance\dimen0 by\dp\z@
  \advance\dimen0 by-0.7\baselineskip
  \ht\z@=0.7\baselineskip
  \dp\z@=\dimen\z@
  \par\leavevmode
  \kern\DT@offset
  \kern\DT@width
  \box\z@
  \endgraf
  \DT@countii=\@ne
  \DT@countiii=\z@
  \dimen3=\dimen\z@
  \@namedef{DT@lastlevel@1}{-0.7\baselineskip}%
  \loop
  \ifnum\DT@countii<\DT@counti
    \advance\DT@countii \@ne
    \advance\DT@countiii \@ne
    \dimen\z@=\@nameuse{DT@level@\the\DT@countii}\DT@all
    \advance\dimen\z@ by\DT@offset
    \advance\dimen\z@ by-\DT@all
    \leavevmode
    \kern\dimen\z@
    \DT@countiv=\DT@countii
    \count@=\z@
    \DT@loop
      \advance\DT@countiv \m@ne
      \ifnum\@nameuse{DT@level@\the\DT@countiv} >
        \@nameuse{DT@level@\the\DT@countii}\relax
      \else
        \count@=\@ne
      \fi
    \ifnum\count@=\z@
    \DT@repeat
    \edef\DT@hsize{\the\hsize}%
    \count@=\@nameuse{DT@level@\the\DT@countii}\relax
    \dimen\z@=\count@\DT@all
    \advance\hsize by-\dimen\z@
    \setbox\z@=\vbox{\@nameuse{DT@body@\the\DT@countii}}%
    \hsize=\DT@hsize
    \dimen\z@=\ht\z@
    \advance\dimen\z@ by\dp\z@
    \advance\dimen\z@ by-0.7\baselineskip
    \ht\z@=0.7\baselineskip
    \dp\z@=\dimen\z@
    \@nameedef{DT@lastlevel@\the\DT@countii}{\the\dimen3}%
    \advance\dimen3 by\dimen\z@
    \advance\dimen3 by0.7\baselineskip
    \dimen\z@=\@nameuse{DT@lastlevel@\the\DT@countii}\relax
    \advance\dimen\z@ by-\@nameuse{DT@lastlevel@\the\DT@countiv}\relax
    \advance\dimen\z@ by0.3\baselineskip
    \ifnum\@nameuse{DT@level@\the\DT@countiv} <
        \@nameuse{DT@level@\the\DT@countii}\relax
      \advance\dimen\z@ by-0.5\baselineskip
    \fi
    \kern-0.5\DT@rulewidth
    \hbox{\vbox to\z@{\vss\hrule width\DT@rulewidth height\dimen\z@}}%
    \kern-0.5\DT@rulewidth
    \kern-0.5\DT@dotwidth
    \vrule width\DT@dotwidth height0.5\DT@dotwidth depth0.5\DT@dotwidth
    \kern-0.5\DT@dotwidth
    \vrule width\DT@width height0.5\DT@rulewidth depth0.5\DT@rulewidth
    \kern\DT@sep
    \box\z@
    \endgraf
  \repeat
  \parindent=\DT@indent
  \parskip=\DT@parskip
  \baselineskip=\DT@baselineskip
  \let\strut\DT@strut
}
\def\DT@readarg.#1 #2. #3\@nil{%
  \advance\DT@counti \@ne
  \@namedef{DT@level@\the\DT@counti}{#1}%
  \@namedef{DT@body@\the\DT@counti}{\strut{\DTstyle{#2}\strut}}%
  \ifx\relax#3\relax
    \let\next\@gobble
  \fi
  \next#3\@nil
}
\catcode`\@=\DTAtCode\relax
 